\documentclass{article}

\usepackage{titlesec}
\usepackage[left=3cm, right=2cm, top=2.5cm, bottom= 2.5cm]{geometry}
\usepackage{setspace} 
\usepackage{amsmath}
\usepackage{graphicx}
\usepackage[colorlinks=true, allcolors=blue]{hyperref}
\usepackage{enumerate}
\usepackage{enumitem}
\usepackage{tabularx}
\usepackage[round]{natbib}
\usepackage{accents}
\usepackage{parskip}
\newcommand{\indep}{\perp \!\!\!\perp}
\usepackage{xcolor, color, soul}
\sethlcolor{yellow}
\usepackage{amsfonts}
\usepackage{bbm}
\usepackage{float}
\graphicspath{ {./images/} }
\usepackage{authblk}
\usepackage{accents}
\usepackage{graphicx}
\usepackage{booktabs}
\usepackage{hyperref}

\DeclareMathOperator{\logit}{logit}
\DeclareMathOperator{\E}{\mathbb{E}}

\DeclareMathOperator{\Var}{Var}

\title{Robust Covariate Adjustment in Multi-Center Randomized Trials}
\author[1,2]{Muluneh Alene}
\author[1]{Stijn Vansteelandt}
\author[1,3]{Kelly Van Lancker}
\affil[1]{Department of Mathematics, Computer Science and Statistics, Ghent University, Ghent, Belgium}
\affil[2]{Department of Public Health, Debre Markos University, Debre Markos, Ethiopia}
\affil[3]{Department of Mathematics and Data Science, Vrije Universiteit Brussel, Brussels, Belgium}

\begin{document}
	\maketitle

\begin{center}
\textbf{Abstract}\\
\end{center}
 
\begin{spacing}{1}
Augmented inverse probability weighting and G-computation with canonical generalized linear models have become increasingly popular for estimating the average treatment effect (ATE) in randomized experiments. These methods leverage outcome prediction models to adjust for imbalances in baseline covariates across treatment arms, improving power compared to unadjusted analyses, while controlling Type I error, even when models are misspecified. In multi-center trials they are often implemented without accounting for clustering by centers. We investigate how ignoring center-level correlation can impair estimation, degrade coverage of confidence intervals, and obscure interpretation. We find these issues to be especially acute for estimators of counterfactual means, as shown through simulations and clarified via theoretical arguments. To address these challenges, we develop semiparametric efficient estimators of counterfactual means and ATE defined for a randomly sampled center and patient. These estimators leverage outcome prediction models to improve efficiency yet retain large-sample unbiasedness under model misspecification. We further introduce an inference framework, inspired by random-effects meta-analysis, tailored to settings with many small centers. Incorporating center effects into the prediction models yields substantial efficiency gains, particularly when treatment effects vary across centers. Simulations and application to the WASH Benefits Bangladesh trial illustrate strong finite-sample performance of the proposed methods.
\\
        
	\end{spacing}
	
\clearpage

\section{Introduction}\label{sec:introduction_ch2}
In response to the U.S. Food and Drug Administration guidance for industry on ``Adjustment for Covariates in Randomized Clinical Trials for Drugs and Biological Products''\citep{FDA2023}, there has been an increased interest in covariate adjustment within the analysis of randomized controlled trials. The main driver behind this is the potential for enhanced efficiency and power \citep{tsiatis2008covariate,kahan2014risks,benkeser2021improving}. It is additionally influenced by advancements in the causal inference literature, which demonstrate that treatment effect estimators utilizing augmented inverse probability weighting (AIPW) are both efficient and robust to model misspecification when applied to randomized controlled trial data \citep{tsiatis2008covariate,rosenblum2009using,van2024covariate}. The practical considerations of covariate adjustment in randomized trials, including the number of baseline covariates to choose, how to handle missing data for baseline covariates, and how to decide the required sample size, have been discussed extensively (e.g., \cite{tackney2023comparison,ye2023toward,van2024covariate}). 

Multi-center designs are common in contemporary randomized trials, particularly in confirmatory and pragmatic trials where recruitment from a single center is often insufficient \citep{meinert2012clinicaltrials,senn2019treatment}. For example, the LEADER trial \citep{marso2016liraglutide} evaluated the effect of liraglutide versus placebo on time to major cardiovascular events based on data from a randomized clinical trial conducted across 410 centers in 32 countries, with an average of nearly 23 patients per center. More generally, the number of published multi-center randomized controlled trials (RCTs) has increased substantially over time \citep{danielsen2014number}, and such trials now represent a major part of the clinical trial literature. Despite this, the multi-center design is commonly ignored in practical implementations of robust covariate-adjusted estimators.

Ignoring center can have at least two important consequences. First, it may reduce efficiency by failing to exploit between-center variation in outcome. Second, patients treated within the same center may have correlated outcomes because they are treated by the same clinicians, receive care under similar institutional protocols and resource constraints, and may be drawn from more similar patient populations. If this within-center dependence is ignored, as is common, standard errors may be underestimated, confidence intervals may have poor coverage, and Type I error rates may be inflated \citep{roberts2005design,chu2011comparing,kahan2013analysis,kahan2013assessing,kahan2014accounting,abadie2023should,benitez2023defining}. For example, in linear mixed models with center-specific intercept, \cite{kahan2013assessing} showed that failing to account for clustering in ordinary least squares estimators of treatment effect can lead to Type I error rates exceeding 20\%, despite a nominal significance level of 5\%, when also treatment assignments are correlated within clusters. As we show later, analogous inflation can arise when treatment effects vary between centers, and may occur more generally under non-linear data-generating mechanisms.

These issues are especially relevant for estimators of treatment-specific counterfactual mean outcomes. Unlike treatment contrasts on a particular scale, counterfactual means are directly interpretable, not tied to a chosen effect scale, and therefore less prone to scale-dependent interpretation. For this reason, we recommend reporting them alongside treatment-effect estimates. However, because counterfactual means do not benefit from the same cancellation of center effects that may occur in treatment contrasts, they may be particularly sensitive to whether and how the multi-center structure is accounted for.

Addressing these challenges is complicated by the fact that many multi-center trials include a large number of small centers. For example, the MISTIE III trial \citep{hanley2019efficacy}, which our simulations partially mimic, included 78 hospitals with an average of approximately six patients per hospital, while the LEADER trial \citep{marso2016liraglutide} included 410 centers with an average of nearly 23 patients per center. In such settings, fixed-effect adjustment for center is often unattractive: it can consume a large number of degrees of freedom, lead to unstable fitted values, and overfit center-specific outcome patterns, especially when outcomes or events are sparse within centers. Mixed-effects models provide a natural alternative by shrinking center-specific estimates toward a common mean, for instance through empirical BLUPs. These shrinkage-based predictions also appear well suited for use within AIPW estimators, since AIPW procedures can remove first-order bias induced by misspecification of the outcome regression \citep{hines2022demystifying,kennedy2024semiparametric}. However, as we show in this paper, this intuition can fail under a small-center asymptotic regime in which the number of centers tends to infinity while the number of patients per center remains bounded. Under this regime, the estimation error and shrinkage bias affecting the outcome predictions may be too large to ignore for robust inference, even when treatment is randomized. This motivates the estimators and inferential procedures developed below, which are designed specifically for multi-center trials with many small centers.

The main contributions of this paper can be summarized as follows:
\begin{enumerate}
    \item We provide insights into the consequences of ignoring clustering in AIPW estimators of counterfactual means and the ATE under both linear and nonlinear models. In linear models, center-specific deviations from the intercept cancel out for the ATE, but not for counterfactual means. The issue becomes more nuanced in nonlinear models or when treatment effects vary across centers. 
    
    \item We develop efficient estimators of the counterfactual means and ATE for a random center and a random patient that leverage baseline covariates and between-center heterogeneity through outcome prediction models while ensuring unbiasedness when the number of centers is  large.

    \item We develop an accompanying inferential framework inspired by the random-effects meta-analysis literature, specifically showing how to account for clustering-induced correlation. This framework is tailored to settings with many small centers, where standard asymptotic arguments may no longer directly apply (e.g., when using empirical BLUPs for outcome prediction).
    \end{enumerate}

Related work has been done in the context of cluster-randomized controlled trials \citep{balzer2019new,kahan2024demystifying,kahan2023estimands, benitez2023defining,wang2024model}. In particular, \cite{kahan2024demystifying,kahan2023estimands} discuss different estimands in cluster-randomized studies and show how the choice of estimand affects interpretation and dictates the selection of an appropriate estimator. \cite{balzer2019new} and \cite{benitez2023defining} propose targeted maximum likelihood estimators for the cluster-level mean outcome, with related developments for AIPW estimators in \cite{wang2024model}. \cite{balzer2019new} and \cite{benitez2023defining} primarily focus on settings with a small number of clusters, as is common in cluster-randomized trials, and assume that between-cluster variation is explained by observed cluster-level covariates. In contrast, our focus is on simple randomization in multi-center studies with either many small centers or a few large centers, for both continuous and binary outcomes. Moreover, we avoid assuming that between-center variation is fully explained by observed center-level covariates (i.e., covariates that are fixed within a center), as violations of this assumption may lead to efficiency loss. 

Traditionally, mixed-effects models and generalized estimating equations are also commonly used in these contexts (e.g., \cite{kahan2013analysis,kahan2014accounting,huang2016generalized,benitez2023defining,wang2016covariance,wang2024model}). However, these approaches are model-dependent, generally delivering biased estimators when the model is misspecified. This makes them less suitable for use in regulatory context. Canonical effect measures under these models (e.g., odds ratios) can also be difficult to interpret. In particular, they may differ in  magnitude and meaning between both model classes (mixed and marginal models) \citep{li2023comparing}, and continue to change magnitude and meaning when adjusting for more and more covariates, rendering a coherent theory on covariate adjustment complicated. Interpretation is especially subtle when the cluster sizes and treatment effects are dependent \citep{wang2022two}. In this paper, we instead focus on model-free estimands, such as the counterfactual means and ATE for a random center and a random patient. These  measures of treatment effect are well-defined without relating to some statistical model, thereby aligning with the International Council for Harmonization E9 (R1) addendum on estimands and sensitivity analysis in clinical trials \citep{ICH2020}. 

The structure of this paper is as follows. Section \ref{sec:estimation_ch2} introduces the estimands and provides the estimation procedures for both counterfactual means and the ATE. In Section \ref{sec:inference_ch2}, we introduce a corresponding inference framework. Section \ref{sec:simulations_ch2} describe the Monte Carlo simulation studies, while Section \ref{sec:data_analysis_ch2} presents the data analysis. Section \ref{sec:discussion_ch2} ends with a discussion and conclusion. 

\section{Estimation}\label{sec:estimation_ch2}
\subsection{Notation}
For a random patient $i$ in center $c$, randomly drawn from an infinite population of centers $\mathcal{C}$, we collect observations on a randomized treatment indicator $A_{ic}$ (where $A_{ic}=1$ represents the treatment group and $A_{ic}=0$ the placebo or active control group), a vector of baseline covariates $X_{ic}$ and a primary outcome of interest $Y_{ic}$, which may take continuous or discrete values. We may drop the indices to refer a generic patient. We denote the number of patients in each center $c$ by $n_c$ and the total number of sampled centers by $k$. We consider a setting in which treatment is allocated by simple randomization, independent of baseline covariates. Observations are assumed independent across centers, and independent within centers conditional on the center effect.

\subsection{``Na\"ive'' approach}\label{sub_sec:naive_approach}
We begin by introducing the now standard ``na\"ive'' estimation strategy that disregards clustering when estimating the counterfactual mean under treatment $\E\!\left(Y^1\right)$ and ATE $\E\!\left(Y^1\right)-\E\!\left(Y^0\right)$, where $Y^1$ and $Y^0$ represent the potential (counterfactual) outcomes that would be observed for a random subject if assigned to the treatment or control, respectively. The expectation is taken over patients randomly sampled from the population of interest. The ``na\"ive'' AIPW estimators, which ignore clustering, are then computed as follows.

\begin{itemize}[leftmargin=4em]
\item[\textbf{Step 1:}] \textbf{Model fitting}\\
Fit a generalized linear model with canonical link via maximum likelihood that regresses the outcome $Y_{ic}$ on pre-specified baseline covariates $X_{ic}$ and randomized treatment indicator $A_{ic}$, across all patients in all centers. For example, for a continuous outcome $Y_{ic}$, we could model $\E\left(Y_{ic}|A_{ic},X_{ic}\right)$ as $m(A_{ic}, X_{ic};\alpha, \beta, \gamma)=\alpha+\beta A_{ic}+\gamma'X_{ic}$, where $\alpha$, $\beta$ and $\gamma$ denote an intercept, a coefficient for treatment and a vector of coefficients for baseline covariates, respectively. For a binary outcome $Y_{ic}$, we could model $P\left(Y_{ic}=1|A_{ic},X_{ic}\right)$ as $m(A_{ic}, X_{ic};\alpha, \beta, \gamma)=\logit^{-1}\left(\alpha+\beta A_{ic}+\gamma'X_{ic}\right)$. For both examples, a model can also be fitted separately for each treatment arm.

\item[\textbf{Step 2:}] \textbf{Predicting}\\
For each patient, use the fitted outcome working model in Step 1 to compute a prediction of the response under treatment, $\hat{m}_{1}(X_{ic})$, and control, $\hat{m}_{0}(X_{ic})$, for patient $i$ in center $c$, using the patient's baseline covariates. In our example, $\hat{m}_{1}(X_{ic})=\hat{\alpha}+\hat{\beta}+\hat{\gamma}'X_{ic}$ and $\hat{m}_{0}(X_{ic})=\hat{\alpha}+\hat{\gamma}'X_{ic}$ for a continuous outcome, and $\hat{m}_{1}(X_{ic})=\logit^{-1}(\hat{\alpha}+\hat{\beta}+\hat{\gamma}'X_{ic})$ and $\hat{m}_{0}(X_{ic})=\logit^{-1}(\hat{\alpha}+\hat{\gamma}'X_{ic})$ for a binary outcome. 

\item[\textbf{Step 3:}] \textbf{Averaging}\\
The counterfactual mean under treatment and ATE can be estimated as:
	 	\begingroup
	 	\setlength{\abovedisplayskip}{0pt}
	 	\setlength{\belowdisplayskip}{0pt}
	 	\begin{align*}
	 		\hat{\mu}_{1,\text{na\"{i}ve}}=\frac{1}{n}\sum_{c=1}^{k}\sum_{i=1}^{n_c}\left[\frac{A_{ic}}{\hat{p}(X_{ic})}\left\{Y_{ic}-\hat{m}_{1}(X_{ic})\right\}+\hat{m}_{1}(X_{ic})\right], \ \text{and}
	 	\end{align*}
	 	\endgroup
	 	\begin{align*}
	 		\hat{\tau}_{\text{na\"{i}ve}}=\frac{1}{n}\sum_{c=1}^{k}\sum_{i=1}^{n_c}\left[\frac{A_{ic}}{\hat{p}(X_{ic})}\left\{Y_{ic}-\hat{m}_{1}(X_{ic})\right\}+\hat{m}_{1}(X_{ic})-\frac{1-A_{ic}}{1-\hat{p}(X_{ic})} \left\{Y_{ic}-\hat{m}_{0}(X_{ic})\right\}-\hat{m}_{0}(X_{ic})\right],
	 	\end{align*}
where $n$ denotes the number of patients across all centers and $\hat{p}(X_{ic})$ is the estimated randomization probability conditional on baseline covariates; this can be obtained for instance by fitting a logistic regression model. In a clinical trial with randomization independent of covariates, these AIPW estimators are equivalent to the G-computation/standardization estimators discussed in the FDA guidance on covariate adjustment \citep{FDA2023} when using canonical generalized linear models fitted via maximum likelihood and with $\hat{p}(X_{ic})$ given by the (estimated) marginal randomization probability $\hat{p}$. That is, $\hat{\tau}_{\text{na\"{i}ve}}=\frac{1}{n}\sum_{c=1}^{k}\sum_{i=1}^{n_c}\left\{\hat{m}_{1}(X_{ic})-\hat{m}_{0}(X_{ic})\right\}$ as the score equations for canonical generalized linear models (evaluated at the MLE) imply the constraints\\$\frac{1}{n}\sum_{c=1}^{k}\sum_{i=1}^{n_c}\frac{A_{ic}}{\hat{p}}\left\{Y_{ic}-\hat{m}_{1}(X_{ic})\right\}=0$ and $\frac{1}{n}\sum_{c=1}^{k}\sum_{i=1}^{n_c}\frac{1-A_{ic}}{1-\hat{p}}\left\{Y_{ic}-\hat{m}_{0}(X_{ic})\right\}=0$.
\end{itemize}

``Na\"ive" inference for AIPW estimators ignores the correlation present in multi-center studies. The ``na\"ive" (i.e., ignoring clustering) variances of $\hat{\mu}_{1,\text{na\"{i}ve}}$ and $\hat{\tau}_{\text{na\"{i}ve}}$ are calculated as $1/n$ times the sample variances of $\frac{A_{ic}}{\hat{p}(X_{ic})}\left\{Y_{ic}-\hat{m}_{1}(X_{ic})\right\}+\hat{m}_{1}(X_{ic})$, and $\frac{A_{ic}}{\hat{p}(X_{ic})}\left\{Y_{ic}-\hat{m}_{1}(X_{ic})\right\}+\hat{m}_{1}(X_{ic})-\frac{1-A_{ic}}{1-\hat{p}(X_{ic})} \left\{Y_{ic}-\hat{m}_{0}(X_{ic})\right\}-\hat{m}_{0}(X_{ic})$, respectively. Appendix A.3 outlines the conditions under which this leads to bias in estimated standard errors. For example, the ``na\"ive" standard errors of the ATE (counterfactual means) are not valid with continuous outcomes obeying linear models, unless there is no variation in the treatment effect (and no variation in the intercept and covariate effects) between centers. They are generally invalid for the ATE when outcomes are binary and follow a Bernoulli distribution with probabilities obeying a logistic regression model.

\subsection{Proposed approach}\label{sub_sec:proposed_approch}
Given known center-specific weights $w(\cdot)$ for which $\sum_{c\in\mathcal{C}}w(c)=1$, our primary focus will be on estimands of the form
	\begingroup
	\setlength{\abovedisplayskip}{0pt}
	\setlength{\belowdisplayskip}{0pt}
	\begin{align}\label{eq:countMean}
	\mu_1=\sum_{c\in\mathcal{C}}\E\left(Y_c^1\right)w(c),
	\end{align}
	\endgroup
	and
	\begingroup
	\setlength{\abovedisplayskip}{0pt}
	\setlength{\belowdisplayskip}{0pt}
	\begin{align}\label{eq:ATE}
	\tau=\sum_{c\in\mathcal{C}}\E\left(Y_c^1\right)w(c)-\sum_{c\in\mathcal{C}}\E\left(Y_c^0\right)w(c),
	\end{align}
	\endgroup
where $Y_c^1$ and $Y_c^0$ correspond to the potential outcomes that would be observed for a random subject in center $c$ if assigned to treatment or control, respectively. A uniform choice of weights $w(c)=1/k$ results in estimands that assign equal weights to all centers, thereby expressing the counterfactual mean outcome or treatment effect for a random center. Alternatively, the choice $w(c)=n_c/\sum_{c^{\star}\in\mathcal{C}}n_{c^{\star}}$ assigns equal weight to patients, thereby expressing the counterfactual mean outcome or treatment effect for a random patient, as considered in Section \ref{sub_sec:naive_approach}. The choice between these estimands depends on the context of the study; see Section \ref{sec:discussion_ch2} for further discussion.

We introduce the following procedure to address clustering, explaining it in the context of binary endpoints, though it can be readily adapted to continuous endpoints (see Appendix A.1 for further details).
      \begin{itemize}[leftmargin=4em]
	\item[\textbf{Step 1:}] \textbf{Model fitting}\\
Fit a mixed-effects logistic regression model of the form
\begin{align*}
    \E\left(Y_{ic}|A_{ic},X_{ic},b_{0c},b_{1c}\right)=\logit^{-1}\left\{\alpha+b_{0c}+(\beta+b_{1c})A_{ic}+\gamma'X_{ic}\right\},
\end{align*}
where $b_{0c}$ and $b_{1c}$ denote center-specific deviations to the intercept (i.e., random intercept) and treatment effect (i.e., random slope), respectively. We assume that $b_{0c}$ and $b_{1c}$ are independent and follow a normal distribution with mean zero and variance $\sigma^2_{b0}$ and $\sigma^2_{b1}$, respectively, independently of treatment and covariates. One can also allow for variation in covariate effects, as we explore in the simulations. Alternatively, when the number of patients per center is relatively large, a fixed-effects logistic regression model $\E\left(Y_{ic}|A_{ic},X_{ic}\right)=\logit^{-1}\left(\gamma_0+\psi A_{ic}+\eta'X_{ic}+\sum_{j=1}^{k-1} \gamma_j \mathit{I}_{ij}\right)$ can be fitted. Here, $\mathit{I}_{ij}$ is an indicator variable for patient $i$ in center $j$, which is 1 for patients in center $j$ and 0 otherwise. 
		
\item[\textbf{Step 2:}] \textbf{Predicting}\\
When a mixed-effects logistic regression model is used, we predict the outcome on treatment as $\hat{m}_{1c}(X_{ic})=\logit^{-1}(\hat{\alpha}+\hat{b}_{0c}+\hat{\beta}+\hat{b}_{1c}+\hat{\gamma}'X_{ic})$ and on control as $\hat{m}_{0c}(X_{ic})=\logit^{-1}(\hat{\alpha}+\hat{b}_{0c}+\hat{\gamma}'X_{ic})$. Here, the random-effects $\hat{b}_{0c}$ and $\hat{b}_{1c}$ are repeatedly sampled (e.g., 1000 draws) from their estimated normal distributions,
\begin{align*}
\hat{b}_{0c} \sim \text{Normal}(0, \hat{\sigma}^2_{b0}), \qquad
\hat{b}_{1c} \sim \text{Normal}(0, \hat{\sigma}^2_{b1}),
\end{align*}
where $\hat{\sigma}^2_{b0}$ and $\hat{\sigma}^2_{b1}$ are REML estimates under the mixed-effects model. The resulting predictions are then averaged over these repeated draws. They will be denoted $\hat{m}_{1c}(X_{ic})$ and $\hat{m}_{0c}(X_{ic})$ with a slight abuse of notation. Alternatively, $\hat{b}_{0c}$ and $\hat{b}_{1c}$ can be substituted by best linear unbiased predictors (BLUPs), or 
$\hat{m}_{1c}(X_{ic})$ and $\hat{m}_{0c}(X_{ic})$ can be obtained as the predictions  $\hat{m}_{1c}(X_{ic})=\logit^{-1}\left(\hat{\gamma}_0+\hat{\psi}+\hat{\eta}'X_{ic}+\sum_{j=1}^{k-1} \hat{\gamma}_j \mathit{I}_{ij}\right)$ and $\hat{m}_{0c}(X_{ic})=\logit^{-1}\left(\hat{\gamma}_0+\hat{\eta}'X_{ic}+\sum_{j=1}^{k-1} \hat{\gamma}_j \mathit{I}_{ij}\right)$ under 
a fixed-effects logistic regression model. However, as our theoretical results will show, the use of BLUPs or fixed-effects estimates is discouraged when (some) cluster sizes are small. 
		
\item[\textbf{Step 3:}] \textbf{Averaging}\\
Estimators of $\mu_1$ and $\tau$ are obtained by averaging center-specific estimates across centers: $\hat{\mu}_{1}=\sum_{c=1}^{k}w(c)\hat{\mu}_{1c},\ \text{and}\ \hat{\tau}=\sum_{c=1}^{k}w(c)\hat{\tau}_{c}$.
For these center-specific estimates, we use AIPW estimators. For the counterfactual mean outcome under treatment in center $c$, we have:\[\hat{\mu}_{1c}=\frac{1}{n_c}\sum_{i=1}^{n_c}\left[\frac{A_{ic}}{\hat{p}_{c}(X_{ic})}\left\{Y_{ic}-\hat{m}_{1c}(X_{ic})\right\}+\hat{m}_{1c}(X_{i})\right],\] and for the average treatment effect in center $c$
\[\hat{\tau}_{c}=\frac{1}{n_c}\sum_{i=1}^{n_c}\left[\frac{A_{ic}}{\hat{p}_{c}(X_{ic})}\left\{Y_{ic}-\hat{m}_{1c}(X_{ic})\right\}+\hat{m}_{1c}(X_{ic})-\frac{1-A_{ic}}{1-\hat{p}_{c}(X_{ic})} \left\{Y_{ic}-\hat{m}_{0c}(X_{ic})\right\}-\hat{m}_{0c}(X_{ic})\right],\] 
with $\hat{p}_{c}(X_{ic})$ the estimated randomization probability of patient $i$ in center $c$. Here, $\hat{p}_{c}(X_{ic})$ are obtained as the conditional predictions (i.e., based on empirical BLUPs for the random-effect) under a mixed-effects logistic regression model with treatment assignment as the dependent variable, incorporating centers as random-effects and (optionally) baseline covariates as fixed-effects. While the use of empirical BLUPs was previously discouraged for the outcome predictions, it will result in better tailored center-specific randomization probabilities, thereby increasing efficiency, without compromising validity when center sizes are small (see Appendix A.2). In clinical trials with simple randomization, a further precision gain may be achieved by including baseline covariates in this mixed-effects model \citep{balzer2019new}, though we typically expect the gains to be small. Note that the treatment model is guaranteed to be correctly specified, since treatment assignment is independent of baseline covariates and center.
\end{itemize}

In Appendix A.6, we also consider a cluster-randomized design, assuming that observations are independent between clusters and within clusters, conditional on the cluster effect.

\section{Inference}\label{sec:inference_ch2}
In this section, we discuss the validity of the proposed procedure along with estimation of the variance of the estimated ATE ($\hat{\tau}$); results are analogous for the estimated counterfactual mean ($\hat{\mu}_{1}$). Since many clinical trials involve numerous small centers, we will assume a small-center asymptotic framework where the number of centers approaches infinity while the sample sizes $n_c$ for centers $c = 1, ..., k$ remain fixed. This brings subtleties, which we will discuss here and of which the details are relegated to Appendix A.2. In this Appendix, we show that the proposed procedure delivers estimators with oracle behavior (i.e., which behave the same in many centers as when the randomization probabilities and outcome predictions were known), which are therefore efficient. This is so despite the use of empirical BLUPs for the estimation of the randomization probabilities, which is the case because there are no systematic differences in randomization probability between centers in simple randomized experiments (see Appendix A.2). This implies in particular that $\hat{\tau}$ is asymptotically normal, satisfying $\sqrt{k}(\hat{\tau}-\tau) \xrightarrow{d} N(0,\sigma^2_{\tau})$ as $k$ goes to infinity under standard regularity conditions. Here, the variance is given by $\sum_{c=1}^k w^2(c) \text{Var}(\hat{\tau}_{c})$, where 
\begin{align}\label{eq:total_variance}
\text{Var}\left(\hat{\tau}_{c}\right)=\E\left\{\text{Var}(\hat{\tau}_{c}|\tau_{c})\right\}+\text{Var}\{\E\left(\hat{\tau}_{c}|\tau_{c}\right)\}.
\end{align}
Equation \eqref{eq:total_variance} expresses that the variation in $\hat{\tau}_{c}$ comes from two sources: the difference between the estimated average treatment effect in center $c$ and its true average treatment effect (i.e., the within-center variance), and the difference between the true average treatment effect in center $c$ and the true average treatment effect across all centers (i.e., the between-center variance). As in meta-analyses, this total variance can be estimated using a random-effects model:
\begin{align}
\hat{\tau}_{c}=\tau+\epsilon_{c}+u_{c},
\end{align}
where $\epsilon_{c}\sim N(0,\sigma^2_{c})$ and $u_{c}\sim N(0,\sigma^2_{u})$ are assumed to be independent. Here, $\sigma^2_{c}$ and $\sigma^2_{u}$ denote the variance of the estimated ATE in center $c$ (i.e., the within-center variance of the estimated ATE) and the heterogeneity (between-center) variance of the estimated ATE, respectively. As a result, we have $\hat{\tau}_{c}\sim N(\tau,\sigma^2_{c}+\sigma^2_{u})$, and the variance of $\hat{\tau}$ can be approximated as $\sum_{c=1}^k w^2(c)\left(\sigma^2_{c}+\sigma^2_{u}\right)$.
Here, the within-center variance $\sigma^2_{c}$ can be estimated as 1 over $n_{c}$ times the sample variance of \[\frac{A_{ic}}{\hat{p}_{c}(X_{ic})}\left\{Y_{ic}-\hat{m}_{1c}(X_{ic})\right\}+\hat{m}_{1c}(X_{ic})-\frac{1-A_{ic}}{1-\hat{p}_{c}(X_{ic})} \left\{Y_{ic}-\hat{m}_{0c}(X_{ic})\right\}-\hat{m}_{0c}(X_{ic}),\]
over all patients $i$ in center $c$. The heterogeneity variance $\sigma^2_{u}$ can be estimated as in a random-effects meta-analysis of center-specific estimates of the ATE (e.g., through the \texttt{rma} function from the \texttt{metafor} package; see Appendix A.5 for details). Specifically, estimation can be carried out using the method of moments (also known as the DerSimonian and Laird method \citep{dersimonian1986meta}) or, alternatively, using restricted maximum likelihood \citep{whitehead2002meta}. These results continue to be valid when the center sizes approach infinity but the number of centers remains fixed (see also Appendix A.2).

\textit{Remark 1.} When the average treatment effect in a center is associated with its size, then the DL and REML heterogeneity variance estimators may be biased. In this case, $\sigma^2_{u}$ can be estimated by debiasing the sample heterogeneity variance $\frac{1}{k}\sum_{c=1}^{k}\left(\hat{\tau}_{c}-\hat{\tau}\right)^2$ via the debiased heterogeneity variance estimator
\begin{align*}
\hat{\sigma}^2_{u,DB}=\max\left\{0,\frac{1}{k}\sum_{c=1}^{k}\left(\hat{\tau}_{c}-\hat{\tau}\right)^2-\frac{k-1}{k^2}\sum_{c=1}^{k}\hat{\sigma}^2_{c}\right\},
\end{align*}
which we show to be asymptotically unbiased in Appendix A.5, where $\hat{\sigma}^2_{c}$ denotes the estimated within-center variance of the estimated ATE.  
This estimator removes variance inflation coming from the use of imprecise estimates $\hat{\tau}_{c}$ of $\tau_c$. 

When the outcome predictions are based on empirical BLUPs (or fixed-effects regression), then the suggested oracle behaviour may no longer be attained. We show this in Appendix A.2 in the context of linear mixed models. In particular, we show that the so-called empirical process term in the asymptotic expansion adds a non-negligible contribution, which implies suboptimal performance. This is so because the empirical BLUPs are not consistent in the considered small-center asymptotic regime, thereby leaving non-negligible excess variability. This excess variability can be quantified in the context of linear mixed models, but formal justification in the context of non-linear mixed models cannot be guaranteed because of potential additional concerns with the so-called remainder terms in the asymptotic expansion (see Appendix A.2). We therefore discourage the use of empirical BLUPs and fixed-effects regression when (some) center sizes are small.  

Finally, confidence intervals (based on all variance estimators) are constructed using a $t$-distribution with estimated degrees of freedom. The estimated degrees of freedom builds directly on the effective sample size approach for clustered data, which can also be used for non-Gaussian data more generally \citep{faes2009effective}. Specifically, the degrees of freedom are estimated as
\begin{align*}
    \sum_{c=1}^{k}\frac{n_c}{1+(n_{c}-1)\hat{\rho}}-1,
\end{align*}
where $\hat{\rho}=\frac{\hat{\sigma}^2_{u}}{\hat{\sigma}^2_{u}+\hat{\sigma}^2}$. Here, $\hat{\sigma}^2=\frac{1}{k}\sum_{c=1}^{k}\hat{\sigma}^2_{c}$, is the average of the estimated variances of the AIPW estimates. In \cite{faes2009effective}, $\hat{\rho}$ is calculated based on the outcome variances. Since the considered estimators behave the same when using true versus estimated nuisance parameters in large samples, we expect this result to be approximately valid for these estimators based on influence functions.

The estimated degrees of freedom incorporate the strength of correlation among observations within the same center. Because the correlation may differ for the counterfactual means and the ATE, their respective estimated degrees of freedom might not be the same. When $\rho=0$, the (true) degrees of freedom are $\sum_{c=1}^{k}n_c-1$, while for $\rho=1$, they are $k-1$. For a large number of centers (e.g., $k\geq30$), confidence intervals using approximate degrees of freedom and the standard normal distribution are nearly identical.

\section{Simulation studies}\label{sec:simulations_ch2}
\subsection{Data-generating mechanisms}\label{subsec:datagen_ch2}
We consider simulation settings with continuous as well as binary endpoints under a range of data-generating mechanisms.

\subsubsection*{Continuous outcomes}
For continuous outcomes, we resample patients with replacement from the ACTG 175 trial dataset \citep{hammer1996trial}, which can be accessed in R from the \texttt{BART} package using the command \texttt{data(ACTG175)}. In particular, we resample data for the following baseline covariates, which will be included in the data-generating models: zidovudine use in the 30 days prior to treatment initiation ($z30$: $0$=no, $1$=yes), age in years at baseline ($age$), CD4 cell count at baseline ($CD40$), and weight in kg ($wt$). Our study includes two treatment arms, generated from a Bernoulli distribution with probability 0.5: zidovudine (denoted by $A=0$) and other therapies (didanosine, zidovudine combined with didanosine, and zidovudine combined with zalcitabine; denoted by $A=1$). 

We consider the log CD4 cell count at week 96 as the primary outcome ($Y$), based on 1,342 observations after excluding 797 missing values in the outcome, which will be generated from a linear mixed model. The fixed-effects coefficients of treatment and baseline covariates (mentioned above) are obtained from the trained outcome model in the ACTG 175 trial. The residual error $\epsilon$ is drawn from a standard normal distribution. In each simulation run, each resampled patient is randomly assigned to one out of a fixed number of centers according to a prespecified vector of center sizes. Within-center correlation is then induced by incorporating center-specific random-effects $b_{0c}$, $b_{1c}$, and $b_{2c}$ to the intercept, treatment effect, and covariate ($CD40$) effect, into the data-generating model. These random-effects introduce between-center heterogeneity and increase the total outcome variance relative to that observed in the original ACTG 175 trial. These effects are assumed to follow a normal distribution with mean zero and variances $\sigma^2_{b0}$, $ \sigma^2_{b1}$, and $\sigma^2_{b2}$, ensuring correlation within clusters. 
We consider different values of the random-effects variances as presented in Table B.1.1 in Appendix B.1. These are selected to ensure that the intraclass correlation coefficient for the outcome ranges from a minimum of 0 to a maximum of 0.29, in line with findings from \citep{kul2014intraclass}. The outcome for patient $i$ in center $c$ is then generated as 
\begin{align*}
Y_{ic} &= 4.06 + b_{0c}+ \bigl(2.9\times 10^{-1} + b_{1c}\bigr)\cdot A_{ic}+ \bigl(4\times 10^{-3} + b_{2c}\bigr)\cdot CD40_{ic}
- \bigl(1.5\times 10^{-1}\bigr)\cdot z30_{ic} \\
&\quad - \bigl(4\times 10^{-4}\bigr)\cdot age_{ic} + \bigl(5\times 10^{-3}\bigr)\cdot wt_{ic}
+ \epsilon_{ic}.
\end{align*}
We also consider an outcome model that includes a covariate transformation and an interaction term expressed as
\begin{align*}
Y_{ic}&= 3.71 + b_{0c}+ \bigl(2.9\times 10^{-1} + b_{1c}\bigr)\cdot A_{ic}+ \bigl(8\times 10^{-2} + b_{2c}\bigr)\cdot \sqrt{CD40_{ic}}-\bigl(1.4\times 10^{-1}\bigr)\cdot z30_{ic} \\
&\quad - \bigl(2\times 10^{-4}\bigr)\cdot age_{ic}-\bigl(1\times 10^{-3}\bigr)\cdot wt_{ic}+\bigl(1\times 10^{-5}\bigr)\cdot CD40_{ic}\cdot wt_{ic}+\epsilon_{ic}.
\end{align*}
In the analysis, the outcome model is specified with only the main effects of the covariates in both cases, resulting in a misspecified outcome model in the second scenario.

We consider different scenarios regarding the numbers of centers and numbers of patients per center (i.e., center size); see Figure B.1.1 in Appendix B.1. The focus of this paper is on settings with many small centers. We also examine a setting with a relatively many large centers to illustrate how the estimators behave under such conditions. Specifically, our analysis explores three distinct settings:  
(1) $k = 100$, with center sizes averaging 5 (Min = 1, Max = 24), resembling the MISTIE III trial;  
(2) $k = 50$, with an average $n_c$ of 10 (Min = 2, Max = 48);  
(3) $k = 100$, with an average $n_c$ of 100 (Min = 50, Max = 145).  

\subsubsection*{Binary outcomes}
For binary outcomes, we perform resampling with replacement from the MISTIE III trial simulated dataset generated by \cite{betz2023}.  Specifically, we resample data from this synthetic dataset for the following baseline covariates: the Glasgow Coma Scale, categorized as \textit{severe} (scores between 3 and 8), \textit{moderate} (scores between 9 and 12), and \textit{mild} (scores between 13 and 15), as well as age in years (\textit{age}) and the intracerebral hemorrhage volume recorded on the stability scan (\textit{ich.v}).  
Treatment $A$ is generated by a Bernoulli distribution with probability 0.5. 

The primary outcome ($Y$) is the modified Rankin Scale score at 365 days, defined as 1 if the score falls between 0 and 3, and 0 otherwise. We generate these outcomes via a mixed effects logistic regression model. The fixed-effects coefficients of treatment and baseline covariates (mentioned above) are obtained from the trained outcome model in the MISTIE III trial simulated dataset. We introduce clustering by adding random-effects $b_{0c}$, $b_{1c}$ and $b_{2c}$ in the trained model, generated from a normal distribution with zero mean and variance $\sigma^2_{b0}$, $\sigma^2_{b1}$, and $\sigma^2_{b2}$, respectively. We consider different combinations of random-effects variances as shown in Table B.1.1 in the Appendix B.1. The values of these variances are chosen to give the intraclass correlation coefficient for the outcome on a logistic scale that ranges from a minimum of 0 to a maximum of 0.65. Subsequently, a binary outcome for patient $i$ in center $c$ is generated as
$Y_{ic} \sim \mathrm{Binomial}(\pi_{ic})$, where
\begin{align*}
\pi_{ic}&=\operatorname{logit}^{-1}\Big\{3.22 + b_{0c}
+ \big(2.8\times 10^{-1} + b_{1c}\big)\cdot A_{ic}-\big(1.71 + b_{2c}\big)\cdot severe_{ic}-\big(7.2\times 10^{-1}\big)\cdot moderate_{ic} \\
&\qquad - \big(4\times 10^{-2}\big)\cdot age_{ic}-\big(7\times 10^{-3}\big)\cdot ichv_{ic}\Big\}.
\end{align*}
We additionally generate $Y_{ic}$, where
\begin{align*}
\pi_{ic}&= \operatorname{logit}^{-1}\Big\{5.52 + b_{0c}+\big(2.9\times 10^{-1} + b_{1c}\big)\cdot A_{ic}-\big(1.72 + b_{2c}\big)\cdot severe_{ic}-\big(7.2\times 10^{-1}\big)\cdot moderate_{ic} \\
&\qquad-\big(1.2\times 10^{-1}\big)\cdot age_{ic}+\big(7\times 10^{-4}\big)\cdot age_{ic}^{2}-\big(7\times 10^{-3}\big)\cdot ichv_{ic}\Big\}.
\end{align*}
As for continuous outcomes, in the analysis, we use only the main effects of covariates in the outcome model, thereby misspecifying the outcome model in the second case. We consider the same scenarios regarding the numbers of centers and numbers of patients per center as for continuous outcomes.

\subsection{Estimators}\label{sub_sec:Simulation_estimators_ch2}
The following estimators (for the counterfactual mean on treatment and ATE) are considered: 
\begin{enumerate}
    \item \textbf{``Na\"ive'' AIPW Estimators} as described in Section \ref{sub_sec:naive_approach}: Uses a generalized linear model (linear for continuous outcomes, logistic for binary), including only the treatment indicator or additionally including baseline covariates as fixed-effects (referred to as \textit{Na\"ive}).
    \item \textbf{Fixed-Effects AIPW Estimators} as described in Section \ref{sub_sec:proposed_approch}: These estimators are based on a fixed-effects model that includes center indicators and the treatment indicator, and may additionally include baseline covariates as fixed-effects (referred to as \textit{Fixed}).
  \item \textbf{Mixed-Effects AIPW Estimators} as described in Section \ref{sub_sec:proposed_approch}: Including only the treatment indicator as a fixed-effect, with random center-specific intercept (referred to as \textit{Mixed(1$\mid$c)} when the random-effects are estimated using empirical BLUPs, and as \textit{Mixed(1$\mid$c) Sam} when they are obtained by sampling from their (estimated) normal distributions) and random center-specific treatment effect (referred to as \textit{Mixed(1+A$\mid$c)} when the random-effects are estimated using empirical BLUPs, and as \textit{Mixed(1+A$\mid$c) Sam} when they are obtained by sampling from their (estimated) normal distributions), and may additionally incorporating baseline covariates as fixed-effects. We consider 1000 draws for the sampling method.
\end{enumerate}
The variance of each estimator is computed as discussed in Section \ref{sec:inference_ch2}.

\subsection{Results}
\subsubsection*{Continuous outcomes}
We present the estimated average standard errors and coverage probabilities for both the counterfactual mean on treatment and the ATE with continuous outcomes for different settings that define the number of patients per center (Setting 1 to Setting 3; see Section \ref{subsec:datagen_ch2}), while varying the random-effects variances. \autoref{table:k_100_nc_5_cont} and \autoref{table:k_100_nc_100_cont} present the results based on equal weighting of centers under Settings 1 and 3, respectively. Additional results can be found in Appendix B.1. 

We see from \autoref{table:k_100_nc_5_cont} that the na\"ive confidence intervals for the counterfactual mean on treatment are not valid unless there is no variation in the intercept, treatment effect and covariate effects between centers. For the ATE, we see this lack of validity of confidence intervals only when there is variation in treatment effect between centers (see \autoref{table:k_100_nc_5_cont}), as supported by the theoretical results in Appendix A.3. For example, the coverage of the 95\%-confidence intervals of the ATE is not affected by variation in covariate effects between centers (i.e., $\sigma_{b2}^2\neq 0$), unlike for the counterfactual mean on treatment.

When data are generated with center-specific deviations to the intercept and treatment effect, the proposed estimation method that includes center and treatment as random-effects preserves the nominal coverage levels for the ATE best. Lower coverage probabilities are obtained when including center as fixed-effect, likely due to the small center sizes. Estimating random-effects by sampling from their normal distributions provides slightly better coverage probabilities compared to using empirical BLUPs with many small centers (Setting 1), particularly for the average treatment effect when the random-effects variances are relatively large ($\sigma^2_{b0}$=0.15, $\sigma^2_{b1}$=0.15, $\sigma^2_{b2}=4\times10^{-6})$. This finding is consistent with theoretical expectations, although the results obtained using empirical BLUPs also perform well. Empirical BLUP-based estimators tend to be more efficient than those based on sampling (see \autoref{table:k_100_nc_5_cont}), which is the result of shrinkage. For both the counterfactual mean on treatment and ATE, the REML, DL and DB variance estimators work comparably. Thus, although obtaining valid center-specific estimates is challenging in a setting of small centers, the proposed AIPW estimators perform well.

\begin{table}
	\caption{Simulation results: Monte Carlo standard deviation, average standard errors and coverage probabilities of 95\% confidence intervals for both counterfactual mean on treatment and ATE based on 1000 simulations. Setting 1: Continuous outcome, $k$=100, $n_c$: Avg=5, Min=1, Max=24, and different values of random-effects variance. Results are based on weighting centers equally.}
	\begin{center}
		\resizebox{\textwidth}{!}{
			\begin{tabular}{>{\raggedright\arraybackslash}p{3.5cm}cccccccccccccccccc}
				\toprule
				& \multicolumn{9}{c}{\textbf{Counterfactual mean on treatment}} &\multicolumn{9}{c}{\textbf{ATE}} \\
				\cmidrule(lr){2-10} \cmidrule(lr){11-19}
				&& \multicolumn{4}{c}{\textbf{SE}} & \multicolumn{4}{c}{\textbf{Coverage (\%)}} &&\multicolumn{4}{c}{\textbf{SE}}& \multicolumn{4}{c}{\textbf{Coverage (\%)}} \\
				\cmidrule(lr){3-6} \cmidrule(lr){7-10} \cmidrule(lr){12-15} \cmidrule(lr){16-19}
				Method &  SD & Na\"ive & REML & DL & DB & Na\"ive & REML & DL & DB  & SD & Na\"ive & REML & DL & DB & Naive & REML & DL & DB \\
				\midrule
				\multicolumn{19}{c}{\textit{Random-effects variances}: $\sigma^2_{b0}$=0, $\sigma^2_{b1}$=0, $\sigma^2_{b2}$=0: $\bar{\rho}_1=0.0008$, $\bar{\rho}=-0.0002$} \\
                \multicolumn{19}{l}{\textbf{Unadjusted}}\\
				Na\"ive & 0.069 & 0.070 &  &  &  & 95.2 &  &  &  & 0.097 & 0.099 &  &  &  & 95.4 &  &  &  \\ 
				Fixed & 0.084 &  & 0.083 & 0.083 & 0.082 &  & 95.2 & 95.2 & 94.9 & 0.114 &  & 0.111 & 0.110 & 0.109 &  & 93.1 & 93.0 & 92.9 \\ 
				Mixed(1$\mid$c) & 0.087 &  & 0.087 & 0.087 & 0.090 &  & 94.9 & 95.1 & 96.0 & 0.121 &  & 0.123 & 0.123 & 0.125 &  & 95.6 & 95.5 & 95.7 \\ 
                Mixed(1$\mid$c) Sam. & 0.107 &  & 0.106 & 0.106 & 0.105 &  & 94.4 & 94.6 & 94.0 & 0.138 &  & 0.136 & 0.137 & 0.138 &  & 94.8 & 94.9 & 95.2 \\
				Mixed(1+A$\mid$c) & 0.087 &  & 0.087 & 0.087 & 0.089 &  & 94.9 & 95.1 & 96.0 & 0.121 &  & 0.122 & 0.123 & 0.125 &  & 95.6 & 95.5 & 95.5 \\ 
                Mixed(1+A$\mid$c) Sam. & 0.107 &  & 0.106 & 0.106 & 0.105 &  & 94.5 & 94.6 & 94.0 & 0.138 &  & 0.136 & 0.137 & 0.138 &  & 94.8 & 95.0 & 95.1 \\ 
                 \multicolumn{19}{l}{\textbf{Adjusted}}\\
				Na\"ive & 0.065 & 0.067 &  &  &  & 95.9 &  &  &  & 0.087 & 0.089 &  &  &  & 95.2 &  &  &  \\ 
				Fixed & 0.079 &  & 0.080 & 0.079 & 0.079 &  & 95.6 & 95.2 & 94.9 & 0.102 &  & 0.100 & 0.099 & 0.098 &  & 94.2 & 93.8 & 93.5 \\ 
				Mixed(1$\mid$c)  & 0.082 &  & 0.083 & 0.083 & 0.085 &  & 94.9 & 95.1 & 96.0 & 0.110 &  & 0.110 & 0.110 & 0.112 &  & 95.1 & 95.2 & 95.4 \\ 
                Mixed(1$\mid$c) Sam. & 0.104 &  & 0.102 & 0.102 & 0.102 &  & 94.5 & 94.6 & 94.4 & 0.128 &  & 0.125 & 0.125 & 0.126 &  & 94.6 & 94.4 & 94.6 \\
				Mixed(1+A$\mid$c) & 0.082 &  & 0.083 & 0.083 & 0.085 &  & 94.9 & 95.1 & 96.0 & 0.110 &  & 0.110 & 0.110 & 0.112 &  & 95.6 & 95.5 & 95.5 \\
                Mixed(1+A$\mid$c) Sam. & 0.104 &  & 0.102 & 0.102 & 0.102 &  & 94.7 & 94.7 & 94.3 & 0.128 &  & 0.125 & 0.125 & 0.126 &  & 94.7 & 94.5 & 94.8 \\ 
				\bottomrule						
				\multicolumn{19}{c}{\textit{Random-effects variances}: $\sigma^2_{b0}$=0.15, $\sigma^2_{b1}$=0, $\sigma^2_{b2}$=0: $\bar{\rho}_1=0.0603$, $\bar{\rho}= -0.0007$}\\
                 \multicolumn{19}{l}{\textbf{Unadjusted}}\\
				Na\"ive & 0.083 & 0.074 &  &  &  & 92.4 &  &  &  & 0.102 & 0.104 &  &  &  & 96.2 &  &  &  \\ 
				Fixed & 0.092 &  & 0.092 & 0.092 & 0.088 &  & 94.5 & 94.5 & 93.7 & 0.115 &  & 0.111 & 0.110 & 0.109 &  & 94.6 & 93.9 & 93.7 \\ 
				Mixed(1$\mid$c) & 0.095 &  & 0.095 & 0.095 & 0.095 &  & 94.7 & 95.0 & 94.9 & 0.119 &  & 0.119 & 0.120 & 0.121 &  & 95.2 & 95.2 & 95.7 \\ 
                Mixed(1$\mid$c) Sam. & 0.106 &  & 0.105 & 0.105 & 0.105 &  & 95.0 & 95.1 & 94.7 & 0.138 &  & 0.136 & 0.136 & 0.138 &  & 94.3 & 94.2 & 94.8 \\
				Mixed(1+A$\mid$c) & 0.095 &  & 0.095 & 0.095 & 0.095 &  & 94.8 & 94.9 & 95.1 & 0.120 &  & 0.120 & 0.121 & 0.122 &  & 95.3 & 95.4 & 95.8 \\ 
                Mixed(1+A$\mid$c) Sam. & 0.106 &  & 0.105 & 0.105 & 0.105 &  & 95.1 & 95.2 & 94.7 & 0.138 &  & 0.136 & 0.136 & 0.138 &  & 94.3 & 94.2 & 94.8 \\ 
                 \multicolumn{19}{l}{\textbf{Adjusted}}\\
				Na\"ive & 0.081 & 0.071 &  &  &  & 91.3 &  &  &  & 0.094 & 0.095 &  &  &  & 94.8 &  &  &  \\ 
				Fixed & 0.089 &  & 0.090 & 0.089 & 0.086 &  & 95.1 & 95.2 & 94.4 & 0.104 &  & 0.099 & 0.098 & 0.098 &  & 94.3 & 94.1 & 93.8 \\ 
				Mixed(1$\mid$c) & 0.092 &  & 0.092 & 0.092 & 0.092 &  & 95.2 & 95.3 & 94.6 & 0.107 &  & 0.107 & 0.107 & 0.108 &  & 94.7 & 94.5 & 94.9\\ 
                Mixed(1$\mid$c) Sam. & 0.102 &  & 0.102 & 0.102 & 0.102 &  & 95.0 & 95.0 & 94.9 & 0.126 &  & 0.125 & 0.125 & 0.127 &  & 95.1 & 95.1 & 95.2\\ 
				Mixed(1+A$\mid$c) & 0.092 &  & 0.091 & 0.091 & 0.091 &  & 94.8 & 94.9 & 95.1 & 0.108 &  & 0.108 & 0.108 & 0.109 &  & 95.3 & 95.4 & 95.8 \\	
                Mixed(1+A$\mid$c) Sam. & 0.102 &  & 0.102 & 0.102 & 0.102 &  & 95.1 & 95.2 & 94.7 & 0.126 &  & 0.125 & 0.126 & 0.127 &  & 94.9 & 95.0 & 95.1 \\ 
				\bottomrule				
				\multicolumn{19}{c}{\textit{Random-effects variances}: $\sigma^2_{b0}$=0.15, $\sigma^2_{b1}$=0.15, $\sigma^2_{b2}$=0: $\bar{\rho}_1=0.1155$, $\bar{\rho}= 0.0331$}\\
                 \multicolumn{19}{l}{\textbf{Unadjusted}}\\
				Na\"ive & 0.099 & 0.078 &  &  &  & 88.4 &  &  &  & 0.116 & 0.107 &  &  &  & 92.9 &  &  &  \\ 
				Fixed & 0.101 &  & 0.101 & 0.100 & 0.096 &  & 94.9 & 94.6 & 93.4 & 0.124 &  & 0.116 & 0.113 & 0.110 &  & 92.7 & 91.6 & 91.1 \\ 
				Mixed(1$\mid$c) & 0.104 &  & 0.105 & 0.105 & 0.105 &  & 94.8 & 95.2 & 95.1 & 0.129 &  & 0.125 & 0.124 & 0.125 &  & 94.7 & 94.6 & 94.6 \\ 
                Mixed(1$\mid$c) Sam. & 0.107 &  & 0.105 & 0.105 & 0.105 &  & 95.2 & 95.1 & 94.5 & 0.139 &  & 0.136 & 0.136 & 0.137 &  & 94.5 & 94.4 & 94.1\\
				Mixed(1+A$\mid$c) & 0.104 &  & 0.104 & 0.104 & 0.103 &  & 94.7 & 94.6 & 94.7 & 0.130 &  & 0.126 & 0.126 & 0.127 &  & 94.9 & 95.0 & 94.6 \\ 
                Mixed(1+A$\mid$c) Sam. & 0.107 &  & 0.105 & 0.105 & 0.105 &  & 95.2 & 95.1 & 94.4 & 0.139 &  & 0.136 & 0.136 & 0.137 &  & 94.4 & 94.1 & 94.2 \\
                 \multicolumn{19}{l}{\textbf{Adjusted}}\\
				Na\"ive & 0.097 & 0.075 &  &  &  & 87.9 &  &  &  & 0.109 & 0.098 &  &  &  & 92.3 &  &  &  \\ 
				Fixed & 0.099 &  & 0.098 & 0.098 & 0.094 &  & 94.5 & 94.2 & 93.5 & 0.116 &  & 0.105 & 0.103 & 0.100 &  & 92.5 & 92.1 & 91.4 \\ 
				Mixed(1$\mid$c) & 0.101 &  & 0.102 & 0.102 & 0.101 &  & 95.3 & 95.1 & 94.9 & 0.118 &  & 0.113 & 0.112 & 0.112 &  & 93.4 & 93.5 & 93.3 \\ 
                Mixed(1$\mid$c) Sam. & 0.103 &  & 0.102 & 0.102 & 0.101 &  & 94.8 & 94.8 & 94.6 & 0.129 &  & 0.125 & 0.125 & 0.126 &  & 93.9 & 94.0 & 93.9 \\ 
				Mixed(1+A$\mid$c) & 0.101 &  & 0.100 & 0.100 & 0.100 &  & 94.7 & 94.6 & 94.7 & 0.119 &  & 0.114 & 0.114 & 0.114 &  & 94.9 & 95.0 & 94.6 \\  
                Mixed(1+A$\mid$c) Sam. & 0.103 &  & 0.102 & 0.102 & 0.101 &  & 94.9 & 95.0 & 94.8 & 0.129 &  & 0.125 & 0.125 & 0.126 &  & 93.9 & 93.9 & 93.9 \\ 
				\bottomrule	
				\multicolumn{19}{c}{\textit{Random-effects variances}: $\sigma^2_{b0}$=0.15, $\sigma^2_{b1}$=0.15, $\sigma^2_{b2}=4\times10^{-6}$: $\bar{\rho}_1=0.2190$, $\bar{\rho}=0.0280$}\\
                 \multicolumn{19}{l}{\textbf{Unadjusted}}\\
			Na\"ive & 0.135 & 0.091 &  &  &  & 80.9 &  &  &  & 0.137 & 0.126 &  &  &  & 92.8 &  &  &  \\ 
			Fixed & 0.126 &  & 0.130 & 0.130 & 0.126 &  & 95.4 & 95.5 & 94.2 & 0.130 &  & 0.118 & 0.116 & 0.113 &  & 93.0 & 92.6 & 91.2 \\ 
			Mixed(1$\mid$c) & 0.127 &  & 0.132 & 0.132 & 0.131 &  & 95.9 & 95.7 & 94.6 & 0.136 &  & 0.123 & 0.123 & 0.121 &  & 93.8 & 93.6 & 93.0 \\ 
            Mixed(1$\mid$c) Sam. & 0.109 &  & 0.106 & 0.106 & 0.105 &  & 93.7 & 93.8 & 94.4 & 0.139 &  & 0.136 & 0.136 & 0.138 &  & 94.7 & 95.0 & 95.1 \\ 
			Mixed(1+A$\mid$c) & 0.127 &  & 0.131 & 0.131 & 0.130 &  & 95.6 & 95.6 & 94.8 & 0.137 &  & 0.125 & 0.124 & 0.122 &  & 94.0 & 93.7 & 93.0 \\ 
            Mixed(1+A$\mid$c) Sam. & 0.109 &  & 0.106 & 0.106 & 0.105 &  & 93.8 & 93.8 & 94.4 & 0.139 &  & 0.136 & 0.136 & 0.138 &  & 94.8 & 95.1 & 95.0 \\ 
             \multicolumn{19}{l}{\textbf{Adjusted}}\\
			Na\"ive & 0.134 & 0.088 &  &  &  & 80.2 &  &  &  & 0.130 & 0.118 &  &  &  & 92.0 &  &  &  \\ 
			Fixed & 0.123 &  & 0.128 & 0.128 & 0.124 &  & 95.9 & 95.9 & 94.9 & 0.119 &  & 0.107 & 0.105 & 0.102 &  & 92.4 & 91.8 & 90.7 \\ 
			Mixed(1$\mid$c) & 0.125 &  & 0.130 & 0.130 & 0.128 &  & 95.3 & 95.4 & 94.9 & 0.125 &  & 0.112 & 0.111 & 0.109 &  & 91.7 & 91.4 & 91.2 \\ 
            Mixed(1$\mid$c) Sam. & 0.106 &  & 0.103 & 0.102 & 0.102 &  & 94.0 & 93.8 & 93.4 & 0.127 &  & 0.125 & 0.125 & 0.126 &  & 95.4 & 95.9 & 96.0 \\ 
			Mixed(1+A$\mid$c) & 0.125 &  & 0.129 & 0.129 & 0.128 &  & 95.6 & 95.6 & 94.8 & 0.126 &  & 0.113 & 0.112 & 0.110 &  & 94.0 & 93.7 & 93.0 \\ 
            Mixed(1+A$\mid$c) Sam. & 0.106 &  & 0.103 & 0.103 & 0.102 &  & 94.0 & 93.9 & 93.3 & 0.127 &  & 0.125 & 0.126 & 0.127 &  & 95.4 & 95.5 & 95.8 \\ 
				\bottomrule	
			\end{tabular}%
		}
	\end{center}
	\tiny
	\vspace{0.01cm} 
	$\bar{\rho}_1$ and $\bar{\rho}$ represent the mean values of intra-class correlation in the influence function of the counterfactual mean and the treatment effect, respectively, across 1000 simulations.
  \label{table:k_100_nc_5_cont}
\end{table}

Although settings with many large centers are not the primary focus of this study, we assess the performance of the estimators under such conditions in the most difficult case with high values of the random-effects variances (see \autoref{table:k_100_nc_100_cont}). The consequence of not accounting for clustering is higher in large centers than in small centers, regardless of the number of centers (e.g., see \autoref{table:k_100_nc_5_cont}, \autoref{table:k_100_nc_100_cont} and Table B.1.2 in the Appendix B.1), but adjusting for center is more subtle when having (many) small centers (e.g., non-negligible estimation error in shrinkage predictions). The reason is that in a setting with small centers, a high within-center variation dominates the total variance, making clustering effects less significant. In case of a higher number of large centers (e.g., Setting 5), the REML, DL and DB variance estimators provide coverage probabilities close to the nominal level. Furthermore, in this setting, the results of fixed-effects and mixed-effects are very similar. Moreover, in this setting, estimating random-effects by sampling from their normal distributions and using empirical BLUPs yield equivalent results (see \autoref{table:k_100_nc_100_cont}).

\begin{table}
\caption{Simulation results: Monte Carlo standard deviation, average standard errors and coverage probabilities of 95\% confidence intervals for both counterfactual mean on treatment and ATE based on 1000 simulations. Setting 3: Continuous outcome, $k$=100, $n_c$: Avg=100, Min=50, Max=145, and at higher values of random-effects variance. Results are based on weighting centers equally.}
	\begin{center}
		\resizebox{\textwidth}{!}{
			\begin{tabular}{>{\raggedright\arraybackslash}p{3.5cm}cccccccccccccccccc}
				\toprule
				& \multicolumn{9}{c}{\textbf{Counterfactual mean on treatment}} &\multicolumn{9}{c}{\textbf{ATE}} \\
				\cmidrule(lr){2-10} \cmidrule(lr){11-19}
				&& \multicolumn{4}{c}{\textbf{SE}} & \multicolumn{4}{c}{\textbf{Coverage (\%)}} &&\multicolumn{4}{c}{\textbf{SE}}&\multicolumn{4}{c}{\textbf{Coverage (\%)}} \\
				\cmidrule(lr){3-6} \cmidrule(lr){7-10} \cmidrule(lr){12-15} \cmidrule(lr){16-19}
				Method &  SD & Na\"ive & REML & DL & DB & Na\"ive & REML & DL & DB  & SD & Na\"ive & REML & DL & DB & Naive & REML & DL & DB \\ \midrule
				\multicolumn{19}{c}{\textit{Random-effects variances}: $\sigma^2_{b0}$=0.15, $\sigma^2_{b1}$=0.15, $\sigma^2_{b2}=4\times10^{-6}$: $\bar{\rho}_1=0.2185$, $\bar{\rho}=0.0237$}\\
                \multicolumn{19}{l}{\textbf{Unadjusted}}\\
				Na\"ive & 0.060 & 0.017 &  &  &  & 43.1 &  &  &  & 0.047 & 0.024 &  &  &  & 69.4 &  &  &  \\ 
				Fixed & 0.057 &  & 0.057 & 0.057 & 0.057 &  & 95.9 & 95.9 & 95.5 & 0.046 &  & 0.045 & 0.045 & 0.045 &  & 94.2 & 94.2 & 94.3 \\ 
				Mixed(1$\mid$c) & 0.057 &  & 0.057 & 0.057 & 0.057 &  & 95.9 & 95.9 & 95.6 & 0.046 &  & 0.045 & 0.045 & 0.045 &  & 94.3 & 94.3 & 94.2 \\ 
                Mixed(1$|$c) Sam. & 0.058 &  & 0.057 & 0.057 & 0.057 &  & 95.0 & 94.9 & 94.9 & 0.045 &  & 0.046 & 0.046 & 0.046 &  & 96.0 & 96.0 & 95.9 \\ 
				Mixed(1+A$\mid$c) & 0.057 &  & 0.057 & 0.057 & 0.057 &  & 95.6 & 96.0 & 95.5 & 0.046 &  & 0.045 & 0.045 & 0.045 &  & 94.3 & 94.3 & 94.5 \\ 
                Mixed(1+A$|$c) Sam. & 0.058 &  & 0.057 & 0.057 & 0.057 &  & 95.0 & 94.9 & 95.0 & 0.045 &  & 0.046 & 0.046 & 0.046 &  & 96.0 & 96.0 & 95.8 \\ 
                \multicolumn{19}{l}{\textbf{Adjusted}}\\
				Na\"ive & 0.060 & 0.017 &  &  &  & 42.2 &  &  &  & 0.046 & 0.022 &  &  &  & 64.5 &  &  &  \\ 
				Fixed & 0.057 &  & 0.057 & 0.057 & 0.056 &  & 95.7 & 95.6 & 95.5 & 0.045 &  & 0.044 & 0.044 & 0.043 &  & 95.2 & 95.3 & 95.1 \\ 
				Mixed(1$\mid$c) & 0.057 &  & 0.057 & 0.057 & 0.056 &  & 95.7 & 95.7 & 95.6 & 0.045 &  & 0.044 & 0.044 & 0.043 &  & 95.2 & 95.5 & 95.1 \\ 
                Mixed(1$|$c) Sam. & 0.058 &  & 0.057 & 0.057 & 0.057 &  & 95.3 & 95.1 & 95.1 & 0.045 &  & 0.045 & 0.045 & 0.044 &  & 95.5 & 95.5 & 95.3 \\ 
				Mixed(1+A$\mid$c) & 0.057 &  & 0.057 & 0.057 & 0.057 &  & 95.6 & 96.0 & 95.5 & 0.045 &  & 0.044 & 0.044 & 0.044 &  & 94.3 & 94.3 & 94.5 \\ 
                Mixed(1+A$|$c) Sam. & 0.058 &  & 0.057 & 0.057 & 0.057 &  & 95.3 & 95.2 & 95.1 & 0.044 &  & 0.045 & 0.045 & 0.044 &  & 95.5 & 95.5 & 95.3 \\ 
				\midrule												
			\end{tabular}%
		}
	\end{center}
	\tiny
	\vspace{0.01cm} 
$\bar{\rho}_1$ and $\bar{\rho}$ represent the mean values of intra-class correlation in the influence function of the counterfactual mean and the treatment effect, respectively, across 1000 simulations.
  \label{table:k_100_nc_100_cont}
\end{table}

Furthermore, we evaluate the performance of estimators by assigning equal weights to patients. Under the assumption that treatment effects are independent of center size, the center-weighted and patient-weighted estimands coincide. However, patient-weighted estimators are generally more efficient than center-weighted estimators (see, for example, \autoref{table:k_100_nc_5_cont} and Table B.1.23 in the Appendix B.1). 

Finally, the proposed estimators perform well under model misspecification, that is, when the outcome model includes only the main effects of the covariates while the true model involves a covariate transformation and an interaction term. For example, see Tables B.1.5 and B.1.11 in Appendix B.1, where results are based on weighting centers equally. This follows from the robustness properties of AIPW estimators.

\subsubsection*{Binary outcomes}
We only show the results (based on weighting centers equally) under Setting 1 and Setting 3 (see \autoref{table:k_100_nc_5_bin} and \autoref{table:k_100_nc_100_bin}); other results can be found in Appendix B.1. 

When only the intercept varies between centers, the coverage of confidence intervals for the ATE using the na\"ive variance estimators remains at the nominal level.
While theory does not support this (see Appendix A.3), it is likely driven by the small intraclass correlation coefficient (ICC).
For example, over 1000 simulations, the average ICC value is -0.0010 with parameters: $\E\left(b_{0c}\right)=0, \sigma^2_{b0}=0.5$, $\E\left(b_{1c}\right)=0, \sigma^2_{b1}=0$, $\E\left(b_{2c}\right)=0$, and $\sigma^2_{b2}=0$. Similarly, over 1000 simulations, the average ICC value is -0.0009 with parameters: $\E\left(b_{0c}\right)=2, \sigma^2_{b0}=1$, $\E\left(b_{1c}\right)=0, \sigma^2_{b1}=0$, $\E\left(b_{2c}\right)=0$, and $\sigma^2_{b2}=0$. However, when there is variation in treatment effect between centers, the confidence intervals for ATE obtained from the na\"ive variance estimators are no longer valid (see \autoref{table:k_100_nc_5_bin}). We further observe that the confidence intervals for the counterfactual mean on treatment based on the na\"ive variance estimators are not valid unless there is no variation in the intercept, treatment effect, and covariate effects between centers (see \autoref{table:k_100_nc_5_bin}).

For Setting 1, where there are relatively many small centers, fixed-effects estimators lead to biased estimates of both the counterfactual mean on treatment and the ATE (see the bias estimates in Table B.1.17 in the Appendix B.1 corresponding to the simulation setting of \autoref{table:k_100_nc_5_bin}). This is a consequence of the fixed-effects model containing numerous parameters relative to the small number of subjects. Whether only center is included as random-effects or both center and treatment are included, the performance remains similar across different data generation processes. As for continuous outcomes, in this setting, estimating random-effects by sampling from their normal distributions yields slightly better coverage probabilities than using empirical BLUPs for the average treatment effect, particularly when the random-effects variances are high ($\sigma^2_{b0}=0.75$, $\sigma^2_{b1}=0.5$, $\sigma^2_{b2}=0.5$). Moreover, the REML, DL and DB variance estimators perform comparably well (see \autoref{table:k_100_nc_5_bin}). 

\begin{table}
	\caption{Simulation results: Monte Carlo standard deviation, average standard errors and coverage probabilities of 95\% confidence intervals for both counterfactual mean on treatment and ATE based on 1000 simulations. Setting 1: Binary outcome, $k$=100, $n_c$: Avg=5, Min=1, Max=24, and at different values of random-effects variance. Results are based on weighting centers equally.}
	\begin{center}
		\resizebox{\textwidth}{!}{
			\begin{tabular}{>{\raggedright\arraybackslash}p{3.5cm}cccccccccccccccccc}
				\toprule
				& \multicolumn{9}{c}{\textbf{Counterfactual mean on treatment}} &\multicolumn{9}{c}{\textbf{ATE}} \\
				\cmidrule(lr){2-10} \cmidrule(lr){11-19}
				&& \multicolumn{4}{c}{\textbf{SE}} & \multicolumn{4}{c}{\textbf{Coverage (\%)}} &&\multicolumn{4}{c}{\textbf{SE}}&\multicolumn{4}{c}{\textbf{Coverage (\%)}} \\
				\cmidrule(lr){3-6} \cmidrule(lr){7-10} \cmidrule(lr){12-15} \cmidrule(lr){16-19}
				Method &  SD & Na\"ive & REML & DL & DB & Na\"ive & REML & DL & DB  & SD & Na\"ive & REML & DL & DB & Naive & REML & DL & DB \\ \midrule
				\multicolumn{19}{c}{\textit{Random-effects variances}: $\sigma^2_{b0}$=0, $\sigma^2_{b1}$=0, $\sigma^2_{b2}$=0: $\bar{\rho_1}=0.0007$, $\bar{\rho}=0.0000$}\\
                 \multicolumn{19}{l}{\textbf{Unadjusted}}\\
				Na\"ive & 0.031 & 0.031 &  &  &  & 94.8 &  &  &  & 0.043 & 0.044 &  &  &  & 94.7 &  &  &  \\ 
				Fixed & 0.036 &  & 0.037 & 0.037 & 0.037 &  & 95.0 & 94.8 & 94.6 & 0.047 &  & 0.050 & 0.049 & 0.049 &  & 95.4 & 95.1 & 95.0 \\ 
				Mixed(1$\mid$c) & 0.039 &  & 0.039 & 0.039 & 0.040 &  & 95.7 & 95.6 & 95.6 & 0.054 &  & 0.055 & 0.055 & 0.055 &  & 95.2 & 95.3 & 95.2 \\ 
                Mixed(1$\mid$c) Sam. & 0.038 &  & 0.039 & 0.039 & 0.040 &  & 95.8 & 95.7 & 95.7 & 0.054 &  & 0.055 & 0.055 & 0.055 &  & 95.4 & 95.5 & 95.5 \\
				Mixed(1+A$\mid$c) & 0.039 &  & 0.039 & 0.039 & 0.039 &  & 95.8 & 95.6 & 95.4 & 0.054 &  & 0.055 & 0.055 & 0.055 &  & 95.2 & 95.3 & 95.3 \\ 
                Mixed(1+A$\mid$c) Sam. & 0.038 &  & 0.039 & 0.039 & 0.040 &  & 95.7 & 95.7 & 95.7 & 0.054 &  & 0.055 & 0.055 & 0.055 &  & 95.4 & 95.5 & 95.5 \\ 
                 \multicolumn{19}{l}{\textbf{Adjusted}}\\
				Na\"ive & 0.030 & 0.030 &  &  &  & 94.6 &  &  &  & 0.042 & 0.041 &  &  &  & 94.5 &  &  &  \\ 
				Fixed & 0.036 &  & 0.036 & 0.036 & 0.036 &  & 94.9 & 94.8 & 94.5 & 0.045 &  & 0.046 & 0.045 & 0.045 &  & 94.9 & 94.4 & 94.4 \\ 
				Mixed(1$\mid$c) & 0.038 &  & 0.038 & 0.038 & 0.038 &  & 94.6 & 94.7 & 94.9 & 0.052 &  & 0.051 & 0.051 & 0.052 &  & 94.8 & 94.6 & 94.6 \\ 
                Mixed(1$\mid$c) Sam. & 0.037 &  & 0.038 & 0.038 & 0.038 &  & 95.5 & 95.5 & 95.4 & 0.050 &  & 0.051 & 0.051 & 0.052 &  & 95.7 & 95.5 & 95.6 \\ 
				Mixed(1+A$\mid$c) & 0.038 &  & 0.038 & 0.038 & 0.038 &  & 95.8 & 95.6 & 95.4 & 0.052 &  & 0.051 & 0.051 & 0.052 &  & 95.2 & 95.3 & 95.3 \\
                Mixed(1+A$\mid$c) Sam. & 0.037 &  & 0.038 & 0.038 & 0.038 &  & 95.6 & 95.6 & 95.6 & 0.050 &  & 0.051 & 0.051 & 0.052 &  & 95.6 & 95.5 & 95.6 \\ 
				\bottomrule
				\multicolumn{19}{c}{\textit{Random-effects variances}: $\sigma^2_{b0}$=0.5, $\sigma^2_{b1}$=0, $\sigma^2_{b2}$=0: $\bar{\rho}_1=0.04228$, $\bar{\rho}= -0.0010$}\\
                 \multicolumn{19}{l}{\textbf{Unadjusted}}\\
				Na\"ive & 0.035 & 0.031 &  &  &  & 92.6 &  &  &  & 0.043 & 0.044 &  &  &  & 95.4 &  &  &  \\ 
				Fixed & 0.038 &  & 0.039 & 0.038 & 0.037 &  & 95.6 & 95.5 & 94.1 & 0.045 &  & 0.048 & 0.047 & 0.047 &  & 96.0 & 95.8 & 95.7 \\ 
				Mixed(1$\mid$c) & 0.040 &  & 0.040 & 0.040 & 0.040 &  & 95.2 & 95.2 & 95.2 & 0.052 &  & 0.052 & 0.052 & 0.052 &  & 94.5 & 94.4 & 94.5 \\ 
                Mixed(1$\mid$c) Sam. & 0.041 &  & 0.040 & 0.041 & 0.041 &  & 94.9 & 94.9 & 94.6 & 0.055 &  & 0.055 & 0.055 & 0.055 &  & 95.7 & 95.6 & 95.6 \\ 
				Mixed(1+A$\mid$c) & 0.040 &  & 0.040 & 0.040 & 0.040 &  & 95.1 & 95.1 & 95.0 & 0.052 &  & 0.052 & 0.052 & 0.052 &  & 94.5 & 94.4 & 94.6 \\ 
                Mixed(1+A$\mid$c) Sam. & 0.041 &  & 0.040 & 0.041 & 0.041 &  & 94.9 & 94.9 & 94.6 & 0.055 &  & 0.055 & 0.055 & 0.055 &  & 95.6 & 95.6 & 95.6 \\ 
                 \multicolumn{19}{l}{\textbf{Adjusted}}\\
				Na\"ive & 0.034 & 0.031 &  &  &  & 92.1 &  &  &  & 0.041 & 0.042 &  &  &  & 95.3 &  &  &  \\ 
				Fixed & 0.037 &  & 0.038 & 0.037 & 0.036 &  & 95.7 & 95.5 & 94.6 & 0.042 &  & 0.044 & 0.043 & 0.043 &  & 95.7 & 95.6 & 95.5 \\ 
				Mixed(1$\mid$c) & 0.039 &  & 0.039 & 0.039 & 0.039 &  & 95.6 & 95.3 & 94.7 & 0.049 &  & 0.048 & 0.048 & 0.049 &  & 94.9 & 95.1 & 95.0 \\
                Mixed(1$\mid$c) Sam. & 0.040 &  & 0.039 & 0.039 & 0.040 &  & 94.2 & 94.3 & 94.2 & 0.051 &  & 0.052 & 0.052 & 0.052 &  & 95.0 & 95.0 & 95.3 \\ 
				Mixed(1+A$\mid$c) & 0.039 &  & 0.039 & 0.039 & 0.039 &  & 95.1 & 95.1 & 95.0 & 0.049 &  & 0.048 & 0.048 & 0.049 &  & 94.5 & 94.4 & 94.6 \\   
                Mixed(1+A$\mid$c) Sam. & 0.040 &  & 0.040 & 0.040 & 0.040 &  & 94.3 & 94.4 & 94.2 & 0.051 &  & 0.052 & 0.052 & 0.052 &  & 95.2 & 95.1 & 95.1 \\ 
				\bottomrule
				\multicolumn{19}{c}{\textit{Random-effects variances}: $\sigma^2_{b0}$=0.5, $\sigma^2_{b1}$=0.5, $\sigma^2_{b2}$=0: $\bar{\rho}_1=0.0796$, $\bar{\rho}=0.0214$}\\
                 \multicolumn{19}{l}{\textbf{Unadjusted}}\\
				Na\"ive & 0.037 & 0.031 &  &  &  & 89.4 &  &  &  & 0.047 & 0.044 &  &  &  & 92.8 &  &  &  \\ 
				Fixed & 0.037 &  & 0.040 & 0.040 & 0.038 &  & 95.9 & 95.5 & 94.4 & 0.046 &  & 0.048 & 0.047 & 0.047 &  & 95.6 & 94.9 & 94.6 \\ 
				Mixed(1$\mid$c) & 0.040 &  & 0.042 & 0.042 & 0.042 &  & 95.8 & 95.8 & 95.6 & 0.054 &  & 0.053 & 0.053 & 0.052 &  & 94.5 & 94.3 & 94.1 \\ 
                Mixed(1$\mid$c) Sam. & 0.042 &  & 0.042 & 0.042 & 0.042 &  & 94.5 & 94.4 & 94.2 & 0.055 &  & 0.056 & 0.056 & 0.056 &  & 94.4 & 94.5 & 94.9 \\ 
				Mixed(1+A$\mid$c) & 0.040 &  & 0.041 & 0.041 & 0.041 &  & 95.6 & 95.6 & 95.4 & 0.054 &  & 0.053 & 0.053 & 0.053 &  & 94.4 & 94.2 & 93.8 \\ 
                Mixed(1+A$\mid$c) Sam. & 0.042 &  & 0.042 & 0.042 & 0.042 &  & 94.5 & 94.4 & 94.2 & 0.055 &  & 0.056 & 0.056 & 0.056 &  & 94.4 & 94.4 & 94.9 \\ 
                 \multicolumn{19}{l}{\textbf{Adjusted}}\\
				Na\"ive & 0.037 & 0.031 &  &  &  & 89.1 &  &  &  & 0.045 & 0.042 &  &  &  & 92.8 &  &  &  \\ 
				Fixed & 0.037 &  & 0.039 & 0.039 & 0.037 &  & 96.1 & 96.0 & 94.6 & 0.044 &  & 0.045 & 0.044 & 0.043 &  & 95.1 & 94.5 & 94.2 \\ 
				Mixed(1$\mid$c) & 0.040 &  & 0.041 & 0.041 & 0.041 &  & 95.5 & 95.3 & 95.2 & 0.051 &  & 0.049 & 0.049 & 0.049 &  & 93.7 & 93.5 & 93.6 \\ 
                Mixed(1$\mid$c) Sam. & 0.042 &  & 0.041 & 0.041 & 0.041 &  & 94.3 & 94.3 & 94.0 & 0.053 &  & 0.053 & 0.053 & 0.053 &  & 94.6 & 94.6 & 94.5 \\ 
				Mixed(1+A$\mid$c) & 0.040 &  & 0.040 & 0.040 & 0.040 &  & 95.6 & 95.6 & 95.4 & 0.051 &  & 0.049 & 0.049 & 0.049 &  & 94.4 & 94.2 & 93.8 \\ 
                Mixed(1+A$\mid$c) Sam. & 0.042 &  & 0.041 & 0.041 & 0.041 &  & 94.2 & 94.1 & 94.0 & 0.053 &  & 0.053 & 0.053 & 0.053 &  & 94.4 & 94.6 & 94.2 \\ 
				\bottomrule		
				\multicolumn{19}{c}{\textit{Random-effects variances}: $\sigma^2_{b0}$=0.75, $\sigma^2_{b1}$=0.5, $\sigma^2_{b2}$=0.5: $\bar{\rho}_1=0.0974$, $\bar{\rho}=0.0195$}\\
                 \multicolumn{19}{l}{\textbf{Unadjusted}}\\
				Na\"ive & 0.040 & 0.031 &  &  &  & 86.7 &  &  &  & 0.047 & 0.044 &  &  &  & 93.6 &  &  &  \\ 
				Fixed & 0.041 &  & 0.041 & 0.040 & 0.038 &  & 94.6 & 94.2 & 93.0 & 0.046 &  & 0.047 & 0.046 & 0.046 &  & 95.4 & 95.3 & 95.0 \\ 
				Mixed(1$\mid$c) & 0.044 &  & 0.042 & 0.042 & 0.042 &  & 94.3 & 94.4 & 93.7 & 0.054 &  & 0.051 & 0.051 & 0.051 &  & 93.4 & 93.3 & 93.5 \\ 
                Mixed(1$\mid$c) Sam. & 0.042 &  & 0.042 & 0.042 & 0.042 &  & 95.5 & 95.4 & 95.5 & 0.054 &  & 0.056 & 0.056 & 0.056 &  & 95.8 & 95.7 & 95.8 \\ 
				Mixed(1+A$\mid$c) & 0.044 &  & 0.042 & 0.041 & 0.041 &  & 94.2 & 94.1 & 93.4 & 0.054 &  & 0.051 & 0.051 & 0.051 &  & 93.7 & 93.7 & 93.6 \\ 
                Mixed(1+A$\mid$c) Sam. & 0.042 &  & 0.042 & 0.042 & 0.042 &  & 95.4 & 95.3 & 95.4 & 0.054 &  & 0.056 & 0.056 & 0.056 &  & 95.8 & 95.9 & 95.8 \\ 
                 \multicolumn{19}{l}{\textbf{Adjusted}}\\
				Na\"ive & 0.040 & 0.031 &  &  &  & 86.6 &  &  &  & 0.044 & 0.042 &  &  &  & 93.1 &  &  &  \\ 
				Fixed & 0.040 &  & 0.040 & 0.039 & 0.038 &  & 94.1 & 94.0 & 92.6 & 0.043 &  & 0.044 & 0.043 & 0.042 &  & 94.8 & 94.3 & 93.8 \\ 
				Mixed(1$\mid$c) & 0.043 &  & 0.041 & 0.041 & 0.041 &  & 94.0 & 94.0 & 93.2 & 0.050 &  & 0.048 & 0.048 & 0.048 &  & 93.8 & 93.9 & 93.8 \\ 
                Mixed(1$\mid$c) Sam. & 0.042 &  & 0.042 & 0.042 & 0.041 &  & 95.4 & 95.4 & 95.4 & 0.052 &  & 0.053 & 0.053 & 0.053 &  & 95.4 & 95.4 & 95.8 \\ 
				Mixed(1+A$\mid$c) & 0.043 &  & 0.041 & 0.041 & 0.040 &  & 94.2 & 94.1 & 93.4 & 0.050 &  & 0.048 & 0.048 & 0.048 &  & 93.7 & 93.7 & 93.6 \\ 
                Mixed(1+A$\mid$c) Sam. & 0.042 &  & 0.042 & 0.042 & 0.041 &  & 95.5 & 95.5 & 95.5 & 0.052 &  & 0.053 & 0.053 & 0.053 &  & 95.5 & 95.6 & 95.8 \\ 
				\bottomrule	
			\end{tabular}%
		}
	\end{center}
	\tiny
	\vspace{0.01cm} 
	$\bar{\rho}_1$ and $\bar{\rho}$ represent the mean values of intra-class correlation in the influence function of the counterfactual mean and the treatment effect, respectively, across 1000 simulations.
   \label{table:k_100_nc_5_bin}
\end{table}

Unlike in Setting 1 (Table B.1.17 in Appendix B.1), fixed-effects estimators produce unbiased effect estimates for both the counterfactual mean on treatment and the ATE in Setting 3, which features a higher number of large centers (Table B.1.21 in the Appendix B.1). This can be explained by the higher number of subjects in the fixed-effects model relative to the number of parameters. Moreover, in this setting, the coverage probabilities based on the ``na\"ive" variance estimators are significantly lower than those obtained under Setting 1 (see \autoref{table:k_100_nc_5_bin}) and Setting 2 (see Table B.1.13 in Appendix B.1), while the REML, DL and DB variance estimators provide coverage probabilities close to the nominal level of 95\% (see \autoref{table:k_100_nc_100_bin}). A possible reason for this result is the adequate number of centers and patients per center in Settings 1 and 2, allowing for reliable estimation of the heterogeneity variance. The ``na\"ive'' variance estimators lead to lower coverage probabilities because the failure to account for clustering has a greater impact in this setting. 

\begin{table}
	\caption{Simulation results: Monte Carlo standard deviation, average standard errors and coverage probabilities of 95\% confidence intervals for both counterfactual mean on treatment and ATE based on 1000 simulations. Setting 3: Binary outcome, $k$=100, $n_c$: Avg=100, Min=50, Max=145, and at higher values of random-effects variance. Results are based on weighting centers equally.}
	\begin{center}
		\resizebox{\textwidth}{!}{
			\begin{tabular}{>{\raggedright\arraybackslash}p{3.5cm}cccccccccccccccccc}
				\toprule
				& \multicolumn{9}{c}{\textbf{Counterfactual mean on treatment}} &\multicolumn{9}{c}{\textbf{ATE}} \\
				\cmidrule(lr){2-10} \cmidrule(lr){11-19}
				&& \multicolumn{4}{c}{\textbf{SE}} & \multicolumn{4}{c}{\textbf{Coverage (\%)}} &&\multicolumn{4}{c}{\textbf{SE}}&\multicolumn{4}{c}{\textbf{Coverage (\%)}} \\
				\cmidrule(lr){3-6} \cmidrule(lr){7-10} \cmidrule(lr){12-15} \cmidrule(lr){16-19}
				Method &  SD & Na\"ive & REML & DL & DB & Na\"ive & REML & DL & DB  & SD & Na\"ive & REML & DL & DB & Naive & REML & DL & DB \\ \midrule
				 \multicolumn{19}{c}{\textit{Random-effects variances}: $\sigma^2_{b0}$=0.75, $\sigma^2_{b1}$=0.5, $\sigma^2_{b2}$=0.50: $\bar{\rho}_1=0.0987$, $\bar{\rho}=0.0221$} \\
                 \multicolumn{19}{l}{\textbf{Unadjusted}}\\
				Na\"ive & 0.021 & 0.007 & & & & 49.1 &  &  &  & 0.016 & 0.010 &  &  &  & 77.7 &  &  &  \\ 
				Fixed & 0.021 &  & 0.022 & 0.022 & 0.022 &  & 95.7 & 95.7 & 95.7 & 0.016 &  & 0.016 & 0.016 & 0.016 &  & 96.0 & 96.2 & 96.0 \\ 
				Mixed(1$\mid$c) & 0.021 &  & 0.022 & 0.022 & 0.022 &  & 95.7 & 95.7 & 95.7 & 0.016 &  & 0.016 & 0.016 & 0.016 &  & 95.9 & 96.0 & 96.0 \\ 
                Mixed(1$\mid$c) Sam. & 0.022 &  & 0.022 & 0.022 & 0.022 &  & 94.3 & 94.4 & 94.2 & 0.016 &  & 0.017 & 0.017 & 0.016 &  & 96.0 & 96.0 & 95.9 \\ 
				Mixed(1+A$\mid$c) & 0.021 &  & 0.022 & 0.022 & 0.022 &  & 95.7 & 95.7 & 95.7 & 0.016 &  & 0.016 & 0.016 & 0.016 &  & 96.1 & 96.1 & 96.1 \\ 
                Mixed(1+A$\mid$c) Sam. & 0.022 &  & 0.022 & 0.022 & 0.022 &  & 94.3 & 94.4 & 94.2 & 0.016 &  & 0.017 & 0.017 & 0.016 &  & 96.0 & 96.0 & 95.8 \\ 
                \multicolumn{19}{l}{\textbf{Adjusted}}\\
				Na\"ive & 0.021 & 0.007 &  &  &  & 48.2 &  &  &  & 0.016 & 0.009 &  &  &  & 76.3 &  &  &  \\ 
				Fixed & 0.021 &  & 0.022 & 0.022 & 0.021 &  & 95.9 & 95.7 & 95.9 & 0.015 &  & 0.016 & 0.016 & 0.016 &  & 95.5 & 95.6 & 95.4 \\ 
				Mixed(1$\mid$c) & 0.021 &  & 0.022 & 0.022 & 0.021 &  & 96.0 & 95.7 & 95.9 & 0.015 &  & 0.016 & 0.016 & 0.016 &  & 95.8 & 95.8 & 95.8 \\ 
                Mixed(1$\mid$c) Sam. & 0.021 &  & 0.022 & 0.022 & 0.022 &  & 94.0 & 94.0 & 94.0 & 0.015 &  & 0.016 & 0.016 & 0.016 &  & 95.8 & 95.8 & 95.9 \\ 
				Mixed(1+A$\mid$c) & 0.021 &  & 0.022 & 0.022 & 0.022 &  & 95.7 & 95.7 & 95.7 & 0.015 &  & 0.016 & 0.016 & 0.016 &  & 96.1 & 96.1 & 96.1 \\ 
                Mixed(1+A$\mid$c) Sam. & 0.021 &  & 0.022 & 0.022 & 0.022 &  & 94.0 & 94.0 & 94.0 & 0.015 &  & 0.016 & 0.016 & 0.016 &  & 95.8 & 95.9 & 95.9 \\ 
				\midrule												
			\end{tabular}%
		}
	\end{center}
	\tiny
	\vspace{0.01cm} 
$\bar{\rho}_1$ and $\bar{\rho}$ represent the mean values of intra-class correlation in the influence function of the counterfactual mean and the treatment effect, respectively, across 1000 simulations.
\label{table:k_100_nc_100_bin}
\end{table}

We further evaluate the performance of estimators by assigning equal weights to patients. In general, patient-weighted estimators are more efficient than center-weighted estimators (see, for example, \autoref{table:k_100_nc_5_bin} and Table B.1.24 in in Appendix B.1). As for continuous outcomes, the proposed estimators perform well under model misspecification---specifically, when the outcome model includes only the main effects of the covariates, whereas the true model involves a covariate transformation (see; Table B.1.16 and Table B.1.22 in Appendix B.1).

\section{Data analysis}\label{sec:data_analysis_ch2}
\subsection{Data description} \label{sec:data_description_ch2}
We reanalyze the WASH Benefits Bangladesh study \citep{luby2018effects} to demonstrate the proposed methods. The study aimed to assess the impact of different interventions on diarrhoea prevalence and child growth.
Randomization was carried out in 90 geographically defined blocks, referred to as centers throughout, to enhance logistical feasibility and to ensure that the intervention groups are balanced with respect to characteristics and events that vary by location.
To minimize contamination, the study employed a cluster-randomized design, with geographically matched clusters (groups of household compounds in Bangladesh) within each center serving as the unit of randomization. The study included six intervention arms with a double-sized control arm (without intervention activities). Within each center, there is one cluster for each treatment arm and two clusters for the control arm.

We compare the combined water, sanitation, hand washing, and nutrition intervention (WSH+N) with the control group. The outcomes of interest are the length-for-age Z-score (a continuous variable) and diarrhoea prevalence (a binary variable). The sample sizes for these two outcomes differ: diarrhoea prevalence was assessed in both the birth cohort and in children under 36 months at enrollment, whereas the length-for-age Z-score was only measured in the birth cohort.
The distributions of the number of children per center within the intervention and control arms for each outcome are shown in Figure B.2.1 in Appendix B.2. 

As baseline covariates, we consider household food insecurity (\textit{hfi}), child age (\textit{chage}), child sex (\textit{chsex}), mother's age (\textit{momage}), mothers education level (\textit{momeduy}), father's education level (\textit{dadeduy}), number of children younger than 18 years in the household (\textit{Nlt18}), number of individuals living in the compound (\textit{Ncomp}), distance in minutes to the primary water source (\textit{watmin}), and whether household has electricity (\textit{elec}). Our confidence intervals adjust for the use of stratified randomization by conditioning on the stratification factor (i.e., center) in the outcome model \citep{wang2023model}, which we achieved via the use of mixed-effects models.

\subsection{Estimators}\label{DA:estimators_ch2}
We consider estimators analogous to those described in Section \ref{sub_sec:Simulation_estimators_ch2}. Because the data have a hierarchical structure--clusters nested within centers--the outcome models are extended to account for this multilevel dependence (in our proposed methods). 

We first consider a fixed-effects model that includes center indicators and the treatment indicator as fixed-effects. The resulting estimator is denoted \textit{Fixed}. In the adjusted analysis, center indicators, the treatment indicator, and the main effects of the covariates described in Section \ref{sec:data_description_ch2} are included as fixed-effects. 

Next, we consider a mixed-effects model that includes the treatment indicator as a fixed-effect and random intercepts for both centers and clusters. The random-effects are either estimated via empirical BLUPs or obtained by sampling from their estimated normal distributions, as described in Section \ref{sub_sec:proposed_approch}. When empirical BLUPs are used, the corresponding AIPW estimator is denoted \textit{Mixed(1$\mid$c:cl)}, whereas sampling from the estimated normal distributions yields \textit{Mixed(1$\mid$c:cl) Sam}. In the adjusted analysis, the treatment indicator and the main effects of the covariates described in Section \ref{sec:data_description_ch2} are included as fixed-effects, while random intercepts for centers and clusters are retained.

Finally, we consider a mixed-effects specification that additionally allows for random treatment effects at both the center and cluster levels, alongside random intercepts.  When the random-effects are estimated via empirical BLUPs, the resulting estimator is denoted \textit{Mixed(1+A$\mid$c:cl)}; when they are obtained by sampling from their estimated normal distributions, it is denoted \textit{Mixed(1+A$\mid$c:cl) Sam}. In the adjusted analysis, the treatment indicator and the main effects of the covariates (described in Section \ref{sec:data_description_ch2}) are included as fixed-effects, together with random intercepts and random treatment effects at both hierarchical levels.

For comparison, we also report results from an AIPW estimator based on a simple generalized linear model that includes only the treatment indicator as a fixed effect and ignores the hierarchical structure. This estimator is referred to as \textit{Na\"{\i}ve}. In the adjusted version, the treatment indicator and the main effects of the covariates are included as fixed effects, but no center- or cluster-level effects are modeled.

To clarify the construction of the AIPW estimator in this setting, consider estimation of the counterfactual mean under treatment (the same steps apply to the average treatment effect). Consistent with the original study \citep{luby2018effects}, we target an estimand that assigns equal weight to each center. Alternatively, one could define an estimand that weights all patients equally, irrespective of center.

First, for each cluster $j$ within center $c$, we compute the cluster-level mean defined as the average, over all individuals $i$ in that cluster, of
\begin{align*}
    \frac{A_{ijc}}{\hat{p}_{j}(X_{ijc})}\left\{Y_{ijc}-\hat{m}_{1jc}(X_{ijc})\right\}+\hat{m}_{1jc}(X_{ijc}).
\end{align*}
Here, $A_{ijc}$ denotes the treatment indicator for patient $i$ in cluster $j$ of center $c$, and $\hat{p}_{j}(X_{ijc})$ is the estimated probability of treatment for that patient. This probability is obtained from a mixed-effects logistic regression model with treatment assignment as the outcome, including center-specific random effects and baseline covariates as fixed effects. Furthermore, $Y_{ijc}$ denotes the observed outcome, and $\hat{m}_{1jc}(X_{ijc})$ is the predicted outcome under treatment from the specified outcome model for that patient. 
Finally, the overall mean outcome under treatment is obtained by averaging these cluster-level means uniformly across clusters within centers and then uniformly across centers, thereby ensuring equal weighting at the center level.

To estimate the variance of the estimated mean outcome under treatment, we consider a different approach to that in Equation \eqref{eq:total_variance} in order to accommodate the hierarchical structure of the data. In particular, we fit a multilevel model with the influence function under treatment as the outcome and include random-effects for centers and for clusters nested within centers. The fitted model provides variance components at the center, the cluster-within-center, and individual levels, which are then used to construct the variance estimator. Further details are given in Appendix A.7.

While our primary goal is to estimate a single intervention's effect versus control, we also evaluate the potential benefits of incorporating data from all treatment arms. In particular, the outcome model (e.g., a mixed-effects model) is fitted using observations from all arms across clusters and centers. Consequently, we define two groups of analyses, with each group including a na\"ive analysis (i.e., one that ignores correlation at both levels): analyses either limited to the WSH+N and control arms, or based on data from all arms.

\subsection{Results}
\autoref{table:WSH_N_ATE} presents the results of the WSH+N intervention compared to the control group on the length-for-age Z-score and diarrhoea prevalence. The corresponding counterfactual means are reported in Table B.2.1 in Appendix B.2. 
 
The confidence intervals obtained from the ``na\"ive" variance estimators are shorter than those obtained from the proposed variance estimators, particularly when random-effects are estimated by sampling from their normal distributions. This holds for both counterfactual mean on treatment and average treatment effect, as well as for both the continuous and binary endpoints. 

For example, consider mixed-effects estimators in which random-effects are estimated via sampling. In the unadjusted analysis using data from the two arms only, confidence intervals for the ATE on length-for-age $Z$-scores based on the proposed variance estimators are, relatively speaking, 23.27\%--23.76\% longer than those based on the ``na\"ive'' variance estimator (see \autoref{table:WSH_N_ATE}). For the corresponding counterfactual mean under treatment (see Table~B.2.1 in Appendix~B.2), the confidence intervals are 28.66\%--29.27\% longer than those obtained using the na\"ive variance estimator.
Similarly, in the adjusted analysis using data from the two arms only, confidence intervals for the ATE on length-for-age $Z$-scores based on the proposed variance estimators are 13.99\%--14.51\% longer than those based on the ``na\"ive'' variance estimator (see \autoref{table:WSH_N_ATE}). For the counterfactual mean under treatment, the corresponding increases are 23.57\%--24.20\% (see Table~B.2.1 in Appendix~B.2).

Similar results are obtained for the binary endpoint, diarrhoea prevalence. In the unadjusted two-arm analysis, confidence intervals for the ATE based on the proposed variance estimators are 27.27\% to 31.82\% longer than those based on the ``na\"ive" estimator (see \autoref{table:WSH_N_ATE}), while for the counterfactual mean under treatment they are 25\% to 37.5\% wider (see Table B.2.1 in Appendix B.2). For adjusted analyses, confidence intervals for the ATE are 27.27\% longer, and for the counterfactual mean under treatment are 29.41\% wider, relative to those based on the ``na\"ive" variance estimator.

Analyses that use data from all arms have smaller confidence intervals than analyses based on only two arms, particularly for mixed-effects estimators in which random-effects are estimated through sampling. Overall, the data analyses show that the confidence intervals obtained from the proposed variance estimators are wider than those from the ``na\"ive'' variance estimators for both the counterfactual mean under treatment and the average treatment effect, across both continuous and binary outcomes, reflecting and capturing the additional uncertainty.

\begin{table}[!h]
	\caption{Average treatment effect estimates and 95\% confidence intervals for the WSH+N intervention compared to the control group on length-for-age Z-score (continuous outcome) and the diarrhoea prevalence (binary outcome) in the WASH Benefits Bangladesh study. Results are based on weighting centers equally.}
	\begin{center}
		\resizebox{\textwidth}{!}{
			\begin{tabular}{>{\raggedright\arraybackslash}p{3.5cm}ccccc}
				\toprule
				& \multicolumn{2}{c}{\textbf{Length-for-age Z-score (continuous outcome)}} &\multicolumn{2}{c}{\textbf{Diarrhoea prevalence (binary outcome)}} \\
				\cmidrule(lr){2-3} \cmidrule(lr){4-5}
				Method & Estimate & Confidence interval &  Estimate &  Confidence interval\\
				\midrule
				\multicolumn{5}{c}{\textit{Analysis based on data from the WSH+N intervention and control arms.}}\\
                 \multicolumn{5}{l}{\textbf{Unadjusted}}\\
				Na\"ive & 0.114 & (0.013, 0.215) &  -0.022 & (-0.033, -0.011) \\ 
				Fixed & 0.128 & (0.022, 0.233) & -0.022 & (-0.034, -0.009)  \\ 
				Mixed(1$\mid$c:cl) & 0.119 & (0.021, 0.217) & -0.021 & (-0.032, -0.010) \\ 
                Mixed(1$\mid$c:cl) Sam. & 0.113 & (-0.011, 0.238) & -0.022 & (-0.036, -0.008) \\ 
				Mixed(1+A$\mid$c:cl) & 0.119 & (0.018, 0.22) & -0.022 & (-0.034, -0.009)\\ 
                Mixed(1+A$\mid$c:cl) Sam. & 0.114 & (-0.011, 0.239) & -0.022 & (-0.036, -0.007) \\ 
                 \multicolumn{5}{l}{\textbf{Adjusted}}\\
				Na\"ive & 0.124 & (0.027, 0.220) & -0.021 & (-0.032, -0.010)  \\ 
				Fixed &  0.139 & (0.038, 0.240) & -0.022 & (-0.035, -0.009) \\ 
				Mixed(1$\mid$c:cl) & 0.127 & (0.032, 0.221) & -0.020 & (-0.031, -0.009) \\ 
                Mixed(1$\mid$c:cl) Sam. & 0.123 & (0.013, 0.234) & -0.021 & (-0.035, -0.007) \\ 
				Mixed(1+A$\mid$c:cl) & 0.126 & (0.028, 0.224) & -0.020 & (-0.033, -0.008) \\ 
                Mixed(1+A$\mid$c:cl) Sam. & 0.123 & (0.013, 0.233) & -0.021 & (-0.035, -0.007) \\ 
				\midrule
				\multicolumn{5}{c}{\textit{Analysis based on data from all arms.}}\\
                 \multicolumn{5}{l}{\textbf{Unadjusted}}\\
				Na\"ive & 0.114 & (0.013, 0.215) &  -0.022 & (-0.033, -0.011)\\
				Fixed & 0.116 & (0.011, 0.220) & -0.021 & (-0.034, -0.009)\\
				Mixed(1$\mid$c:cl) & 0.115 & (0.017, 0.212) & -0.020 & (-0.031, -0.009)\\ 
                Mixed(1$\mid$c) Sam. & 0.110 & (-0.004, 0.224) & -0.021 & (-0.034, -0.008)\\ 
				Mixed(1+A$\mid$c:cl) & 0.114 & (0.016, 0.213) & -0.021 & (-0.034, -0.009)\\
                Mixed(1+A$\mid$c:cl) Sam. & 0.109 & (-0.004, 0.223) & -0.021 & (-0.034, -0.009)\\
                 \multicolumn{5}{l}{\textbf{Adjusted}}\\
				Na\"ive & 0.124 & (0.027, 0.221) &  -0.022 & (-0.032, -0.011)\\
				Fixed & 0.125 & (0.024, 0.226) & -0.021 & (-0.033, -0.008)\\
				Mixed(1$\mid$c:cl) & 0.123 & (0.028, 0.217) & -0.019 & (-0.030, -0.009)\\
                Mixed(1$\mid$c:cl) Sam. & 0.121 & (0.016, 0.225) & -0.021 & (-0.034, -0.008)\\
				Mixed(1+A$\mid$c:cl) & 0.122 & (0.021, 0.224) & -0.021 & (-0.033, -0.008)\\
                Mixed(1+A$\mid$c:cl) Sam. & 0.120 & (0.015, 0.225) & -0.021 & (-0.034, -0.008)\\
				\bottomrule	
			\end{tabular}%
		}
	\end{center}
	\tiny
	\vspace{0.01cm} 
	\label{table:WSH_N_ATE}
\end{table}

\section{Discussion}\label{sec:discussion_ch2}
There are two primary reasons for conducting a multi-center randomized trial. First, it may be the most practical way to enroll enough patients to achieve the trial's objectives within a reasonable timeframe. Second, it strengthens the ability to generalize the results \citep{meinert2012clinicaltrials}. In this article, we studied the implications of ignoring center-level clustering in AIPW estimators when analyzing multi-center randomized studies. We then demonstrated how to best analyze such data by accounting for both clustering and baseline covariates. We considered both continuous and binary outcomes.

Our proposed variance estimators outperform the na\"ive variance estimators in all considered scenarios. The proposed methods remain valid (i.e., correct Type I error) in the absence of clustering (e.g., see \autoref{table:k_100_nc_5_cont} and \autoref{table:k_100_nc_5_bin} when $\sigma^2_{b0}$=0, $\sigma^2_{b1}$=0, $\sigma^2_{b2}$=0), consistent with the results of \cite{abadie2023should}. However, when there are many small centers, the estimation method that includes center as fixed-effects with binary outcome results in biased effect estimates for both counterfactual mean on treatment and ATE as a result of overfitting bias \citep{kahan2014accounting,sur2019modern}; this can be resolved via random-effects modeling. Estimating random-effects based on their distribution generally yields better coverage probabilities and is theoretically justified, unlike the use of empirical BLUPs when there are many small centers. When center sizes are relatively large, fixed-effects and mixed-effects models deliver comparable results.

As in previous studies (e.g., \cite{tsiatis2008covariate,aronow2013class,kahan2014risks,benkeser2021improving}), the proposed covariate-adjusted estimators are more efficient than unadjusted estimators. Our proposed estimators also yield efficiency gains by taking into account center effects (e.g., see Table B.1.3 in Appendix B.1, which considers settings with an average of 50 patients per center), particularly when center effects on the outcome are substantial. Moreover, in observational studies, adjusting for center effects offers the additional advantage of accounting for both measured and unmeasured center-level confounders. This substantially weakens the assumption in \cite{balzer2019new} that observed cluster-level covariates, together with individually measured covariates, are sufficient to control for confounding.

We considered both estimands that weigh centers versus patients equally. These estimands are identical when the center size is random and independent of the treatment effect, in which case they can more efficiently be estimated using weights $w(c)$ set to the inverse variance in center $c$. Interpretation of these estimands becomes more subtle when treatment effects and center sizes are dependent \citep{senn2000many}. Estimands that weigh centers versus patients equally may then take different values. 

The choice between patient-weighted and center-weighted estimands should be determined by the unit to which the scientific conclusion is intended to generalize. When the aim is to quantify the expected benefit or harm for patients receiving care, the patient-weighted estimand is the natural target, since larger centers contribute in proportion to the number of patients they treat. This estimand is therefore most relevant for patient-focused, clinical, public-health, regulatory, or cost-effectiveness questions. By contrast, the center-weighted estimand targets the effect in a typical center and is most relevant when centers themselves are the units of generalization or decision-making, for example when evaluating whether an implementation strategy performs well across hospitals or when predicting the effect in a newly adopting center.

However, when treatment effects depend on center size, the two estimands may differ substantially. For example, the RESTORE trial compared protocolized sedation with usual care in critically ill children \citep{curley2015protocolized}. Although 31 centers were randomized, center sizes varied substantially, ranging from 12 to 272 patients per center. Because treatment effects appeared to depend on center size, estimators of the patient-weighted and center-weighted estimands differed \citep{kahan2024demystifying}. The larger patient-weighted estimate suggests that children in larger centers benefited more from protocolized sedation. Such discrepancies should not merely be resolved by choosing one estimand over the other; they indicate clinically or operationally important heterogeneity and may suggest that a uniform decision across centers is inappropriate. In such settings, reporting both estimands, and investigating how treatment effects vary with center size or related center characteristics, can be more informative than either marginal summary alone. The appropriate choice and interpretation of estimands are further discussed by \cite{kahan2023estimands}.

Because patients are individually randomized to the treatment and control arms, some centers, particularly small centers, which are the focus of this study, may include only treated or only control patients, or may be highly unbalanced. We addressed this by estimating propensity across all centers using mixed-effects models. As confirmed by our simulation studies, this does not introduce (practical) positivity violations as both the true and estimated randomization probabilities are bounded away from 0 and 1.

When the number of centers is small, our method tends to yield lower coverage, indicating that further correction may be needed in this setting. Corresponding results for both continuous and binary outcomes (e.g., $k$=5 and 10) are presented in Appendix B.1.

Our sampling-based approach for the random-effects was based on 1,000 draws in our simulation studies; however, a larger number of draws is preferable in practice. In our first data analysis with continuous outcomes (see Section \ref{DA:estimators_ch2}), the Monte Carlo error---computed as the standard deviation of the ATE estimates across Monte Carlo replications divided by the square root of the number of replications---was approximately $2\times10^{-4}$ with 1000 replications, but would decrease to about $5.6\times10^{-5}$ with 10000 replications.

 In the data analysis, heterogeneity variances were estimated under the assumption that center size and center-specific deviations in the intercept and treatment effect are independent. When center sizes are dependent on these deviations, it becomes necessary to adapt existing heterogeneity variance estimators, which should be explored in future research. The DB variance estimator appears promising in this context (e.g., see Table B.1.12 in the Appendix B.1). This heterogeneity variance estimator works comparably to $\hat{\sigma}^2_{u,DL}$ and $\hat{\sigma}^{2}_{u,REML}$ in a setting where random-effects and center sizes are independent, but does not rely on this independence assumption. Further, though we relied on results for stratified randomization in \cite{wang2023model}, it remains to be seen if these extend to mixed-model-based adjustment for the stratification factor (i.e., center). 
 
 Finally, studying the problem of clustering for other types of estimands, such as the risk ratio, may also be an important direction for future research.

\textbf{Declaration of interest statement}\\
The authors report that there are no computing interests to declare.

\clearpage
\bibliographystyle{plainnat}
\bibliography{bibilio}

\appendix
\renewcommand{\thesection}{\Alph{section}}
\renewcommand{\thesubsection}{\thesection.\arabic{subsection}}

\renewcommand{\thefigure}{\thesubsection.\arabic{figure}}
\renewcommand{\thetable}{\thesubsection.\arabic{table}}

\pretocmd{\subsection}{\setcounter{figure}{0}\setcounter{table}{0}}{}{}

\section{Theoretical Derivations}
\subsection{Proposed AIPW estimators with continuous endpoints}\label{appe_proposed_AIPW_ch2}
To acknowledge the fact that the outcomes are affected by clustering, we consider AIPW estimators that account for this. The counterfactual mean on treatment and the average treatment effect estimates with continuous endpoints can then be obtained by:
\begin{itemize}[leftmargin=4em]
\item[\textbf{Step 1:}] \textbf{Model fitting}\\
Fit a linear mixed-effects model of the form $$Y_{ic}=\alpha+b_{0c}+\left(\beta+b_{1c}\right)A_{ic}+\gamma'X_{ic}+\epsilon_{ic},$$ 
where $b_{0c}$ and $b_{1c}$ denote center-specific deviations to the intercept (i.e., random intercept) and treatment effect (i.e., random slope), respectively. Furthermore, $\epsilon_{ic}$ denotes the residual error term. We assume that $b_{0c}$ and $b_{1c}$ are independent and follow a normal distribution with mean zero and variance $\sigma^2_{b0}$ and $\sigma^2_{b1}$, respectively. We can also allow for variation in covariate effects, as we explore in the simulations. Alternatively, a fixed-effects model that includes an indicator variable for each center, $Y_{ic} = \gamma_0+\psi A_{ic}+\eta'X_{ic} + \sum_{j=1}^{k-1} \gamma_j \mathit{I}_j+\epsilon_{ic}$, can be fitted. Here, $\mathit{I}_j$ is an indicator variable for center $j$, which is 1 for patients in center $j$ and 0 otherwise.

\item[\textbf{Step 2:}] \textbf{Predicting}\\
When a linear mixed-effects model is used, we predict the outcome on treatment as $\hat{m}_{1c}(X_{i})=\hat{\alpha}+\hat{b}_{0c}+\hat{\beta}+\hat{b}_{1c}+\hat{\gamma}'X_{ic}$ and on control as $\hat{m}_{0c}(X_{i})=\hat{\alpha}+\hat{b}_{0c}+\hat{\gamma}'X_{ic}$. The random-effect estimates, $\hat{b}_{0c}$ and $\hat{b}_{1c}$, are obtained either as best linear unbiased predictors (BLUPs) or by sampling from their corresponding normal distributions, $\hat{b}_{0c} \sim \text{Normal}(0,\hat{\sigma}^2_{b0})$ and $\hat{b}_{1c} \sim \text{Normal}(0,\hat{\sigma}^2_{b1})$. The estimators $\hat{\sigma}^2_{b0}$ and $\hat{\sigma}^2_{b1}$ can be obtained from the mixed-effects model (i.e., restricted maximum likelihood estimators). Similarly, when a linear fixed-effects model is used, we predict the outcome on treatment as $\hat{m}_{1c}(X_{i})=\hat{\gamma}_0+\hat{\psi}+\hat{\eta}'X_{ic} + \sum_{j=1}^{k-1} \hat{\gamma}_j \mathit{I}_j$ and on control as $\hat{m}_{0c}(X_{i})=\hat{\gamma}_0+\hat{\eta}'X_{ic}+\sum_{j=1}^{k-1} \hat{\gamma}_j \mathit{I}_j$.

\item[\textbf{Step 3:}] \textbf{Averaging}\\
The AIPW estimators of the counterfactual mean on treatment and the ATE based on center $c$ only can be expressed as \[\hat{\mu}_{1c}=\frac{1}{n_c}\sum_{i=1}^{n_c}\left[\frac{A_{ic}}{\hat{p}_{c}(X_{ic})}\left\{Y_{ic}-\hat{m}_{1c}(X_{ic})\right\}+\hat{m}_{1c}(X_{ic})\right],\] and
\[\hat{\tau}_{c}=\frac{1}{n_c}\sum_{i=1}^{n_c}\left[\frac{A_{ic}}{\hat{p}_{c}(X_{ic})}\left\{Y_{ic}-\hat{m}_{1c}(X_{ic})\right\}+\hat{m}_{1c}(X_{ic})-\frac{1-A_{ic}}{1-\hat{p}_{c}(X_{ic})} \left\{Y_{ic}-\hat{m}_{0c}(X_{ic})\right\}-\hat{m}_{0c}(X_{ic})\right],\] respectively. Here, $\hat{p}_{c}(X_{ic})$ are obtained as the conditional predictions (i.e., based on empirical BLUPs for the random-effect) under a mixed-effects logistic regression model with treatment assignment as the dependent variable, incorporating centers as random-effects and (optionally) baseline covariates as fixed-effects. Finally, the AIPW estimators of $\mu_{1}$ and $\tau$ can be obtained as $\hat{\mu}_{1}=\sum_{c=1}^{k}w(c)\hat{\mu}_{1c},\ \text{and}\ \hat{\tau}=\sum_{c=1}^{k}w(c)\hat{\tau}_{c}$.
\end{itemize}

\subsection{Small center asymptotics based on efficient influence curves}\label{appe_small_center_aysmptotics_ch2}
We consider the problem of estimating the mean outcome on treatment based on a randomized experiment carried out in a large number of centers, when the sample size per center is small. We will therefore study behavior under asymptotic regimes where the number of centers goes to infinity, but the sample sizes $n_c$ for centers $c=1,...,k$ are fixed.

Let $\hat{\mu}_{1c}$ be the AIPW estimator 
\[\hat{\mu}_{1c}=\frac{1}{n_c}\sum_{i=1}^{n_c}
\frac{A_{ic}}{\hat{p}_c}\left\{Y_{ic}-\hat{m}_{1c}(X_{ic})\right\}+\hat{m}_{1c}(X_{ic}),
\]
where $\hat{p}_c$ and $\hat{m}_{1c}(X_{ic})$ are estimates of the percentage of treated patients and the mean outcome on treatment given $X_{ic}$, respectively, in center $c$. Since $\hat{\mu}_{1c}$ is a one-step estimator, it obeys the standard expansion
\begin{align}\label{aysmptotic_expanision}
    \hat{\mu}_{1c}-\mu_{1c}=\frac{1}{n_c}\sum_{i=1}^{n_c} \phi(O_{ic},\mu_{1c},p_c,m_{1c})+R_c^{(1)}(\hat{P}_{N},P)+R_c^{(2)}(\hat{P}_{N},P),
\end{align}
where $O_{ic}=(Y_{ic},A_{ic},X_{ic})$, $\phi(O_{ic},\mu_{1c},p_c,m_{1c})$ is the efficient influence curve
\[\phi(O_{ic},\mu_{1c},p_c,m_{1c})=\frac{A_{ic}}{{p}_c}\left\{Y_{ic}-m_{1c}(X_{ic})\right\}+m_{1c}(X_{ic})-\mu_{1c},\]
$\hat{P}_{N}$ and $P$ refer to the estimated and true distribution of the observed data across all centers, and $R_c^{(1)}(\hat{P}_{N},P)$ and $R_c^{(2)}(\hat{P}_{N},P)$ are empirical process and second order remainder terms that will be studied later. It then follows for 
\[\mu_{1}=\E(\mu_{1C}),\]
where the mean is taken uniformly over the population of centers, that 
\begin{align*}
	\frac{1}{k}\sum_{c=1}^k \hat{\mu}_{1c}-\E(\mu_{1C})
	&=\frac{1}{k}\sum_{c=1}^k (\hat{\mu}_{1c}-\mu_{1c})+\frac{1}{k}\sum_{c=1}^k \left\{\mu_{1c}-\E(\mu_{1C})\right\}\\
	&=\frac{1}{k}\sum_{c=1}^k\frac{1}{n_c}\sum_{i=1}^{n_c} \left\{\phi(O_{ic},\mu_{1c},p_c,m_{1c})+\mu_{1c}-\E(\mu_{1C})\right\}\\
	&+\frac{1}{k}\sum_{c=1}^kR_c^{(1)}(\hat{P}_{N},P)+\frac{1}{k}\sum_{c=1}^kR_c^{(2)}(\hat{P}_{N},P).
\end{align*}
Here, the term on the first line has a standard asymptotic behavior when $k$ goes to infinity, that is governed by the variance of 
\[\frac{1}{n_c}\sum_{i=1}^{n_c} \left\{\phi(O_{ic},\mu_{1c},p_c,m_{1c})+\mu_{1c}-\E(\mu_{1C})\right\}.\]
We next study the term
\begin{align}\label{eq:secod_order} 
   	\frac{1}{k}\sum_{c=1}^k R_c^{(2)}(\hat{P}_{N},P)
	=\frac{1}{k}\sum_{c=1}^k \E\left[
	\frac{\left\{p_c-\hat{p}_c\right\}}{\hat{p}_c}\left\{m_{1c}(X_c)-\hat{m}_{1c}(X_c)\right\}\Bigm| C=c\right].
\end{align}
By the Cauchy-Schwarz inequality and assuming that $\hat{p}_c$ is bounded away from zero (which is a very weak restriction in randomized experiments), this is $o_p(k^{-1/2})$ when
\[\left(\frac{1}{k}\sum_{c=1}^k \E\left[
\left\{p_c-\hat{p}_c\right\}^2\bigm|C=c\right]\right)^{1/2}\left(\frac{1}{k}\sum_{c=1}^k \E\left[\left\{m_{1c}(X_c)-\hat{m}_{1c}(X_c)\right\}^2\bigm|C=c\right]\right)^{1/2}=o_p(k^{-1/2}).\]
This is a realistic requirement when $\hat{m}_{1c}(X_c)$ is obtained as the average of the predictions from a mixed model for subject $i$ in center $c$, averaging over the estimated random-effects distribution, as the fixed-effects parameters and variance components are root-$k$ consistent under correct model specification. In that case, we merely need consistent estimators of the randomization probabilities $p_c$, which is trivial in a simple randomized experiment.  Unfortunately, this may not be a realistic requirement when these predictions are based on fixed-effects models or based on empirical BLUPs in mixed-effects models. The reason is that the center sizes do not increase with sample size, so that the empirical BLUPs (and likewise the fixed center effect estimates) have a variation that does not shrink with sample size. For instance, suppose that 
\[Y_{ic}=\theta' W_{ic}+u_c+\epsilon_{ic},\]
with $W_{ic}=(A_{ic},X_{ic})'$. Further, note that the empirical BLUP is given by
\[\hat{u}_c=\frac{\hat{\sigma}_u^2}{\hat{\sigma}_u^2+\hat{\sigma}^2/n_c}(\overline{Y}_c-\hat{\theta}' \overline{W}_{c}).\]
Here, $\overline{Y}_c$ denotes the average outcome within center $c$, while $\overline{W}_{c}$ denotes the average values of the treatment and covariates for that center.
It then follows that 
\begin{align}\label{eq:outcome_reg_error}
    m_{1c}(X_c)-\hat{m}_{1c}(X_c)&=(\theta-\hat{\theta})' W_{ic}+u_c-\frac{\hat{\sigma}_u^2}{\hat{\sigma}_u^2+\hat{\sigma}^2/n_c}(\overline{Y}_c-\hat{\theta}' \overline{W}_{c}) \notag \\ 
&=(\theta-\hat{\theta})' W_{ic}+u_c-\frac{\hat{\sigma}_u^2}{\hat{\sigma}_u^2+\hat{\sigma}^2/n_c}(\theta'\overline{W}_c+u_c-\hat{\theta}' \overline{W}_{c}) \notag \\ 
&=(\theta-\hat{\theta})' \left(W_{ic}-\frac{\hat{\sigma}_u^2}{\hat{\sigma}_u^2+\hat{\sigma}^2/n_c}\overline{W}_c\right)+\frac{\hat{\sigma}^2/n_c}{\hat{\sigma}_u^2+\hat{\sigma}^2/n_c}u_c.
\end{align}
Upon squaring and taking the mean, we see that the term $\E\left[\left\{m_{1c}(X_c)-\hat{m}_{1c}(X_c)\right\}^2|C=c\right]$ is partly dominated by the variance of $u_c$ divided by $n_c^2$, which does not shrink with sample size. Let us therefore revisit term \eqref{eq:secod_order}. Suppose further that $\hat{p}_c$ is the fitted value $\hat{p}+\hat{v}_c$ from fitting a random effect model
\[p_c=p+v_c;\]
note that in actual fact, $p=0.5$ and $v_c=0$. Then, we have that 
\[\hat{v}_c=\frac{\hat{\sigma}_v^2}{\hat{\sigma}_v^2+p(1-p)/n_c}(\overline{A}_c-\hat{p}).\]
We then have that 
\begin{align}\label{eq:propensity_error}
    p_c-\hat{p}_c=p-\hat{p}-\frac{\hat{\sigma}_v^2}{\hat{\sigma}_v^2+p(1-p)/n_c}(\overline{A}_c-\hat{p}).
\end{align}

Substituting the expressions in \eqref{eq:outcome_reg_error} and \eqref{eq:propensity_error} into \eqref{eq:secod_order} yields
\begin{eqnarray*}
\frac{1}{k}\sum_{c=1}^k \E\left[
	\frac{p-\hat{p}-\frac{\hat{\sigma}_v^2}{\hat{\sigma}_v^2+p(1-p)/n_c}(\overline{A}_c-\hat{p})}{\hat{p}+\frac{\hat{\sigma}_v^2}{\hat{\sigma}_v^2+p(1-p)/n_c}(\overline{A}_c-\hat{p})}\left\{(\theta-\hat{\theta})' \left(W_{ic}-\frac{\hat{\sigma}_u^2}{\hat{\sigma}_u^2+\hat{\sigma}^2/n_c}\overline{W}_c\right)+\frac{\hat{\sigma}^2/n_c}{\hat{\sigma}_u^2+\hat{\sigma}^2/n_c}u_c\right\}\Bigm|C=c\right].
\end{eqnarray*}
Here, the term involving $u_c$ averages to zero and can therefore be ignored. Further, we have that $p-\hat{p}=O_p(n^{-1/2})$, $\theta-\hat{\theta}=O_p(n^{-1/2})$ (where $\theta$ can be redefined to be the probability limit of $\hat{\theta}$ under model misspecification \citep{van2026automated}) and $ \hat{\sigma}_v^2=O_p(n^{-1/2})$ (since $\sigma_v^2=0$), from which we conclude that the above expression is $O_p(n^{-1})$, which meets the required condition of being $o_p(k^{-1/2})$ (since $n$ is proportional to $k$).

We next study $\sqrt{k}$ times the empirical process term, i.e. 
\begin{eqnarray*}
	\frac{1}{\sqrt{k}}\sum_{c=1}^kR_c^{(1)}(\hat{P}_{N},P)
	&=&\frac{1}{\sqrt{k}}\sum_{c=1}^k 
    \frac{1}{n_c}\sum_{i=1}^{n_c}
\frac{A_{ic}}{\hat{p}_c}\left\{Y_{ic}-\hat{m}_{1c}(X_{ic})\right\}+\hat{m}_{1c}(X_{ic})\\
&&-\frac{1}{\sqrt{k}}\sum_{c=1}^k 
    \frac{1}{n_c}\sum_{i=1}^{n_c}
\frac{A_{ic}}{{p}_c}\left\{Y_{ic}-{m}_{1c}(X_{ic})\right\}+{m}_{1c}(X_{ic})\\
&&-\frac{1}{\sqrt{k}}\sum_{c=1}^k 
    \frac{1}{n_c}\sum_{i=1}^{n_c}
\frac{{p}_c}{\hat{p}_c}\left\{{m}_{1c}(X_{ic})-\hat{m}_{1c}(X_{ic})\right\}+\hat{m}_{1c}(X_{ic})\\
	&=&\frac{1}{\sqrt{k}}\sum_{c=1}^k 
    \frac{1}{n_c}\sum_{i=1}^{n_c}
A_{ic}\left\{\frac{1}{\hat{p}_c}-\frac{1}{{p}_c}\right\}\left\{Y_{ic}-\hat{m}_{1c}(X_{ic})\right\}\\
&&+\frac{1}{\sqrt{k}}\sum_{c=1}^k 
    \frac{1}{n_c}\sum_{i=1}^{n_c}
\left\{\frac{A_{ic}}{{p}_c}-\frac{{p}_c}{\hat{p}_c}\right\}\left\{{m}_{1c}(X_{ic})-\hat{m}_{1c}(X_{ic})\right\}\\
	&=&\frac{1}{\sqrt{k}}\sum_{c=1}^k \left\{\frac{1}{\hat{p}_c}-\frac{1}{{p}_c}\right\}
    \frac{1}{n_c}\sum_{i=1}^{n_c}
A_{ic}\left\{Y_{ic}-{m}_{1c}(X_{ic})\right\}\\
&&+\frac{1}{\sqrt{k}}\sum_{c=1}^k 
    \frac{1}{n_c}\sum_{i=1}^{n_c}
\frac{1}{\hat{p}_c}\left\{A_{ic}-{p}_c\right\}\left\{{m}_{1c}(X_{ic})-\hat{m}_{1c}(X_{ic})\right\}.
\end{eqnarray*}
It follows from (\ref{eq:propensity_error}) that 
\[\frac{1}{\hat{p}_c}-\frac{1}{p_c}=
2\frac{p_c-\hat{p}_c}{\hat{p}_c}=\frac{2}{\hat{p}_c}(p-\hat{p})-\frac{2\hat{\sigma}_v^2}{\hat{p}_c\hat{\sigma}_v^2+\hat{p}_cp(1-p)/n_c}(\overline{A}_c-\hat{p}).\]
Here, 
\[\sup_c|p_c-\hat{p}_c|\leq |p-\hat{p}|+\sup_c\left|\frac{\hat{\sigma}_v^2}{\hat{\sigma}_v^2+p(1-p)/n_c}\right|\sup_c|\overline{A}_c-\hat{p}|.\]
Here, the first term on the right hand side is $O_p(k^{-1/2})$ and so is the second, since $\hat{\sigma}^2_v=O_p(k^{-1/2})$ when $\sigma^2_v=0$. Since $\overline{A}_c-\hat{p}$ is bounded, we conclude that $\sup_c|p_c-\hat{p}_c|=O_p(k^{-1/2})$. Further, since 
\[\hat{p}_c\geq 0.5-|\hat{p}_c-0.5|\]
we also have 
\[\inf_c\hat{p}_c\geq 0.5-\sup_c|\hat{p}_c-0.5|.\]
Since the latter converges to zero, we have that $\inf_c\hat{p}_c$ is bounded away from zero in probability. We conclude based on the above expression that 
\[\sup_c\left|\frac{1}{\hat{p}_c}-\frac{1}{p_c}\right|=O_p(k^{-1/2}).\]
This enables us to rewrite the first term in the above derivation as 
\[\sup_c\left|\frac{1}{\hat{p}_c}-\frac{1}{p_c}\right|\frac{1}{\sqrt{k}}\sum_{c=1}^k 
    \frac{1}{n_c}\sum_{i=1}^{n_c}
A_{ic}\left\{Y_{ic}-{m}_{1c}(X_{ic})\right\},\]
which is $O_p(k^{-1/2})O_p(1)=o_p(1)$, as desired. When $\hat{m}_{1c}(X_c)$ is obtained as the average of the predictions from a mixed model for subject $i$ in center $c$, averaging over the estimated random-effects distribution, then it follows from a Taylor expansion of $\hat{m}_{1c}(X_{ic})$ around 
${m}_{1c}(X_{ic})$ that the second term in the above derivation can be written as
\[\frac{1}{\sqrt{k}}\sum_{c=1}^k \frac{O_p(n^{-1/2})}{\hat{p}_c}
    \frac{1}{n_c}\sum_{i=1}^{n_c}
\left\{A_{ic}-{p}_c\right\}h(X_{ic})\]
for some known function $h(X_{ic})$. This is also $o_p(1)$, as desired. Note that no sample splitting was required to justify the above reasoning. However, this second term is not guaranteed when $\hat{m}_{1c}(X_c)$ is instead obtained using empirical BLUPs as the difference ${m}_{1c}(X_{ic})-\hat{m}_{1c}(X_{ic})$ is then not guaranteed to shrink with sample size; the use of sample splitting does not overcome this concern. We therefore discourage the use of empirical BLUPs for making outcome predictions.  

Although our primary focus is on settings with many small centers, we also study the validity of results when the center sizes $n_c$ for centers $c=1,...,k$, diverge while the number of centers $k$ remains fixed. Here, we first study the asymptotic behaviour of center-specific estimators (i.e., $\hat{\mu}_{1c}$). In this case, the first term on the right hand side of \eqref{aysmptotic_expanision} has a standard large sample behavior as $n_c$ goes to infinity, characterized by the variance of 
\begin{align*}
    \frac{1}{n_c}\sum_{i=1}^{n_c}\phi(O_{ic},\mu_{1c},p_c,m_{1c}).
\end{align*}
Further, the second order remainder term $R_c^{(2)}(\hat{P}_{N},P) = o_p(n_c^{-1/2})$ by Cauchy–Schwarz since
\begin{align*}
\left(\E\left[\left\{p_c-\hat{p}_c\right\}^2\bigm|C=c\right]\right)^{1/2}\left(\E\left[\left\{m_{1c}(X_c)-\hat{m}_{1c}(X_c)\right\}^2\bigm|C=c\right]\right)^{1/2}=o_p(n_c^{-1/2}).
\end{align*}
Moreover, the empirical process term $R_c^{(1)}(\hat{P}_{N},P)$ converges in probability to zero when the conditional variance of
\begin{align*}
\frac{1}{n_c}\sum_{i=1}^{n_c}
\big[\phi(O_{ic},\hat{\mu}_{1c},\hat{p}_{c},\hat{m}_{1c}) - \phi(O_{ic},\mu_{1c},p_c,m_{1c})\big],
\end{align*}
given the nuisance estimators, converges to zero in probability.  

Finally, note that
\begin{align*}
    \hat{\mu}_{1}=\frac{1}{k}\sum_{c=1}^k \hat{\mu}_{1c},
\end{align*}
is consistent and asymptotically normal, provided that center-specific estimators $\hat{\mu}_{1c}$ are themselves consistent and asymptotically normal. This shows the results are also valid when the center sizes approach infinity but the number of centers remains fixed.

When the outcome model is misspecified, the above results continue to hold \citep{van2026automated}, with the definition of $\phi(\cdot)$ modified to
\begin{align*}
    \frac{A_{ic}}{p_c}\left\{Y_{ic}-m_{1c}(X_{ic})\right\}+m_{1c}(X_{ic})-\E\left[\frac{A_{ic}}{p^2_c}\left\{Y_{ic}-m_{1c}(X_{ic})\right\}\right](A_{ic}-p_c)-\mu_{1c}.
\end{align*}

\subsection{When to account for clustering?}\label{appe_when_to_account_clustering_ch2}
In this section, we examine when it is necessary to account for clustering when estimating the counterfactual mean on treatment and the ATE for both continuous and binary outcomes.

\subsubsection{Continuous outcomes}
First, we will explain when it is necessary to account for clustering in the estimation of the counterfactual mean under treatment for continuous outcomes. The counterfactual mean on treatment can be expressed as 
\begin{align*}
	&\E\left[\frac{A_{ic}}{p_{c}(X_{ic})}\left\{Y_{ic}-m_{1c}(X_{ic})\right\}+m_{1c}(X_{ic})\right],
\end{align*}
where $Y_{ic}$ denotes a continuous outcome for patient $i$ in center $c$ while $m_{1c}(X_{ic})$ denotes the mean outcome under treatment conditional on baseline covariates in center $c$. Further, $A_{ic}$ and $p_{c}(X_{ic})$ denote the treatment indicator and randomization probability for patient $i$ in center $c$, respectively.

As we discussed in the main paper, the ``na\"ive" variance estimators that ignore the correlation at center-level may underestimate the true variability, as they do not account for the clustering of patients within centers. Here, we evaluate whether the contributions of two different patients from the same center are correlated when estimating the counterfactual mean on treatment for continuous outcomes. To this end, we consider the covariance of $\frac{A_{ic}}{p_{c}(X_{ic})}\left\{Y_{ic}-m_{1c}(X_{ic})\right\}+m_{1c}(X_{ic})$ and $\frac{A_{i'c}}{p_{c}(X_{i'c})}\left\{Y_{i'c}-m_{1c}(X_{i'c})\right\}+m_{1c}(X_{i'c})$, that is, the covariance of the influence functions of the counterfactual mean on treatment for patient $i$ and $i'$ from the same center $c$. This equals
\begin{equation}\label{eq:cov_countmean}
    \begin{aligned}
    &\E\left(\left[\frac{A_{ic}}{p_{c}(X_{ic})}\left\{Y_{ic}-m_{1c}(X_{ic})\right\}+m_{1c}(X_{ic})\right]\left[\frac{A_{i'c}}{p_{c}(X_{i'c})}\left\{Y_{i'c}-m_{1c}(X_{i'c})\right\}+m_{1c}(X_{i'c})\right]\right)\\
    &\quad-\E\left[\frac{A_{ic}}{p_{c}(X_{ic})}\left\{Y_{ic}-m_{1c}(X_{ic})\right\}+m_{1c}(X_{ic})\right]\E\left[\frac{A_{i'c}}{p_{c}(X_{i'c})}\left\{Y_{i'c}-m_{1c}(X_{i'c})\right\}+m_{1c}(X_{i'c})\right].
    \end{aligned}
\end{equation}

As an example, consider a continuous outcome that, in reality, is generated using only a random intercept (but without a random treatment effect),
\begin{align}\label{eq:datagen_cont_int}
    Y_{ic}=\alpha+b_{0c}+\beta A_{ic}+\gamma'X_{ic}+\epsilon_{ic}.
\end{align}
Then, Equation \eqref{eq:cov_countmean} equals
\begin{equation}\label{eq:cov_countmean_datagen_int}
\begin{aligned}
    &\E\left[\left\{\frac{A_{ic}}{p_{c}(X_{ic})}\epsilon_{ic}+\alpha+b_{0c}+\beta+\gamma'X_{ic}\right\}\left\{\frac{A_{i'c}}{p_{c}(X_{i'c})}\epsilon_{i'c}+\alpha+b_{0c}+\beta+\gamma'X_{i'c}\right\}\right]\\
   &\quad-\E\left\{\frac{A_{ic}}{p_{c}(X_{ic})}\epsilon_{ic}+\alpha+b_{0c}+\beta+\gamma'X_{ic}\right\}\E\left\{\frac{A_{i'c}}{p_{c}(X_{i'c})}\epsilon_{i'c}+\alpha+b_{0c}+\beta+\gamma'X_{i'c}\right\}.
\end{aligned}
\end{equation}
The first line in Equation \eqref{eq:cov_countmean_datagen_int} can be written as 
\begin{align*}
   & \E\Bigg\{\frac{A_{ic}}{p_{c}(X_{ic})}\epsilon_{ic}\frac{A_{i'c}}{p_{c}(X_{i'c})}\epsilon_{i'c}+\frac{A_{ic}}{p_{c}(X_{ic})}\epsilon_{ic}(\alpha+b_{0c}+\beta+\gamma'X_{i'c})+\frac{A_{i'c}}{p_{c}(X_{i'c})}\epsilon_{i'c}(\alpha+b_{0c}+\beta+\gamma'X_{ic})\\
    &\quad +(\alpha+b_{0c}+\beta+\gamma'X_{ic})(\alpha+b_{0c}+\beta+\gamma'X_{i'c})\Bigg\}\\
    &=\E\left\{\frac{A_{ic}}{p_{c}(X_{ic})}\epsilon_{ic}\frac{A_{i'c}}{p_{c}(X_{i'c})}\epsilon_{i'c}\right\}+\E\left\{\frac{A_{ic}}{p_{c}(X_{ic})}\epsilon_{ic}(\alpha+b_{0c}+\beta+\gamma'X_{i'c})\right\}+\E\left\{\frac{A_{i'c}}{p_{c}(X_{i'c})}\epsilon_{i'c}(\alpha+b_{0c}+\beta+\gamma'X_{ic})\right\}\\
    &\quad+\E\left\{\left(\alpha+b_{0c}+\beta+\gamma'X_{ic}\right)\left(\alpha+b_{0c}+\beta+\gamma'X_{i'c}\right)\right\}.
\end{align*}
Here, the first three terms are zero because the residuals are independent across individuals conditional on the center random-effect, they are independent of the treatment indicators and covariates given the center effect, and the residuals have mean zero conditional on the center effect. Note, however, that this represents a stronger set of assumptions than strictly necessary. In a weaker, more general formulation, the residuals may depend on \(A\) and \(X\) within each unit, provided they remain conditionally independent across units given the full conditioning set---including \(A\), \(X\), and \(b_{0c}\)---and have mean zero conditional on \(A\), \(X\), and \(b_{0c}\). This weaker version allows for some within-unit correlation of residuals with covariates.

Thus, for the first line in Equation \eqref{eq:cov_countmean_datagen_int}, we have 
\begin{align*}
   & \E\left\{\left(\alpha+b_{0c}+\beta+\gamma'X_{ic}\right)\left(\alpha+b_{0c}+\beta+\gamma'X_{i'c}\right)\right\}\\
   &=\E\left(b_{0c}^2\right)+\E\left\{b_{0c}(\alpha+\beta+\gamma'X_{i'c})\right\}+\E\left\{b_{0c}(\alpha+\beta+\gamma'X_{ic})\right\}+\E\left(\alpha+\beta+\gamma'X_{ic}\right)\E\left(\alpha+\beta+\gamma'X_{i'c}\right),
\end{align*}
where the second-line follows by assuming that $X_{ic}$ and $X_{i'c}$ are independent.

The second line in Equation \eqref{eq:cov_countmean_datagen_int} can also be written as 
\begin{align*}
   &-\left[\left\{\E\left(b_{0c}\right)\right\}^2+\E\left\{b_{0c}\left(\alpha+\beta+\gamma'X_{i'c})\right\}+\E\left\{b_{0c}(\alpha+\beta+\gamma'X_{ic}\right)\right\}+\E\left(\alpha+\beta+\gamma'X_{ic}\right)\E\left(\alpha+\beta+\gamma'X_{i'c}\right)\right],
\end{align*}
by assuming independence of the residual errors with the treatment indicators, the residual terms having expectation zero, and independence of $b_{0c}$ with $\alpha$, $\beta$, $\gamma$, and $X_c$.

Consequently, Equation \eqref{eq:cov_countmean_datagen_int} equals 
\begin{align*}
    &\E\left(b^2_{0c}\right)-\left\{\E\left(b_{0c}\right)\right\}^2=\sigma^2_{b0},
\end{align*}
where $\sigma^2_{b0}$ is the variance of $b_{0c}$. Since the covariance between the influence functions of two patients from center $c$ in estimating the counterfactual mean on treatment is non-zero, it is necessary to consider center-level clustering when estimating counterfactual means.

Now, consider a continuous outcome generated with center-specific deviations to the intercept and treatment effect, 
\begin{align}\label{eq:datagen_cont_int_slope}
    Y_{ic}=\alpha+b_{0c}+(\beta+b_{1c})A_{ic}+\gamma'X_{ic}+\epsilon_{ic}.
\end{align}
Then, Equation \eqref{eq:cov_countmean} equals
\begin{equation}\label{eq:cov_countmean_datagen_int_slope}
  \begin{aligned}
    &\E\left[\left\{\frac{A_{ic}}{p_{c}(X_{ic})}\epsilon_{ic}+\alpha+b_{0c}+b_{1c}+\beta+\gamma'X_{ic}\right\}\left\{\frac{A_{i'c}}{p_{c}(X_{i'c})}\epsilon_{i'c}+\alpha+b_{0c}+b_{1c}+\beta+\gamma'X_{i'c}\right\}\right]\\
   &\quad-\E\left\{\frac{A_{ic}}{p_{c}(X_{ic})}\epsilon_{ic}+\alpha+b_{0c}+b_{1c}+\beta+\gamma'X_{ic}\right\}\E\left\{\frac{A_{i'c}}{p_{c}(X_{i'c})}\epsilon_{i'c}+\alpha+b_{0c}+b_{1c}+\beta+\gamma'X_{i'c}\right\}.
\end{aligned}   
\end{equation}   
In this case, the first line of Equation \eqref{eq:cov_countmean_datagen_int_slope} can be written as 
\begin{align*}
   & \E\Bigg\{\frac{A_{ic}}{p_{c}(X_{ic})}\epsilon_{ic}\frac{A_{i'c}}{p_{c}(X_{i'c})}\epsilon_{i'c}+\frac{A_{ic}}{p_{c}(X_{ic})}\epsilon_{ic}(\alpha+b_{0c}+b_{1c}+\beta+\gamma'X_{i'c})+\frac{A_{i'c}}{p_{c}(X_{i'c})}\epsilon_{i'c}(\alpha+b_{0c}+b_{1c}+\beta+\gamma'X_{ic})\\
    &\quad+(\alpha+b_{0c}+b_{1c}+\beta+\gamma'X_{ic})(\alpha+b_{0c}+b_{1c}+\beta+\gamma'X_{i'c})\Bigg\}\\
& =\E\left\{\frac{A_{ic}}{p_{c}(X_{ic})}\epsilon_{ic}\frac{A_{i'c}}{p_{c}(X_{i'c})}\epsilon_{i'c}\right\}+\E\left\{\frac{A_{ic}}{p_{c}(X_{ic})}\epsilon_{ic}(\alpha+b_{0c}+b_{1c}+\beta+\gamma'X_{i'c})\right\}\\
& \quad+\E\left\{\frac{A_{i'c}}{p_{c}(X_{i'c})}\epsilon_{i'c}(\alpha+b_{0c}+b_{1c}+\beta+\gamma'X_{ic})\right\}+\E\left\{(\alpha+b_{0c}+b_{1c}+\beta+\gamma'X_{ic})(\alpha+b_{0c}+b_{1c}+\beta+\gamma'X_{i'c})\right\}.
\end{align*}
Similar to before, the first three terms are zero because the residuals are independent across individuals conditional on the center random-effect, they are independent of the treatment indicators and covariates given the center effect, and the residuals have mean zero conditional on the center effect.
Thus, for the first line of Equation \eqref{eq:cov_countmean_datagen_int_slope}, we have 
\begin{align*}
   &\E\left\{\left(\alpha+b_{0c}+b_{1c}+\beta+\gamma'X_{ic}\right)\left(\alpha+b_{0c}+b_{1c}+\beta+\gamma'X_{i'c}\right)\right\}\\
   & =\E\left\{\left(b_{0c}+b_{1c}\right)^2\right\}+\E\left\{(b_{0c}+b_{1c})(\alpha+\beta+\gamma'X_{i'c})\right\}+\E\left\{(b_{0c}+b_{1c})(\alpha+\beta+\gamma'X_{ic})\right\}\\
   &\quad+\E\left(\alpha+\beta+\gamma'X_{ic}\right)\E\left(\alpha+\beta+\gamma'X_{i'c}\right),
\end{align*}
with the last term following from the independence of $X_{ic}$ and $X_{i'c}$.
The second line of Equation \eqref{eq:cov_countmean_datagen_int_slope} can also be written as 
\begin{align*}
   & -\Bigg[\left\{\E\left(b_{0c}+b_{1c}\right)\right\}^2+\E\left\{(b_{0c}+b_{1c})(\alpha+\beta+\gamma'X_{i'c})\right\}+\E\left\{(b_{0c}+b_{1c})(\alpha+\beta+\gamma'X_{ic})\right\}\\
   &\quad+\E\left(\alpha+\beta+\gamma'X_{ic}\right)\E\left(\alpha+\beta+\gamma'X_{i'c}\right)\Bigg],
\end{align*}
by assuming independence of the residual errors and the treatment indicators, the residual terms having expectation zero, and independence of $b_{0c}$ and $b_{1c}$ with $\alpha$, $\beta$, $\gamma$, and $X_c$.

Accordingly, Equation \eqref{eq:cov_countmean_datagen_int_slope} equals
\begin{align*}
    &\E\left\{(b_{0c}+b_{1c})^2\right\}-\left\{\E\left(b_{0c}+b_{1c}\right)\right\}^2=\sigma^2_{b0}+\sigma^2_{b1},
\end{align*}
since $b_{0c} \indep b_{1c}$. Here, $\sigma^2_{b1}$ is the variance of $b_{1c}$. As before, since the covariance between the influence functions of two patients of center $c$ in estimating the counterfactual mean on treatment is non-zero, we need to account for center-level clustering when estimating the counterfactual means.

Likewise, we examine whether the correlation at the center level is non-zero in the estimation of the ATE for continuous outcomes. The ATE can be expressed as
\begin{align*}
	&\E\left[\frac{A_{ic}}{p_{c}(X_{ic})}\left\{Y_{ic}-m_{1c}(X_{ic})\right\}+m_{1c}(X_{ic})-\frac{1-A_{ic}}{1-p_{c}(X_{ic})}\left\{Y_{ic}-m_{0c}(X_{ic})\right\}-m_{0c}(X_{ic})\right].
\end{align*}
In order to know whether we can ignore the correlation on center-level clustering when estimating the ATE, we can consider the covariance of $\frac{A_{ic}}{p_{c}(X_{ic})}\left\{Y_{ic}-m_{1c}(X_{ic})\right\}+m_{1c}(X_{ic})-\frac{1-A_{ic}}{1-p_{c}(X_{ic})}\left\{Y_{ic}-m_{0c}(X_{ic})\right\}-m_{0c}(X_{ic})$ and $\frac{A_{i'c}}{p_{c}(X_{i'c})}\left\{Y_{i'c}-m_{1c}(X_{i'c})\right\}+m_{1c}(X_{i'c})-\frac{1-A_{i'c}}{1-p_{c}(X_{i'c})}\left\{Y_{i'c}-m_{0c}(X_{i'c})\right\}-m_{0c}(X_{i'c})$. The covariance equals
\begin{equation}\label{eq:cov_ATE}
    \begin{aligned}
&\E\Bigg(\left[\frac{A_{ic}}{p_{c}(X_{ic})}\left\{Y_{ic}-m_{1c}(X_{ic})\right\}+m_{1c}(X_{ic})-\frac{1-A_{ic}}{1-p_{c}(X_{ic})}\left\{Y_{ic}-m_{0c}(X_{ic})\right\}-m_{0c}(X_{ic})\right]\cdot\\
&\quad \left[\frac{A_{i'c}}{p_{c}(X_{i'c})}\left\{Y_{i'c}-m_{1c}(X_{i'c})\right\}+m_{1c}(X_{i'c})-\frac{1-A_{i'c}}{1-p_{c}(X_{i'c})}\left\{Y_{i'c}-m_{0c}(X_{i'c})\right\}-m_{0c}(X_{i'c})\right]\Bigg)\\
&\quad-\E\left[\frac{A_{ic}}{p_{c}(X_{ic})}\left\{Y_{ic}-m_{1c}(X_{ic})\right\}+m_{1c}(X_{ic})-\frac{1-A_{ic}}{1-p_{c}(X_{ic})}\left\{Y_{ic}-m_{0c}(X_{ic})\right\}-m_{0c}(X_{ic})\right]\cdot\\
&\quad\E\left[\frac{A_{i'c}}{p_{c}(X_{i'c})}\left\{Y_{i'c}-m_{1c}(X_{i'c})\right\}+m_{1c}(X_{i'c})-\frac{1-A_{i'c}}{1-p_{c}(X_{i'c})}\left\{Y_{i'c}-m_{0c}(X_{i'c})\right\}-m_{0c}(X_{i'c})\right].
    \end{aligned}
\end{equation}
For example, considering a continuous outcome generated as in Equation \eqref{eq:datagen_cont_int}, then Equation \eqref{eq:cov_ATE} equals
\begin{align*}
& \E\left[\left\{\beta+\frac{A_{ic}}{p_{c}(X_{ic})}\epsilon_{ic}-\frac{1-A_{ic}}{1-p_{c}(X_{ic})}\epsilon_{ic}\right\}
\left\{\beta+\frac{A_{i'c}}{p_{c}(X_{i'c})}\epsilon_{i'c}-\frac{1-A_{i'c}}{1-p_{c}(X_{i'c})}\epsilon_{i'c}\right\}\right]\\
&\quad-\E\left\{\beta+\frac{A_{ic}}{p_{c}(X_{ic})}\epsilon_{ic}-\frac{1-A_{ic}}{1-p_{c}(X_{ic})}\epsilon_{ic}\right\}
\E\left\{\beta+\frac{A_{i'c}}{p_{c}(X_{i'c})}\epsilon_{i'c}-\frac{1-A_{i'c}}{1-p_{c}(X_{i'c})}\epsilon_{i'c}\right\}\\
&=\E\left(\beta^2\right)+\beta\E\left\{\frac{A_{i'c}}{p_{c}(X_{i'c})}\epsilon_{i'c}-\frac{1-A_{i'c}}{1-p_{c}(X_{i'c})}\epsilon_{i'c}\right\}+\beta\E\left\{\frac{A_{ic}}{p_{c}(X_{ic})}\epsilon_{ic}-\frac{1-A_{ic}}{1-p_{c}(X_{ic})}\epsilon_{ic}\right\}\\
&\quad+\E\left[\left\{\frac{A_{ic}}{p_{c}(X_{ic})}\epsilon_{ic}-\frac{1-A_{ic}}{1-p_{c}(X_{ic})}\epsilon_{ic}\right\}
\left\{\frac{A_{i'c}}{p_{c}(X_{i'c})}\epsilon_{i'c}-\frac{1-A_{i'c}}{1-p_{c}(X_{i'c})}\epsilon_{i'c}\right\}\right]\\
&\quad-\E\left\{\beta+\frac{A_{ic}}{p_{c}(X_{ic})}\epsilon_{ic}-\frac{1-A_{ic}}{1-p_{c}(X_{ic})}\epsilon_{ic}\right\}
\E\left\{\beta+\frac{A_{i'c}}{p_{c}(X_{i'c})}\epsilon_{i'c}-\frac{1-A_{i'c}}{1-p_{c}(X_{i'c})}\epsilon_{i'c}\right\}\\
&=\E\left(\beta^2\right)-\left\{\E(\beta)\right\}^2\\
&=0.
\end{align*}
The second equality holds due to the residuals being independent across individuals conditional on the center random-effect, the residuals being independent of the treatment indicators and covariates given the center effect, and the residuals have mean zero conditional on the center effect. Because the covariance between the influence functions of ATE for two patients within center $c$ is zero, center-level clustering does not need to be accounted for when estimating ATE in a scenario where the continuous outcome is truly generated with only a random intercept. Thus, the ``na\"ive" variance estimators are valid for ATE with continuous outcomes, in this setting. This result aligns with the results of \cite{abadie2023should}.

Furthermore, consider a continuous outcome generated with center-specific deviations to the intercept and treatment effect (see Equation \eqref{eq:datagen_cont_int_slope}), then under similar assumptions as before, Equation \eqref{eq:cov_ATE} can be written as 
\begin{align*}
& \E\left[\left\{\beta+b_{1c}+\frac{A_{ic}}{p_{c}(X_{ic})}\epsilon_{ic}-\frac{1-A_{ic}}{1-p_{c}(X_{ic})}\epsilon_{ic}\right\}
\left\{\beta+b_{1c}+\frac{A_{i'c}}{p_{c}(X_{i'c})}\epsilon_{i'c}-\frac{1-A_{i'c}}{1-p_{c}(X_{i'c})}\epsilon_{i'c}\right\}\right]\\
&\quad-\E\left\{\beta+b_{1c}+\frac{A_{ic}}{p_{c}(X_{ic})}\epsilon_{ic}-\frac{1-A_{ic}}{1-p_{c}(X_{ic})}\epsilon_{ic}\right\}
\E\left\{\beta+b_{1c}+\frac{A_{i'c}}{p_{c}(X_{i'c})}\epsilon_{i'c}-\frac{1-A_{i'c}}{1-p_{c}(X_{i'c})}\epsilon_{i'c}\right\}\\
&=\beta^2+2\beta\E\left(b_{1c}\right)+\E\left(b_{1c}^2\right)-\left[\beta^2+2\beta\E\left(b_{1c}\right)+\left\{\E\left(b_{1c}\right)\right\}^2\right]\\
&=\E\left(b^2_{1c}\right)-\left\{\E\left(b_{1c}\right)\right\}^2\\
&=\sigma^2_{b1}.
\end{align*}
Since the covariance between the influence functions of two patients of center $c$ in estimating the ATE is non-zero, we need to account for center-level clustering when estimating the ATE in a setting where a continuous outcome is generated with center-specific deviations to the intercept and treatment effect.

\subsubsection{Binary outcomes}
As with continuous outcomes, we start by analyzing under what conditions clustering should be accounted for when estimating the counterfactual mean on treatment for binary outcomes. Similar to before, we can express the counterfactual mean on treatment with binary outcomes as
\begin{align*}
	&\E\left[\frac{A_{ic}}{p_{c}(X_{ic})}\left\{Y_{ic}-m_{1c}(X_{ic})\right\}+m_{1c}(X_{ic})\right].
\end{align*}
Here, $Y_{ic}$ denotes a binary outcome for patient $i$ in center $c$ and $m_{1c}(X_{ic})$ denotes the mean outcome under treatment conditional on baseline covariates in center $c$. Similar to continuous outcomes, we can consider the covariance of $\frac{A_{ic}}{p_{c}(X_{ic})}\left\{Y_{ic}-m_{1c}(X_{ic})\right\}+m_{1c}(X_{ic})$ and $\frac{A_{i'c}}{p_{c}(X_{i'c})}\left\{Y_{i'c}-m_{1c}(X_{i'c})\right\}+m_{1c}(X_{i'c})$ in order to know whether we can ignore the correlation on center-level when estimating the counterfactual mean on treatment with binary outcomes. This equals
\begin{equation}\label{eq:cov_countmean_bin}
    \begin{aligned}
    &\E\left(\left[\frac{A_{ic}}{p_{c}(X_{ic})}\left\{Y_{ic}-m_{1c}(X_{ic})\right\}+m_{1c}(X_{ic})\right]\left[\frac{A_{i'c}}{p_{c}(X_{i'c})}\left\{Y_{i'c}-m_{1c}(X_{i'c})\right\}+m_{1c}(X_{i'c})\right]\right)\\
    &\quad-\E\left[\frac{A_{ic}}{p_{c}(X_{ic})}\left\{Y_{ic}-m_{1c}(X_{ic})\right\}+m_{1c}(X_{ic})\right]\E\left[\frac{A_{i'c}}{p_{c}(X_{i'c})}\left\{Y_{i'c}-m_{1c}(X_{i'c})\right\}+m_{1c}(X_{i'c})\right].
    \end{aligned}
\end{equation}
Suppose that a binary outcome $Y_{ic}$ is generated from a Bernoulli distribution with probability
	\begin{align}\label{eq:datagen_bin_int}
		\logit^{-1}\left(\alpha+b_{0c}+\beta A_{ic}+\gamma'X_{ic}\right),
	\end{align}
and thus, $m_{1c}(X_{ic})=\logit^{-1}\left(\alpha+b_{0c}+\beta A_{ic}+\gamma'X_{ic}\right)$. Here, we only consider random intercepts. Then, Equation \eqref{eq:cov_countmean_bin} equals
\begin{align*}
&\E\left[\left\{\logit^{-1}\left(\alpha+b_{0c}+\beta+\gamma'X_{ic}\right)\right\}\left\{\logit^{-1}\left(\alpha+b_{0c}+\beta+\gamma'X_{i'c}\right)\right\}\right]\\
&\quad-\E\left\{\logit^{-1}\left(\alpha+b_{0c}+\beta+\gamma'X_{ic}\right)\right\}\E\left\{\logit^{-1}\left(\alpha+b_{0c}+\beta +\gamma'X_{i'c}\right)\right\}.
\end{align*}
Given that patient $i$ and $i'$ share a random effect $b_{0c}$, this covariance is non-zero (greater-than zero). In this situation, clustering must be considered when estimating the counterfactual means for binary outcomes.

Consider again a binary outcome $Y_{ic}$, which is generated (with center-specific deviations to the intercept and treatment effect) from a Bernoulli distribution with probability
	\begin{align}\label{eq:datagen_bin_int_slope}
		\logit^{-1}\left\{\alpha+b_{0c}+(\beta+b_{1c})A_{ic}+\gamma'X_{ic}\right\}.
	\end{align}
In this case, Equation \eqref{eq:cov_countmean_bin} equals
\begin{align*}
&\E\left[\left\{\logit^{-1}\left(\alpha+b_{0c}+b_{1c}+\beta+\gamma'X_{ic}\right)\right\} \left\{\logit^{-1}\left(\alpha+b_{0c}+b_{1c}+\beta+\gamma'X_{i'c}\right)\right\}\right]\\
&\quad-\E\left\{\logit^{-1}\left(\alpha+b_{0c}+b_{1c}+\beta+\gamma'X_{ic}\right)\right\}\E\left\{\logit^{-1}\left(\alpha+b_{0c}+b_{1c}+\beta +\gamma'X_{i'c}\right)\right\}.
\end{align*}
Similar to the above, this covariance is non-zero (greater-than zero) as the two patients shared a random effects $b_{0c}+b_{1c}$. As a result, clustering needs to be factored in when estimating the counterfactual means for binary outcomes in this context.

The ATE with binary outcomes can be expressed as
\begin{align*}
\E\left[\frac{A_{ic}}{p_{c}(X_{ic})}\left\{Y_{ic}-m_{1c}(X_{ic})\right\}+m_{1c}(X_{ic})-\frac{1-A_{ic}}{1-p_{c}(X_{ic})}\left\{Y_{ic}-m_{0c}(X_{ic})\right\}-m_{0c}(X_{ic})\right].
\end{align*}
In order to know whether we can ignore the correlation on center-level clustering when estimating the ATE, we can consider the covariance of $\frac{A_{ic}}{p_{c}(X_{ic})}\left\{Y_{ic}-m_{1c}(X_{ic})\right\}+m_{1c}(X_{ic})-\frac{1-A_{ic}}{1-p_{c}(X_{ic})}\left\{Y_{ic}-m_{0c}(X_{ic})\right\}-m_{0c}(X_{ic})$ and $\frac{A_{i'c}}{p_{c}(X_{i'c})}\left\{Y_{i'c}-m_{1c}(X_{i'c})\right\}+m_{1c}(X_{i'c})-\frac{1-A_{i'c}}{1-p_{c}(X_{i'c})}\left\{Y_{i'c}-m_{0c}(X_{i'c})\right\}-m_{0c}(X_{i'c})$, that is, the covariance of influence functions of the ATE for patient $i$ and $i'$ from the same center $c$. This equals 
\begin{equation}\label{eq:cov_ATE_bin}
    \begin{aligned}
&\E\Bigg(\left[\frac{A_{ic}}{p_{c}(X_{ic})}\left\{Y_{ic}-m_{1c}(X_{ic})\right\}+m_{1c}(X_{ic})-\frac{1-A_{ic}}{1-p_{c}(X_{ic})}\left\{Y_{ic}-m_{0c}(X_{ic})\right\}-m_{0c}(X_{ic})\right]\\
&\quad \left[\frac{A_{i'c}}{p_{c}(X_{i'c})}\left\{Y_{i'c}-m_{1c}(X_{i'c})\right\}+m_{1c}(X_{i'c})-\frac{1-A_{i'c}}{1-p_{c}(X_{i'c})}\left\{Y_{i'c}-m_{0c}(X_{i'c})\right\}-m_{0c}(X_{i'c})\right]\Bigg)\\
&\quad-\E\left[\frac{A_{ic}}{p_{c}(X_{ic})}\left\{Y_{ic}-m_{1c}(X_{ic})\right\}+m_{1c}(X_{ic})-\frac{1-A_{ic}}{1-p_{c}(X_{ic})}\left\{Y_{ic}-m_{0c}(X_{ic})\right\}-m_{0c}(X_{ic})\right]\\
&\quad\E\left[\frac{A_{i'c}}{p_{c}(X_{i'c})}\left\{Y_{i'c}-m_{1c}(X_{i'c})\right\}+m_{1c}(X_{i'c})-\frac{1-A_{i'c}}{1-p_{c}(X_{i'c})}\left\{Y_{i'c}-m_{0c}(X_{i'c})\right\}-m_{0c}(X_{i'c})\right].
    \end{aligned}
\end{equation}
For example, consider a binary outcome generated as in Equation \eqref{eq:datagen_bin_int}, then Equation \eqref{eq:cov_ATE_bin} equals
\begin{align*}
&\E\Bigg[\left\{\logit^{-1}\left(\alpha+b_{0c}+\beta+\gamma'X_{ic}\right)-\logit^{-1}\left(\alpha+b_{0c}+\gamma'X_{ic}\right)\right\}\cdot\\
& \quad\left\{\logit^{-1}\left(\alpha+b_{0c}+\beta+\gamma'X_{i'c}\right)-\logit^{-1}\left(\alpha+b_{0c}+\gamma'X_{i'c}\right)\right\}\Bigg]\\
&\quad-\E\left\{\logit^{-1}\left(\alpha+b_{0c}+\beta+\gamma'X_{ic}\right)-\logit^{-1}\left(\alpha+b_{0c}+\gamma'X_{ic}\right)\right\}\cdot\\
&\quad\E\left\{\logit^{-1}\left(\alpha+b_{0c}+\beta +\gamma'X_{i'c}\right)-\logit^{-1}\left(\alpha+b_{0c}+\gamma'X_{i'c}\right)\right\}.
\end{align*}
Here, due to the non-linear nature of logistic function, the random effects ($b_{0c}$, in this case) do not cancel out. Consequently, it introduces a correlation between patient $i$ and $i'$. Thus, we need to account for clustering when estimating the ATE with binary outcomes.

Furthermore, consider a binary outcome which is generated with center-specific deviations to the intercept and treatment effect (see Equation \eqref{eq:datagen_bin_int_slope}). Then, Equation \eqref{eq:cov_ATE_bin} can be expressed as 
\begin{align*}
&\E\Bigg[\left\{\logit^{-1}\left(\alpha+b_{0c}+b_{1c}+\beta+\gamma'X_{ic}\right)-
\logit^{-1}\left(\alpha+b_{0c}+\gamma'X_{ic}\right)\right\}\cdot\\
& \quad\left\{\logit^{-1}\left(\alpha+b_{0c}+b_{1c}+\beta+\gamma'X_{i'c}\right)-\logit^{-1}\left(\alpha+b_{0c}+\gamma'X_{i'c}\right)\right\}\Bigg]\\
&\quad-\E\left\{\logit^{-1}\left(\alpha+b_{0c}+b_{1c}+\beta+\gamma'X_{ic}\right)-\logit^{-1}\left(\alpha+b_{0c}+\gamma'X_{ic}\right)\right\}\cdot\\
&\quad\E\left\{\logit^{-1}\left(\alpha+b_{0c}+b_{1c}+\beta+\gamma'X_{i'c}\right)-\logit^{-1}\left(\alpha+b_{0c}+\gamma'X_{i'c}\right)\right\}.
\end{align*}
In this case, the covariance is non-zero as patient $i$ and $i'$ share $b_{0c}+b_{1c}$. Accordingly, we need to account for center-level clustering when estimating the ATE in a setting where a binary outcome is generated with center-specific deviations to the intercept and treatment effect.

\subsection{Equivalence of patient and center weighted estimands}\label{appe_equivalene_estimands_ch2}
In this section, we show that estimands that weigh patients equally and estimands that weigh centers equally are theoretically the same when treatment effects and center sizes are independent. 

The ATE estimand that weighs patients equally can be expressed as
\begin{align*}
                     &\E\{w(C)\E\left(Y^1_C-Y^0_C|C\right)\},
                     \ \text{where}\ w(C)=\frac{n_{C}}{\E(n_{C})}\\
                     &=\E\left\{w(C)\right\}\E\{\E\left(Y^1_C-Y^0_C|C\right)\}, \ \text{follows from} \ w(C)\indep\E\left(Y^1_C-Y^0_C|C\right) \\
                     &=\E\{\E\left(Y^1_C-Y^0_C|C\right)\},
	\end{align*}
where $C$ represents an indicator of the considered center. In that case, the last expression can be rewritten as $\tau$ in Equation (2).

\subsection{Heterogeneity variance estimators}\label{appe_heterogeneity_variance_ch2}
The heterogeneity variance (the between-center variance) $\sigma^2_{u}$ can be estimated in different ways. For example, as in random-effects meta-analysis, it can be estimated using the method of moments, also known as the DerSimonian and Laird method \citep{dersimonian1986meta}, or via a likelihood-based approach \citep{whitehead2002meta}. Using the method of moments ($\hat{\sigma}^2_{u,DL}$), $\sigma^2_{u}$ can be estimated as
	\begin{align*}
		\frac{Q-(k-1)}{\sum_{c=1}^k\frac{1}{\hat{\sigma}^2_c}-\frac{\sum_{c=1}^k\left(\frac{1}{\hat{\sigma}^2_c}\right)^2}{\sum_{c=1}^k\frac{1}{\hat{\sigma}^2_c}}},
	\end{align*}
where $Q=\sum_{c=1}^{k}\frac{1}{\hat{\sigma}^2_c}(\hat{\tau}_{c}-\hat{\tau}^{*})^2$ and $\hat{\tau}^{*}=\frac{\sum_{c=1}^{k}\frac{1}{\hat{\sigma}^2_c}\hat{\tau}_c}{\sum_{c=1}^{k}\frac{1}{\hat{\sigma}^2_c}}$. Alternatively, $\sigma^2_{u}$ can be estimated using a likelihood approach, specifically through the restricted maximum likelihood approach (REML). The REML log-likelihood function (see Section A.7 in the Appendix of \cite{whitehead2002meta}) is given by
	\begin{align*}
		-\frac{k}{2}\log(2\pi)-\frac{1}{2}\sum_{c=1}^{k}\log \left(\sigma^2_{c}+\sigma^2_{u}\right)-\frac{1}{2}\sum_{c=1}^{k}\frac{(\hat{\tau}_{c}-\tau^{*}_{ML_{t+1}})}{\sigma^2_{c}+\sigma^2_{u}}-\frac{1}{2}\log\left(\sum_{c=1}^{k} \frac{1}{\sigma^2_{c}+\sigma^2_{u}}\right),
	\end{align*}
where $\hat{\tau}^{*}_{ML_{t+1}}=\frac{\sum_{c=1}^{k}\frac{\hat{\tau}_{c}}{\hat{\sigma}^2_{c}+\hat{\sigma}^2_{u,ML_{t}}}}{\sum_{c=1}^{k}\frac{1}{\hat{\sigma}^2_{c}+\hat{\sigma}^2_{u,ML_{t}}}}$. Here, $\hat{\sigma}^2_{u,ML_{t}}$ is the ML estimate of $\sigma^2_{u}$ at $t$th cycle of the iteration. The REML estimate of $\sigma^2_{u}$ at the $(t+1)$th cycle of iteration $(\hat{\sigma}^{2}_{u,REML_{t+1}})$ can be approximately estimated by iterating 
	\begin{align*}
		\frac{\sum_{c=1}^{k}\frac{\frac{k}{(k-1)}\left(\hat{\tau}_{c}-\hat{\tau}^{*}_{ML_{t+1}}\right)^2-\hat{\sigma}^2_{c}}{\left(\hat{\sigma}^2_{c}+\hat{\sigma}^2_{u,REML_{t}}\right)^2}}{\sum_{c=1}^{k}\frac{1}{\left(\hat{\sigma}^2_{c}+\hat{\sigma}^2_{u,REML_{t}}\right)^2}},\ t=0,1,2,3...
	\end{align*}

When center sizes and treatment effects are not independent, the DL and REML heterogeneity variance estimators may be biased. In this case, $\sigma^2_{u}$ can be estimated by debiasing the sample heterogeneity variance, which represents the na\"ive estimator of $\sigma^2_{u}$. To estimate $\sigma^2_{u}$, the sample heterogeneity variance, given by $\frac{1}{k}\sum_{c=1}^{k}\left(\hat{\tau}_{c}-\hat{\tau}\right)^2$ is then debiased to account for estimation noise as follows:
\begin{align*}
& \E\left\{\frac{1}{k}\sum_{c=1}^{k}(\hat{\tau}_{c}-\hat{\tau})^2\right\}\\
&= \E\left\{\frac{1}{k}\sum_{c=1}^{k}(\hat{\tau}_{c}-\tau_{c}+\tau_{c}-\tau+\tau-\hat{\tau})^2\right\} \\
&= \E\left\{\frac{1}{k}\sum_{c=1}^{k}(\hat{\tau}_{c}-\tau_{c})^2\right\}+\E\left\{\frac{1}{k}\sum_{c=1}^{k}(\tau_{c}-\tau)^2\right\} + \E\left\{(\tau-\hat{\tau})^2\right\} \\
& + 2\E\left\{\frac{1}{k}\sum_{c=1}^{k}(\hat{\tau}_{c}-\tau_{c})(\tau_{c}-\tau)\right\} + 2\E\left\{\frac{1}{k}\sum_{c=1}^{k}(\hat{\tau}_{c}-\tau_{c})(\tau-\hat{\tau})\right\}+2\E\left\{\frac{1}{k}\sum_{c=1}^{k}(\tau_{c}-\tau)(\tau-\hat{\tau})\right\}\\
&=\E\left\{\frac{1}{k}\sum_{c=1}^{k}(\hat{\tau}_{c}-\tau_{c})^2\right\}+ \E\left\{\frac{1}{k}\sum_{c=1}^{k}(\tau_{c}-\tau)^2\right\} + \E\left\{(\tau-\hat{\tau})^2\right\}+0-2\E\left\{(\tau-\hat{\tau})^2\right\}+0.
\end{align*}
The fourth term in the third equality follows from
\begin{align*}
&2\E\left\{\frac{1}{k}\sum_{c=1}^{k}(\hat{\tau}_{c}-\tau_{c})(\tau_{c}-\tau)\right\} \\
&=2\E\left\{\frac{1}{k}\sum_{c=1}^{k}(\hat{\tau}_{c}-\tau_{c})\tau_{c}\right\}-2\E\left\{\frac{1}{k}\sum_{c=1}^{k}(\hat{\tau}_{c}-\tau_{c})\tau\right\} \\
&=2\E\left[\frac{1}{k}\sum_{c=1}^{k}\E\left\{(\hat{\tau}_{c}-\tau_{c})\tau_{c}|\tau_{c}\right\}\right]-2\E\left[\tau\frac{1}{k}\sum_{c=1}^{k}\E\left\{(\hat{\tau}_{c}-\tau_{c})|\tau_{c}\right\}\right]\\
&=2\E\left[\frac{1}{k}\sum_{c=1}^{k}\tau_{c}\E\left\{(\hat{\tau}_{c}-\tau_{c})|\tau_{c}\right\}\right]-2\E\left[\tau\frac{1}{k}\sum_{c=1}^{k}\E\left\{(\hat{\tau}_{c}-\tau_{c})|\tau_{c}\right\}\right]\\
&=0.
\end{align*}
Here, the last line follows from the fact that $\E\left\{(\hat{\tau}_{c}-\tau_{c})|\tau_{c}\right\}=0$.

Furthermore, the fifth term in the third equality follows from
\begin{align*}
& 2\E\left\{\frac{1}{k}\sum_{c=1}^{k}(\hat{\tau}_{c}-\tau_{c})(\tau-\hat{\tau})\right\}\\
&=2\E\left\{(\tau-\hat{\tau})\frac{1}{k}\sum_{c=1}^{k}(\hat{\tau}_{c}-\tau_{c})\right\}\\
&=-2\E\left\{(\tau-\hat{\tau})^2\right\}.
\end{align*}  
Moreover, the sixth term in the third equality follows from
\begin{align*}
& 2\E\left\{\frac{1}{k}\sum_{c=1}^{k}(\tau_{c}-\tau)(\tau-\hat{\tau})\right\}\\
&=2\E\left\{(\tau-\hat{\tau})\frac{1}{k}\sum_{c=1}^{k}(\tau_{c}-\tau)\right\}\\
&=0.
\end{align*}
Then, from the first term in the third equality, we obtain
\begin{align*}
    \frac{1}{k}\sum_{c=1}^{k}\E(\hat{\tau}_{c}-\tau_{c})^2=\frac{1}{k}\sum_{c=1}^{k}\text{Var}\left(\hat{\tau}_{c}\right).
\end{align*}

Further, the second term in the third equality simplifies to $\text{Var}\left(\tau_{c}\right)$, which represents the between-center variability in the true treatment effects. Moreover, combining the third and fifth terms in the third equality yields
\begin{align*}
    -\E\left(\hat{\tau}-\tau\right)^2=&-\E\left\{\frac{1}{k}\sum_{c=1}^{k}(\hat{\tau}_{c}-\tau_{c})\right\}^2\\
    =&-\frac{1}{k^2}\sum_{c=1}^{k}\E(\hat{\tau}_{c}-\tau_{c})^2\\
    =&-\frac{1}{k^2}\sum_{c=1}^{k}\text{Var}\left(\hat{\tau}_{c}\right).
\end{align*}

Finally, combining all terms gives
\begin{align*}
    &\frac{1}{k}\sum_{c=1}^{k}\text{Var}\left(\hat{\tau}_{c}\right)+\text{Var}\left(\tau_{c}\right)-\frac{1}{k^2}\sum_{c=1}^{k}\text{Var}\left(\hat{\tau}_{c}\right)\\
&=\text{Var}\left(\tau_{c}\right)+\frac{k-1}{k^2}\sum_{c=1}^{k}\text{Var}\left(\hat{\tau}_{c}\right).
\end{align*}
Thus, $\sigma^2_{u}$ can be estimated unbiasedly as
\begin{align*}
\hat{\sigma}^2_{u,DB}=\frac{1}{k}\sum_{c=1}^{k}\left(\hat{\tau}_{c}-\hat{\tau}\right)^2-\frac{k-1}{k^2}\sum_{c=1}^{k}\hat{\sigma}^2_{c},
\end{align*}
where $\hat{\sigma}^2_{c}$ is the variance of $\hat{\tau}_{c}$ just on center $c$.

\subsection{Accounting for clustering in cluster-randomized trials}\label{appe_accounting_clustering_cluter_RCTs_ch2} 
Although this paper primarily focuses on accounting for clustering in individually randomized trials, we here briefly discuss its implications for cluster-randomized trials. Let clusters be indexed by $j = 1, \ldots, J$, with individuals $i = 1, \ldots, n_j$ within each cluster. Define $A_{j} = 1$ if cluster $j$ is assigned to the treatment group, and $A_{j}= 0$ if it is assigned to the control group. 
We focus on the counterfactual mean under treatment
\begin{align*}\label{eq:countMean_c_RCTs}
	\psi_1=\E\left(\overline{Y}_j^1\right),
	\end{align*}
where $\overline{Y}_j^1=\E\left(Y_{ij}^1\right)$ for cluster $j$ and where the averaging is w.r.t. a uniform distribution over the clusters. This represents the expected counterfactual outcome in a random cluster if it were assigned to treatment. 

We introduce the following estimation procedure to account for clustering, illustrating it for the case of a binary outcome. First, we fit a mixed-effects logistic regression model of the form
\begin{align*}
    \mathbb{E}\left(Y_{ij} \mid A_{j}, X_{ij}, b_{j}\right) 
    &= \operatorname{logit}^{-1}\left\{\alpha + b_{j} + \beta A_{j} + \gamma' X_{ij}\right\},
\end{align*}
where $b_{j}$ represents a random intercept for cluster $j$, assumed to follow a normal distribution with mean zero and variance $\sigma_{b}^{2}$. Further, $X_{ij}$ represents a vector of individual-level (and possibly cluster-level) baseline covariates. Next, we predict the outcome on treatment from the fitted mixed-effects model as $\hat{m}_{1j}(X_{ij})=\logit^{-1} (\hat{\alpha}+\hat{b}_{j}+\hat{\beta}+\hat{\gamma}'X_{ij})$, where $\hat{b}_{j}$ is a random draw from the estimated random-effect distribution. Alternatively, a fixed-effects logistic regression model can be fitted, allowing each cluster to have its own intercept and thereby adjusting for unobserved cluster-level differences.
		
The counterfactual mean under treatment estimator based on cluster $j$ can be then expressed as
\begin{align*}
    \hat{\psi}_{1j}=\frac{1}{n_j}\sum_{i=1}^{n_j}\left[\frac{A_{ij}}{\hat{p}_{j}(X_{ij})}\left\{Y_{ij}-\hat{m}_{1j}(X_{ij})\right\}+\hat{m}_{1j}(X_{ij})\right],
\end{align*}
with $\hat{p}_{j}(X_{ij})$ is the estimated randomization probability. Here, with a slight abuse of notation, $\hat{p}_{j}(X_{ij})$ is based solely on cluster-level covariates, thus being constant within each cluster $j$. It can be obtained by fitting a logistic regression for treatment assignment on the cluster-level covariates. Finally, the counterfactual mean under treatment across all clusters can be estimated as $\hat{\psi}_{1}=\frac{1}{J}\sum_{j=1}^{J}\hat{\psi}_{1j}$. The variance of $\hat{\psi}_{1}$ can be approximated as 1 over $J$ times the sample variance of $\frac{A_{j}}{\hat{p}_{j}(X_{j})} \left(\bar{Y}_{j}-\hat{\psi}_{1j}\right)+ \hat{\psi}_{1j}$, where $\bar{Y}_{j}$ represents the mean observed outcome within cluster $j$. For the average treatment effect, the variance is computed as 1 over $J^2$ times the sum of the variances of the mean outcomes under treatment and control, given clusters are independent. This gives results that are equivalent to those in \cite{balzer2019new} when the number of clusters is large. See also related work by \cite{wang2023model}.

\subsection{Standard error calculation for the data analysis}\label{appe_standard_error_data_analysis_ch2}
Let $\omega_c$ denote the weight assigned to center $c$. The mean outcome under treatment can be estimated as
\begin{align*}
\hat{\tau}=\frac{\sum_{c=1}^{k}\sum_{j=1}^{J_c}\sum_{i=1}^{n_{jc}}\omega_cIF_{ijc}}{\sum_{c=1}^{k}\sum_{j=1}^{J_c}\sum_{i=1}^{n_{jc}}\omega_c},
\end{align*}
where $k$ denotes the number of centers, $n_{jc}$ the number of patients in cluster $j$ of center $c$, $J_c$ the number of clusters in center $c$, and $IF_{ijc}$ is the influence function under treatment for patient $i$ in cluster $j$ of center $c$, which admits the decomposition
\begin{align*}
\mu+\alpha_c+b_{jc}+\epsilon_{ijc}.
\end{align*}
Here, $\mu$, $\alpha_c$, $b_{jc}$, and $\epsilon_{ijc}$ denote the overall mean, the center effect, cluster effect and residual term, respectively, and are assumed to be mutually independent. The choice of $\omega_c$ determines the target estimand. We impose the constraint $\sum_{c=1}^k\sum_{j=1}^{J_c}\sum_{i=1}^{n_{jc}}\omega_c=1$, and if the interest lies in estimating patient-weighted estimand, $\omega_c=1/\sum_{c^{\star}=1}^kn_{c^{\star}}$. In contrast, if the interest lies in center-weighted estimand, $\omega_c=1/(kn_c)$, where $n_c=\sum_{j=1}^{J_{c}}n_{jc}$. The variance of $\hat{\tau}$ can be estimated as
\begin{align*}
 \Var\left(\frac{\sum_{c=1}^{k}\sum_{j=1}^{J_c}\sum_{i=1}^{n_{jc}}\omega_cIF_{ijc}}{\sum_{c=1}^{k}\sum_{j=1}^{J_c}\sum_{i=1}^{n_{jc}}\omega_c}\right)
&=\Var\left(\frac{\sum_{c=1}^{k}\sum_{j=1}^{J_c}\sum_{i=1}^{n_{jc}}\omega_cIF_{ijc}}{\sum_{c=1}^{k}\omega_cn_c}\right)\\
&=\frac{1}{\left(\sum_{c=1}^{k}\omega_cn_c\right)^2} \Var\left\{\sum_{c=1}^{k}\sum_{j=1}^{J_c}\sum_{i=1}^{n_{jc}}\omega_c \left(\alpha_c+b_{jc}+\epsilon_{ijc}\right)\right\}\\
&= \frac{1}{\left(\sum_{c=1}^{k}\omega_cn_c\right)^2} \Bigg\{\Var\left(\sum_{c=1}^{k}\sum_{j=1}^{J_c}\sum_{i=1}^{n_{jc}} \omega_c \alpha_c\right)+ \Var\left(\sum_{c=1}^{k}\sum_{j=1}^{J_c}\sum_{i=1}^{n_{jc}} \omega_c b_{jc}\right)+\\
&\qquad\qquad\Var\left(\sum_{c=1}^{k}\sum_{j=1}^{J_c}\sum_{i=1}^{n_{jc}} \omega_c \epsilon_{ijc}\right)\Bigg\}\\
&= \frac{1}{\left(\sum_{c=1}^{k}\omega_cn_c\right)^2} \Bigg\{\Var\left(\sum_{c=1}^{k}\omega_c\alpha_cn_c\right)+ \Var\left(\sum_{c=1}^{k}\omega_c\sum_{j=1}^{J_c}n_{jc}b_{jc}\right)+\\
&\qquad\qquad\Var\left(\sum_{c=1}^{k}\omega_c\sum_{j=1}^{J_c}\sum_{i=1}^{n_{jc}}\epsilon_{ijc}\right)\Bigg\}\\
&=\frac{1}{\left(\sum_{c=1}^{k}\omega_cn_c\right)^2} \Bigg\{\sum_{c=1}^{k}\Var\left(\alpha_c\right)\omega^2_cn^2_c+\sum_{c=1}^{k}\omega^2_c\sum_{j=1}^{J_c}n^2_{jc}\Var\left(b_{jc}\right)+\\
&\qquad\qquad\sum_{c=1}^{k}\Var\left(\epsilon_{ijc}\right) \omega^2_c n_c\Bigg\}\\
&=\frac{1}{\left(\overline{\omega n}\right)^2}\left\{\Var(\alpha_c)\overline{\omega^2 n^2}+\Var(b_{jc})\overline{\omega^2 \tilde{n}^2}
+\Var(\epsilon_{ijc})\,\overline{\omega^2n}\right\}.
\end{align*}
Here, $\overline{\omega n}=\sum_{c=1}^{k}\omega_cn_c$, $\overline{\omega^2n^2}=\sum_{c=1}^{k}\omega^2_cn^2_{c}$, $\overline{\omega^2\tilde{n}^2}=\sum_{c=1}^{k}\omega^2_c\sum_{j=1}^{J_c}n^2_{jc}$ and $\overline{\omega^2n}=\sum_{c=1}^{k}\omega^2_cn_c$. Furthermore, $\Var\left(\alpha_c\right)$, $\Var\left(b_{jc}\right)$, and $\Var\left(\epsilon_{ijc}\right)$ can be obtained by fitting a hierarchical mixed-effects model.

\clearpage
\section{Additional simulation and data application results}\label{appe_sim_data_analysis_results_ch2}
\subsection{Additional simulation results}\label{appe_sim_results_ch2}
\begin{figure}[!h]
    \centering
    \includegraphics[width=1\textwidth]{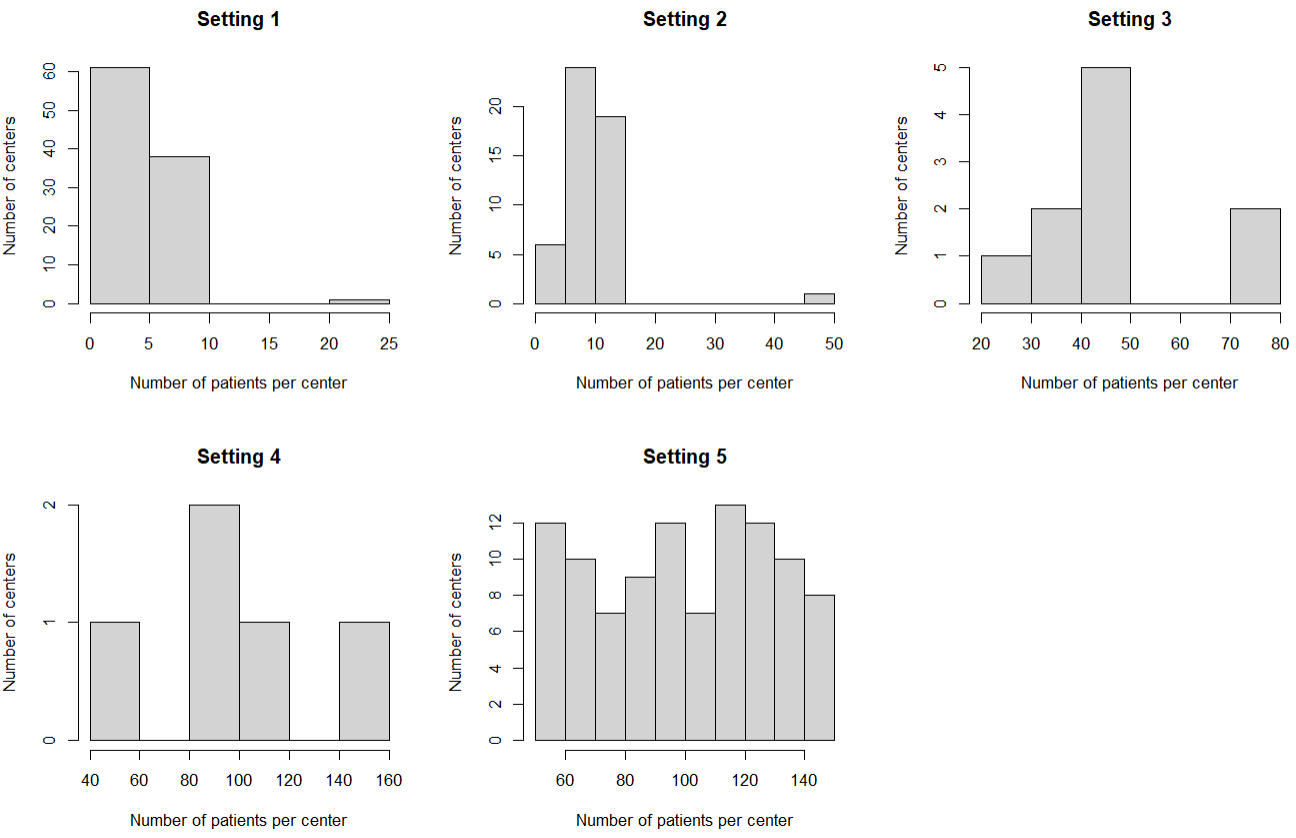}
    \caption{Distribution of the number of patients per center across simulation settings.}
      \label{figure:appe_dist_nc_ch2}
\end{figure}

\autoref{figure:appe_dist_nc_ch2} displays the distribution of center sizes used across the simulation settings, illustrating the substantial heterogeneity in the number of patients per center that reflects realistic multicenter trial. Further, \autoref{table:appe_ran_eff_ch2} summarizes the combinations of random‐effects variances considered for both continuous and binary outcomes, with each column representing a distinct simulation scenario.

\begin{table}[htbp]
	\caption{Different combinations of random-effects variances considered for both continuous and binary outcomes (each column corresponds with a single simulation scenario).}
	\makebox[\textwidth][c]{
		\begin{tabular}{ccccccccccccccc}
			\toprule
             \multicolumn{8}{c}{Continuous outcome} & \multicolumn{7}{c}{Binary outcome} \\
				\cmidrule(lr){2-8} \cmidrule(lr){9-15} 
			$\sigma^2_{b0}$ & 0 & 0.05 & 0.10 & 0.15 & 0.10   & 0.15 & 0.15 & 0 & 0.25 & 0.5 & 0.5 & 0.5 & 0.5 & 0.75   \\
			$\sigma^2_{b1}$ & 0 & 0 & 0 & 0  & 0.10  & 0.15 & 0.15& 0 & 0 & 0 & 0.25 & 0.25 & 0.5 & 0.5   \\
			$\sigma^2_{b2}$ & 0 & 0 & 0  & 0  & 0  & 0  & $4\times10^{-6}$  & 0 & 0 & 0  & 0  & 0.25 & 0 & 0.5    \\
			\bottomrule
		\end{tabular}
	}
  \label{table:appe_ran_eff_ch2}
\end{table}

\clearpage
\autoref{table:appe_k_50_nc_10_cont_ch2} reports the Monte Carlo standard deviation, average standard errors, and 95\% confidence interval coverage for the counterfactual mean under treatment and the ATE for a setting with 50 centers and continuous outcomes.
\begin{table}[!h]
	\caption{Simulation results: Monte Carlo standard deviation, average standard errors and coverage probabilities of 95\% confidence intervals for both counterfactual mean on treatment and ATE based on 1000 simulations. Setting: Continuous outcome, $k$=50, $n_c$(Avg=10, Min=2, Max=48), and different values of random-effects variance. Results are based on weighting centers equally.}
	\begin{center}
		\resizebox{\textwidth}{!}{
			\begin{tabular}{>{\raggedright\arraybackslash}p{3.5cm}cccccccccccccccccc}
				\toprule
				& \multicolumn{9}{c}{\textbf{Counterfactual mean on treatment}} &\multicolumn{9}{c}{\textbf{ATE}} \\
				\cmidrule(lr){2-10} \cmidrule(lr){11-19}
				&& \multicolumn{4}{c}{\textbf{SE}} & \multicolumn{4}{c}{\textbf{Coverage (\%)}} &&\multicolumn{4}{c}{\textbf{SE}}&\multicolumn{4}{c}{\textbf{Coverage (\%)}} \\
				\cmidrule(lr){3-6} \cmidrule(lr){7-10} \cmidrule(lr){12-15} \cmidrule(lr){16-19}
				Method &  SD & Na\"ive & REML & DL & DB & Na\"ive & REML & DL & DB  & SD & Na\"ive & REML & DL & DB & Naive & REML & DL & DB \\
				\midrule
				&&&&& \multicolumn{9}{c}{\textit{Random-effects variances}: $\sigma^2_{b0}$=0, $\sigma^2_{b1}$=0, $\sigma^2_{b2}$=0}  &&&&&\\
                \multicolumn{19}{l}{\textbf{Unadjusted}}\\
				Na\"ive & 0.068 & 0.070 &  &  &  & 95.1 &  &  &  & 0.097 & 0.099 &  &  &  & 95.1 &  &  &  \\ 
				Fixed & 0.079 &  & 0.081 & 0.081 & 0.082 &  & 95.0 & 95.0 & 95.2 & 0.112 &  & 0.112 & 0.112 & 0.112 &  & 94.5 & 94.5 & 94.7\\ 
				Mixed(1$\mid$c) & 0.079 &  & 0.083 & 0.083 & 0.085 &  & 95.1 & 95.4 & 95.5 & 0.112 &  & 0.117 & 0.118 & 0.119 &  & 95.0 & 95.0 & 95.6 \\ 
                Mixed(1$\mid$c) Sam. & 0.111 &  & 0.115 & 0.115 & 0.114 &  & 94.3 & 94.5 & 94.7 & 0.134 &  & 0.134 & 0.134 & 0.135 &  & 95.5 & 95.3 & 95.5 \\ 
				Mixed(1+A$\mid$c) & 0.079 &  & 0.083 & 0.083 & 0.084 &  & 95.4 & 95.3 & 95.5 & 0.112 &  & 0.117 & 0.117 & 0.119 &  & 94.9 & 95.0 & 95.5 \\ 
                Mixed(1+A$\mid$c) Sam. & 0.111 &  & 0.115 & 0.115 & 0.114 &  & 94.2 & 94.6 & 94.7 & 0.134 &  & 0.134 & 0.134 & 0.135 &  & 95.2 & 95.1 & 95.2 \\ 
                \multicolumn{19}{l}{\textbf{Adjusted}}\\
				Na\"ive & 0.065 & 0.067 &  &  &  & 95.1 &  &  &  & 0.088 & 0.089 &  &  &  & 94.5 &  &  &  \\ 
				Fixed & 0.075 &  & 0.078 & 0.078 & 0.078 &  & 95.9 & 95.7 & 95.6 & 0.102 &  & 0.101 & 0.100 & 0.100 &  & 94.0 & 94.0 & 94.4 \\ 
				Mixed(1$\mid$c) & 0.076 &  & 0.079 & 0.079 & 0.081 &  & 95.8 & 96.0 & 96.1 & 0.101 &  & 0.105 & 0.106 & 0.107 &  & 95.4 & 95.5 & 96.0 \\ 
                Mixed(1$\mid$c) Sam. & 0.108 &  & 0.112 & 0.112 & 0.111 &  & 94.8 & 94.9 & 94.8 & 0.124 &  & 0.124 & 0.124 & 0.125 &  & 95.4 & 95.4 & 95.3 \\ 
				Mixed(1+A$\mid$c) & 0.076 &  & 0.079 & 0.079 & 0.080 &  & 95.4 & 95.3 & 95.5 & 0.101 &  & 0.105 & 0.106 & 0.107 &  & 94.9 & 95.0 & 95.5 \\ 
                Mixed(1+A$\mid$c) Sam. & 0.108 &  & 0.113 & 0.113 & 0.111 &  & 94.8 & 95.0 & 94.9 & 0.124 &  & 0.125 & 0.125 & 0.125 &  & 95.5 & 95.5 & 95.3\\
				\midrule		
				&&&&& \multicolumn{9}{c}{\textit{Random-effects variances}: $\sigma^2_{b0}$=0.1, $\sigma^2_{b1}$=0, $\sigma^2_{b2}$=0}  &&&&&\\
                \multicolumn{19}{l}{\textbf{Unadjusted}}\\
				Na\"ive & 0.091 & 0.073 &  &  &  & 88.4 &  &  &  & 0.103 & 0.103 &  &  &  & 94.5 &  &  &  \\ 
				Fixed & 0.094 &  & 0.092 & 0.092 & 0.091 &  & 95.1 & 95.0 & 95.0 & 0.114 &  & 0.112 & 0.112 & 0.112 &  & 94.7 & 94.6 & 94.7 \\ 
				Mixed(1$\mid$c) & 0.095 &  & 0.093 & 0.093 & 0.093 &  & 95.0 & 95.1 & 94.8 & 0.116 &  & 0.115 & 0.116 & 0.117 &  & 95.2 & 95.4 & 95.5 \\ 
                Mixed(1$\mid$c) Sam. & 0.119 &  & 0.114 & 0.114 & 0.114 &  & 93.9 & 94.0 & 94.2 & 0.138 &  & 0.134 & 0.134 & 0.135 &  & 93.9 & 94.1 & 94.0 \\ 
				Mixed(1+A$\mid$c) & 0.095 &  & 0.093 & 0.093 & 0.093 &  & 95.0 & 95.100 & 94.8 & 0.117 &  & 0.117 & 0.117 & 0.119 &  & 95.4 & 95.7 & 95.7 \\
                Mixed(1+A$\mid$c) Sam. & 0.119 &  & 0.114 & 0.114 & 0.114 &  & 93.9 & 94.0 & 94.3 & 0.138 &  & 0.134 & 0.134 & 0.135 &  & 94.0 & 94.0 & 94.1 \\ 
                \multicolumn{19}{l}{\textbf{Adjusted}}\\
				Na\"ive & 0.089 & 0.070 &  &  &  & 87.4 &  &  &  & 0.094 & 0.093 &  &  &  & 94.4 &  &  &  \\ 
				Fixed & 0.091 &  & 0.089 & 0.089 & 0.088 &  & 94.7 & 94.7 & 94.4 & 0.103 &  & 0.101 & 0.101 & 0.101 &  & 94.0 & 93.7 & 94.0 \\ 
				Mixed(1$\mid$c) & 0.091 &  & 0.090 & 0.090 & 0.090 &  & 95.0 & 95.0 & 95.0 & 0.105 &  & 0.104 & 0.104 & 0.105 &  & 94.8 & 94.8 & 95.0 \\ 
                Mixed(1$\mid$c) Sam. & 0.116 &  & 0.112 & 0.112 & 0.111 &  & 94.1 & 94.2 & 93.5 & 0.126 &  & 0.124 & 0.124 & 0.125 &  & 94.8 & 94.7 & 94.6 \\ 
				Mixed(1+A$\mid$c) & 0.091 &  & 0.090 & 0.090 & 0.090 &  & 95.0 & 95.1 & 94.8 & 0.106 &  & 0.105 & 0.106 & 0.107 &  & 95.4 & 95.7 & 95.7 \\ 
                Mixed(1+A$\mid$c) Sam. & 0.116 &  & 0.112 & 0.112 & 0.111 &  & 94.0 & 94.3 & 93.5 & 0.126 &  & 0.124 & 0.124 & 0.125 &  & 94.9 & 95.0 & 94.6 \\ 
				\bottomrule		
				&&&&& \multicolumn{9}{c}{\textit{Random-effects variances}: $\sigma^2_{b0}$=0.15, $\sigma^2_{b1}$=0.15, $\sigma^2_{b2}$=0}  &&&&&\\
                \multicolumn{19}{l}{\textbf{Unadjusted}}\\
				Na\"ive & 0.120 & 0.078 &  &  &  & 78.6 &  &  &  & 0.126 & 0.107 &  &  &  & 89.7 &  &  &  \\ 
				Fixed & 0.116 &  & 0.113 & 0.113 & 0.111 &  & 95.1 & 95.0 & 94.0 & 0.128 &  & 0.122 & 0.121 & 0.118 &  & 93.6 & 93.5 & 92.5 \\ 
				Mixed(1$\mid$c) & 0.116 &  & 0.115 & 0.115 & 0.114 &  & 95.0 & 94.7 & 93.8 & 0.129 &  & 0.125 & 0.125 & 0.124 &  & 93.5 & 93.5 & 93.3 \\ 
                Mixed(1$\mid$c) Sam. & 0.114 &  & 0.115 & 0.115 & 0.113 &  & 95.1 & 95.1 & 95.2 & 0.131 &  & 0.135 & 0.135 & 0.135 &  & 96.0 & 95.9 & 96.0 \\ 
				Mixed(1+A$\mid$c) & 0.117 &  & 0.114 & 0.114 & 0.113 &  & 94.5 & 94.4 & 93.6 & 0.130 &  & 0.127 & 0.127 & 0.127 &  & 94.2 & 94.1 & 93.7 \\ 
                Mixed(1+A$\mid$c) Sam. & 0.114 &  & 0.115 & 0.115 & 0.113 &  & 95.1 & 95.1 & 95.0 & 0.131 &  & 0.135 & 0.135 & 0.135 &  & 96.0 & 96.1 & 96.1 \\ 
                \multicolumn{19}{l}{\textbf{Adjusted}}\\
				Na\"ive & 0.118 & 0.075 &  &  &  & 79.1 &  &  &  & 0.117 & 0.098 &  &  &  & 89.0 &  &  &  \\ 
				Fixed & 0.113 &  & 0.111 & 0.111 & 0.109 &  & 94.8 & 94.7 & 93.9 & 0.118 &  & 0.111 & 0.111 & 0.108 &  & 92.9 & 92.5 & 91.7 \\ 
				Mixed(1$\mid$c) & 0.114 &  & 0.112 & 0.112 & 0.111 &  & 94.3 & 94.4 & 94.2 & 0.119 &  & 0.114 & 0.114 & 0.113 &  & 93.2 & 93.1 & 92.7 \\ 
                 Mixed(1$\mid$c) Sam. & 0.110 &  & 0.112 & 0.112 & 0.111 &  & 95.5 & 95.6 & 94.9 & 0.120 &  & 0.125 & 0.125 & 0.125 &  & 96.4 & 96.5 & 96.2 \\ 
				Mixed(1+A$\mid$c) & 0.114 &  & 0.112 & 0.112 & 0.111 &  & 94.5 & 94.4 & 93.6 & 0.120 &  & 0.117 & 0.117 & 0.116 &  & 94.2 & 94.1 & 93.7 \\ 
                Mixed(1+A$\mid$c) Sam. & 0.110 &  & 0.112 & 0.112 & 0.111 &  & 95.5 & 95.6 & 95.0 & 0.120 &  & 0.125 & 0.125 & 0.125 &  & 96.6 & 96.6 & 96.3 \\ 
				\bottomrule
				&&&&& \multicolumn{9}{c}{\textit{Random-effects variances}: $\sigma^2_{b0}$=0.15, $\sigma^2_{b1}$=0.15, $\sigma^2_{b2}=4\times10^{-6}$} &&&&&\\
                \multicolumn{19}{l}{\textbf{Unadjusted}}\\
				Na\"ive & 0.182 & 0.091 &  &  &  & 67.7 &  &  &  & 0.141 & 0.126 &  &  &  & 93.1 &  &  &  \\ 
				Fixed & 0.161 &  & 0.155 & 0.156 & 0.154 &  & 93.9 & 93.7 & 93.6 & 0.135 &  & 0.124 & 0.123 & 0.120 &  & 93.1 & 93.2 & 92.6 \\ 
				Mixed(1$\mid$c) & 0.162 &  & 0.156 & 0.156 & 0.155 &  & 93.6 & 93.5 & 93.4 & 0.135 &  & 0.125 & 0.125 & 0.124 &  & 93.6 & 93.4 & 93.1 \\ 
                Mixed(1$\mid$c) Sam. & 0.114 &  & 0.114 & 0.114 & 0.113 &  & 95.0 & 94.8 & 94.6 & 0.134 &  & 0.134 & 0.134 & 0.135 &  & 95.6 & 95.5 & 95.6 \\
				Mixed(1+A$\mid$c) & 0.162 &  & 0.156 & 0.156 & 0.155 &  & 93.5 & 93.5 & 93.4 & 0.136 &  & 0.127 & 0.126 & 0.125 &  & 93.8 & 93.6 & 93.4 \\ 
                 Mixed(1+A$\mid$c) Sam. & 0.114 &  & 0.114 & 0.114 & 0.113 &  & 95.0 & 94.8 & 94.5 & 0.134 &  & 0.134 & 0.135 & 0.135 &  & 95.5 & 95.4 & 95.6 \\ 
                \multicolumn{19}{l}{\textbf{Adjusted}}\\
				Na\"ive & 0.181 & 0.088 &  &  &  & 66.4 &  &  &  & 0.135 & 0.118 &  &  &  & 91.3 &  &  &  \\ 
				Fixed & 0.159 &  & 0.153 & 0.154 & 0.152 &  & 94.0 & 93.4 & 93.4 & 0.124 &  & 0.113 & 0.113 & 0.110 &  & 94.1 & 93.6 & 93.2 \\ 
				Mixed(1$\mid$c) & 0.160 &  & 0.154 & 0.154 & 0.152 &  & 93.5 & 93.6 & 93.4 & 0.124 &  & 0.115 & 0.114 & 0.113 &  & 94.3 & 94.2 & 93.2 \\ 
                Mixed(1$\mid$c) Sam. & 0.111 &  & 0.112 & 0.112 & 0.111 &  & 95.0 & 95.0 & 93.8 & 0.123 &  & 0.124 & 0.124 & 0.125 &  & 95.7 & 95.5 & 95.5 \\ 
				Mixed(1+A$\mid$c) & 0.159 &  & 0.154 & 0.154 & 0.153 &  & 93.5 & 93.5 & 93.4 & 0.124 &  & 0.116 & 0.116 & 0.115 &  & 93.8 & 93.6 & 93.4 \\ 
                Mixed(1+A$\mid$c) Sam. & 0.111 &  & 0.112 & 0.112 & 0.111 &  & 94.9 & 94.9 & 94.0 & 0.124 &  & 0.125 & 0.125 & 0.125 &  & 95.6 & 95.6 & 95.5 \\ 
				\bottomrule
			\end{tabular}%
		}
	\end{center}
	\tiny
	\vspace{0.01cm} 
   \label{table:appe_k_50_nc_10_cont_ch2}
\end{table}

\clearpage
\autoref{table:appe_k_10_nc_50_cont_ch2} and \autoref{table:appe_k_5_nc_100_cont_ch2} report the Monte Carlo standard deviation, average standard errors, and 95\% confidence interval coverage for the counterfactual mean under treatment and the ATE under small number of centers ($k=10$ and $k=5$) with continuous outcomes. The results demonstrate that greater between-center heterogeneity and a smaller number of centers lead to undercoverage across the estimators considered. However, estimators that account clustering provide better coverage probabilities than the na\"ive variance estimators. In addition, under these settings, the REML and DL estimators yield better performance than DB. Our method constructs confidence intervals using a $t$-distribution with degrees of freedom correction, which already provides a small sample adjustment. Investigating alternative corrections that bring coverage probabilities closer to the nominal level is an important future extension.

\clearpage
\begin{table}[!h]
	\caption{Simulation results: Monte Carlo standard deviation, average standard errors and coverage probabilities of 95\% confidence intervals for both counterfactual mean on treatment and ATE based on 1000 simulations. Setting: Continuous outcome, $k$=10, $n_c$(Avg=50, Min=25, Max=80), and different values of random-effects variance. Results are based on weighting centers equally.}
	\begin{center}
		\resizebox{\textwidth}{!}{
			\begin{tabular}{>{\raggedright\arraybackslash}p{3.5cm}cccccccccccccccccc}
				\toprule
				& \multicolumn{9}{c}{\textbf{Counterfactual mean on treatment}} &\multicolumn{9}{c}{\textbf{ATE}} \\
				\cmidrule(lr){2-10} \cmidrule(lr){11-19}
				&& \multicolumn{4}{c}{\textbf{SE}} & \multicolumn{4}{c}{\textbf{Coverage (\%)}} &&\multicolumn{4}{c}{\textbf{SE}}&\multicolumn{4}{c}{\textbf{Coverage (\%)}} \\
				\cmidrule(lr){3-6} \cmidrule(lr){7-10} \cmidrule(lr){12-15} \cmidrule(lr){16-19}
				Method &  SD & Na\"ive & REML & DL & DB & Na\"ive & REML & DL & DB  & SD & Na\"ive & REML & DL & DB & Naive & REML & DL & DB \\ 
				\midrule
				&&&&& \multicolumn{9}{c}{\textit{Random-effects variances}: $\sigma^2_{b0}$=0, $\sigma^2_{b1}$=0, $\sigma^2_{b2}$=0} &&&&&\\
                \multicolumn{19}{l}{\textbf{Unadjusted}}\\
				Na\"ive & 0.071 & 0.070 &  &  &  & 94.9 &  &  &  & 0.100 & 0.099 &  &  &  & 94.1 &  &  &  \\ 
				Fixed & 0.075 &  & 0.073 & 0.073 & 0.073 &  & 94.3 & 94.3 & 94.4 & 0.106 &  & 0.103 & 0.103 & 0.103 &  & 94.8 & 94.8 & 94.7 \\ 
				Mixed(1$\mid$c) & 0.074 &  & 0.074 & 0.074 & 0.074 &  & 94.8 & 94.9 & 95.1 & 0.105 &  & 0.104 & 0.104 & 0.104 &  & 95.3 & 95.1 & 95.2 \\ 
                Mixed(1$\mid$c) Sam. & 0.192 &  & 0.185 & 0.184 & 0.177 &  & 94.1 & 94.0 & 93.5 & 0.169 &  & 0.161 & 0.161 & 0.157 &  & 93.5 & 93.6 & 93.2 \\ 
				Mixed(1+A$\mid$c) & 0.074 &  & 0.073 & 0.073 & 0.073 &  & 94.8 & 94.7 & 94.9 & 0.105 &  & 0.104 & 0.104 & 0.104 &  & 95.2 & 95.1 & 95.1 \\ 
                Mixed(1+A$\mid$c) Sam. & 0.192 &  & 0.185 & 0.184 & 0.177 &  & 94.0 & 94.0 & 93.5 & 0.169 &  & 0.161 & 0.161 & 0.157 &  & 93.5 & 94.0 & 93.4 \\ 
               \multicolumn{19}{l}{\textbf{Adjusted}}\\
				Na\"ive   & 0.069 & 0.067 &  &  &  & 95.0 &  &  &  & 0.092 & 0.089 &  &  &  & 95.4 &  &  &  \\ 
				Fixed  & 0.072 &  & 0.070 & 0.070 & 0.070 &  & 95.1 & 95.1 & 95.1 & 0.097 &  & 0.093 & 0.093 & 0.093 &  & 94.4 & 94.5 & 94.5 \\ 
				Mixed(1$\mid$c)  & 0.071 &  & 0.070 & 0.070 & 0.070 &  & 94.8 & 94.9 & 94.9 & 0.096 &  & 0.093 & 0.093 & 0.093 &  & 95.4 & 95.4 & 95.3 \\ 
                Mixed(1$\mid$c) Sam. & 0.190 &  & 0.183 & 0.183 & 0.175 &  & 93.9 & 94.0 & 92.8 & 0.161 &  & 0.153 & 0.153 & 0.149 &  & 92.8 & 92.8 & 92.7 \\ 
				Mixed(1+A$\mid$c)  & 0.071 &  & 0.070 & 0.070 & 0.070 &  & 94.8 & 94.7 & 94.9 & 0.096 &  & 0.093 & 0.093 & 0.093 &  & 95.2 & 95.1 & 95.1 \\ 
                Mixed(1+A$\mid$c) Sam. & 0.190 &  & 0.184 & 0.183 & 0.176 &  & 93.9 & 94.0 & 92.8 & 0.161 &  & 0.154 & 0.154 & 0.150 &  & 93.0 & 92.8 & 92.7 \\ 
				\midrule
				&&&&& \multicolumn{9}{c}{\textit{Random-effects variances}: $\sigma^2_{b0}$=0.1, $\sigma^2_{b1}$=0, $\sigma^2_{b2}$=0} &&&&&\\
                \multicolumn{19}{l}{\textbf{Unadjusted}}\\
				Na\"ive & 0.126 & 0.073 &  &  &  & 75.5 &  &  &  & 0.105 & 0.103 &  &  &  & 94.2 &  &  &  \\ 
				Fixed & 0.122 &  & 0.121 & 0.121 & 0.117 &  & 93.7 & 93.8 & 93.2 & 0.108 &  & 0.103 & 0.103 & 0.103 &  & 94.6 & 94.6 & 94.6 \\ 
				Mixed(1$\mid$c) & 0.122 &  & 0.121 & 0.121 & 0.117 &  & 93.9 & 93.7 & 92.8 & 0.108 &  & 0.103 & 0.103 & 0.103 &  & 94.5 & 94.5 & 94.5 \\ 
                Mixed(1$\mid$c) Sam. & 0.190 &  & 0.182 & 0.182 & 0.174 &  & 93.5 & 93.6 & 93.2 & 0.168 &  & 0.161 & 0.161 & 0.156 &  & 93.0 & 93.3 & 92.5 \\ 
				Mixed(1$\mid$c) & 0.122 &  & 0.121 & 0.121 & 0.117 &  & 94.0 & 93.9 & 93.0 & 0.108 &  & 0.103 & 0.103 & 0.103 &  & 94.3 & 94.3 & 94.4 \\
                Mixed(1+A$\mid$c) Sam. & 0.190 &  & 0.182 & 0.182 & 0.174 &  & 93.5 & 93.6 & 93.1 & 0.168 &  & 0.161 & 0.161 & 0.157 &  & 93.1 & 93.4 & 92.6 \\ 
                \multicolumn{19}{l}{\textbf{Adjusted}}\\
				Na\"ive   & 0.124 & 0.069 &  &  &  & 73.8 &  &  &  & 0.096 & 0.093 &  &  &  & 94.8 &  &  &  \\ 
				Fixed & 0.120 &  & 0.119 & 0.119 & 0.115 &  & 93.9 & 94.0 & 93.2 & 0.096 &  & 0.093 & 0.093 & 0.093 &  & 94.7 & 94.7 & 94.7 \\ 
				Mixed(1$\mid$c)  & 0.120 &  & 0.119 & 0.119 & 0.115 &  & 94.2 & 94.1 & 93.5 & 0.096 &  & 0.093 & 0.093 & 0.093 &  & 95.2 & 95.2 & 95.2 \\
                Mixed(1$\mid$c) Sam. & 0.189 &  & 0.180 & 0.181 & 0.173 &  & 93.1 & 93.2 & 91.9 & 0.163 &  & 0.153 & 0.153 & 0.148 &  & 92.9 & 93.0 & 92.2\\ 
				Mixed(1+A$\mid$c)  & 0.120 &  & 0.119 & 0.119 & 0.115 &  & 94.0 & 93.9 & 93.0 & 0.097 &  & 0.093 & 0.093 & 0.093 &  & 94.3 & 94.3 & 94.4 \\
                Mixed(1+A$\mid$c) Sam. & 0.190 &  & 0.181 & 0.181 & 0.173 &  & 93.3 & 93.2 & 91.9 & 0.163 &  & 0.154 & 0.154 & 0.149 &  & 92.9 & 93.1 & 92.2 \\  
				\midrule
				&&&&& \multicolumn{9}{c}{\textit{Random-effects variances}: $\sigma^2_{b0}$=0.15, $\sigma^2_{b1}$=0.15, $\sigma^2_{b2}$=0} &&&&&\\
               \multicolumn{19}{l}{\textbf{Unadjusted}}\\
				Na\"ive & 0.207 & 0.077 &  &  &  & 54.6 &  &  &  & 0.172 & 0.107 &  &  &  & 76.7 &  &  &  \\ 
				Fixed & 0.198 &  & 0.183 & 0.183 & 0.175 &  & 92.8 & 93.0 & 92.0 & 0.165 &  & 0.155 & 0.155 & 0.150 &  & 92.7 & 92.7 & 92.0 \\ 
				Mixed(1$\mid$c) & 0.198 &  & 0.183 & 0.183 & 0.175 &  & 92.9 & 93.0 & 92.1 & 0.165 &  & 0.155 & 0.155 & 0.150 &  & 92.4 & 92.6 & 91.7 \\ 
                Mixed(1$\mid$c) Sam. & 0.188 &  & 0.184 & 0.184 & 0.176 &  & 95.3 & 95.2 & 94.2 & 0.164 &  & 0.161 & 0.161 & 0.157 &  & 94.1 & 94.1 & 93.7\\ 
				Mixed(1+A$\mid$c) & 0.198 &  & 0.183 & 0.183 & 0.175 &  & 93.3 & 93.0 & 92.0 & 0.165 &  & 0.156 & 0.157 & 0.152 &  & 92.6 & 92.6 & 91.9 \\
                Mixed(1+A$\mid$c) Sam. & 0.189 &  & 0.184 & 0.184 & 0.176 &  & 95.3 & 95.2 & 94.2 & 0.164 &  & 0.161 & 0.161 & 0.157 &  & 94.1 & 94.0 & 93.6 \\ 
                \multicolumn{19}{l}{\textbf{Adjusted}}\\
				Na\"ive  & 0.205 & 0.074 &  &  &  & 53.4 &  &  &  & 0.165 & 0.098 &  &  &  & 75.4 &  &  &  \\ 
				Fixed  & 0.195 &  & 0.181 & 0.181 & 0.173 &  & 93.8 & 93.9 & 92.1 & 0.155 &  & 0.147 & 0.148 & 0.142 &  & 92.0 & 91.9 & 91.1 \\ 
				Mixed(1$\mid$c)  & 0.196 &  & 0.181 & 0.181 & 0.173 &  & 93.6 & 93.7 & 92.1 & 0.156 &  & 0.147 & 0.148 & 0.142 &  & 92.0 & 92.1 & 91.0 \\
                Mixed(1$\mid$c) Sam. & 0.187 &  & 0.182 & 0.182 & 0.174 &  & 95.3 & 95.1 & 94.2 & 0.158 &  & 0.154 & 0.153 & 0.149 &  & 93.8 & 93.9 & 93.1 \\ 
				Mixed(1+A$\mid$c)  & 0.195 &  & 0.182 & 0.182 & 0.174 &  & 93.3 & 93.0 & 92.0 & 0.156 &  & 0.150 & 0.150 & 0.144 &  & 92.6 & 92.6 & 91.9 \\
                Mixed(1+A$\mid$c) Sam. & 0.187 &  & 0.182 & 0.182 & 0.175 &  & 95.3 & 95.0 & 94.3 & 0.158 &  & 0.154 & 0.154 & 0.150 &  & 94.0 & 94.0 & 93.4 \\ 
				\bottomrule		
				&&&&& \multicolumn{9}{c}{\textit{Random-effects variances}: $\sigma^2_{b0}$=0.15, $\sigma^2_{b1}$=0.15, $\sigma^2_{b2}=4\times10^{-6}$} &&&&&\\
                \multicolumn{19}{l}{\textbf{Unadjusted}}\\
				Na\"ive & 0.192 & 0.077 &  &  &  & 58.2 &  &  &  & 0.167 & 0.107 &  &  &  & 79.6 &  &  &  \\ 
				Fixed & 0.184 &  & 0.183 & 0.182 & 0.175 &  & 94.2 & 94.0 & 93.2 & 0.164 &  & 0.155 & 0.155 & 0.151 &  & 93.0 & 92.5 & 92.5 \\ 
				Mixed(1$\mid$c) & 0.184 &  & 0.183 & 0.182 & 0.175 &  & 94.3 & 94.1 & 93.1 & 0.164 &  & 0.156 & 0.155 & 0.151 &  & 92.9 & 92.5 & 92.1 \\ 
                Mixed(1$\mid$c) Sam. & 0.191 &  & 0.183 & 0.183 & 0.176 &  & 94.2 & 94.4 & 92.4 & 0.163 &  & 0.161 & 0.161 & 0.157 &  & 93.9 & 94.2 & 93.5 \\ 
				Mixed(1+A$\mid$c) & 0.184 &  & 0.183 & 0.183 & 0.175 &  & 94.4 & 94.3 & 92.9 & 0.164 &  & 0.157 & 0.157 & 0.152 &  & 92.8 & 92.5 & 92.3 \\ 
                 Mixed(1+A$\mid$c) Sam. & 0.191 &  & 0.183 & 0.184 & 0.176 &  & 94.2 & 94.4 & 92.4 & 0.163 &  & 0.161 & 0.161 & 0.157 &  & 93.9 & 94.2 & 93.5 \\ 
               \multicolumn{19}{l}{\textbf{Adjusted}}\\
				Na\"ive  & 0.190 & 0.074 &  &  &  & 56.8 &  &  &  & 0.161 & 0.098 &  &  &  & 76.9 &  &  &  \\ 
				Fixed  & 0.182 &  & 0.181 & 0.181 & 0.174 &  & 94.6 & 94.0 & 93.8 & 0.157 &  & 0.148 & 0.148 & 0.143 &  & 92.2 & 92.1 & 91.0 \\ 
				Mixed(1$\mid$c)  & 0.182 &  & 0.182 & 0.181 & 0.174 &  & 94.5 & 94.2 & 93.6 & 0.157 &  & 0.148 & 0.148 & 0.143 &  & 92.0 & 92.0 & 91.3 \\ 
                Mixed(1$\mid$c) Sam. & 0.190 &  & 0.182 & 0.182 & 0.174 &  & 93.7 & 93.9 & 92.6 & 0.156 &  & 0.153 & 0.153 & 0.149 &  & 93.3 & 93.4 & 92.7 \\ 
				Mixed(1+A$\mid$c)  & 0.182 &  & 0.182 & 0.182 & 0.174 &  & 94.4 & 94.3 & 92.9 & 0.157 &  & 0.150 & 0.150 & 0.145 &  & 92.8 & 92.5 & 92.3 \\ 
                Mixed(1+A$\mid$c) Sam. & 0.190 &  & 0.182 & 0.182 & 0.174 &  & 94.0 & 94.0 & 92.7 & 0.155 &  & 0.154 & 0.154 & 0.149 &  & 93.5 & 93.4 & 92.8 \\ 
				\bottomrule		
			\end{tabular}%
		}
	\end{center}
	\tiny
	\vspace{0.01cm} 
   \label{table:appe_k_10_nc_50_cont_ch2}
\end{table}

\clearpage
\begin{table}[!h]
	\caption{Simulation results: Monte Carlo standard deviation, average standard errors and coverage probabilities of 95\% confidence intervals for both counterfactual mean on treatment and ATE based on 1000 simulations. Setting: Continuous outcome, $k$=5, $n_c$(Avg=100, Min=50, Max=150), and different values of random-effects variance. Results are based on weighting centers equally.}
	\begin{center}
		\resizebox{\textwidth}{!}{
			\begin{tabular}{>{\raggedright\arraybackslash}p{3.5cm}cccccccccccccccccc}
				\toprule
				& \multicolumn{9}{c}{\textbf{Counterfactual mean on treatment}} &\multicolumn{9}{c}{\textbf{ATE}} \\
				\cmidrule(lr){2-10} \cmidrule(lr){11-19}
				&& \multicolumn{4}{c}{\textbf{SE}} & \multicolumn{4}{c}{\textbf{Coverage (\%)}} &&\multicolumn{4}{c}{\textbf{SE}}&\multicolumn{4}{c}{\textbf{Coverage (\%)}} \\
				\cmidrule(lr){3-6} \cmidrule(lr){7-10} \cmidrule(lr){12-15} \cmidrule(lr){16-19}
				Method &  SD & Na\"ive & REML & DL & DB & Na\"ive & REML & DL & DB  & SD & Na\"ive & REML & DL & DB & Naive & REML & DL & DB \\ 
				\midrule
				&&&&& \multicolumn{9}{c}{\textit{Random-effects variances}: $\sigma^2_{b0}$=0, $\sigma^2_{b1}$=0, $\sigma^2_{b2}$=0} &&&&&\\
                \multicolumn{19}{l}{\textbf{Unadjusted}}\\
				Na\"ive & 0.070 & 0.070 &  &  &  & 94.2 &  &  &  & 0.102 & 0.100 &  &  &  & 94.9 &  &  &  \\ 
				Fixed & 0.075 &  & 0.075 & 0.075 & 0.075 &  & 96.0 & 96.0 & 96.0 & 0.109 &  & 0.106 & 0.106 & 0.106 &  & 95.4 & 95.4 & 95.5 \\ 
				Mixed(1$\mid$c) & 0.075 &  & 0.075 & 0.075 & 0.075 &  & 96.0 & 96.0 & 95.9 & 0.109 &  & 0.106 & 0.106 & 0.106 &  & 95.8 & 95.8 & 95.8 \\ 
                Mixed(1$\mid$c) Sam. & 0.254 &  & 0.241 & 0.241 & 0.219 &  & 94.3 & 93.8 & 91.9 & 0.207 &  & 0.196 & 0.196 & 0.183 &  & 92.2 & 92.4 & 90.5\\
				Mixed(1+A$\mid$c) & 0.075 &  & 0.075 & 0.075 & 0.075 &  & 96.0 & 96.0 & 95.9 & 0.109 &  & 0.106 & 0.106 & 0.106 &  & 95.8 & 95.9 & 95.8 \\ 
                Mixed(1+A$\mid$c) Sam. & 0.254 &  & 0.241 & 0.241 & 0.219 &  & 94.3 & 93.8 & 91.9 & 0.207 &  & 0.197 & 0.196 & 0.184 &  & 92.3 & 92.4 & 90.6 \\ 
                \multicolumn{19}{l}{\textbf{Adjusted}}\\
				Na\"ive & 0.066 & 0.067 &  &  &  & 94.9 &  &  &  & 0.090 & 0.090 &  &  &  & 94.9 &  &  &  \\ 
				Fixed  & 0.071 &  & 0.071 & 0.071 & 0.071 &  & 96.1 & 96.0 & 95.9 & 0.096 &  & 0.095 & 0.095 & 0.095 &  & 95.4 & 95.3 & 95.4 \\ 
				Mixed(1$\mid$c)  & 0.071 &  & 0.071 & 0.071 & 0.071 &  & 96.0 & 95.9 & 95.8 & 0.096 &  & 0.095 & 0.095 & 0.095 &  & 95.4 & 95.5 & 95.6 \\
                Mixed(1$\mid$c) Sam. & 0.252 &  & 0.241 & 0.240 & 0.218 &  & 93.8 & 93.8 & 92.0 & 0.200 &  & 0.190 & 0.190 & 0.177 &  & 92.0 & 92.1 & 90.1 \\ 
				Mixed(1+A$\mid$c)  & 0.071 &  & 0.071 & 0.071 & 0.071 &  & 96.0 & 96.0 & 95.9 & 0.096 &  & 0.095 & 0.095 & 0.095 &  & 95.8 & 95.9 & 95.8 \\ 
                Mixed(1+A$\mid$c) Sam. & 0.252 &  & 0.241 & 0.241 & 0.219 &  & 93.9 & 93.8 & 92.0 & 0.200 &  & 0.192 & 0.191 & 0.178 &  & 92.1 & 92.1 & 90.3 \\ 
				\midrule
				&&&&& \multicolumn{9}{c}{\textit{Random-effects variances}: $\sigma^2_{b0}$=0.1, $\sigma^2_{b1}$=0, $\sigma^2_{b2}$=0} &&&&&\\
                \multicolumn{19}{l}{\textbf{Unadjusted}}\\
				Na\"ive  & 0.161 & 0.072 &  &  &  & 62.0 &  &  &  & 0.101 & 0.102 &  &  &  & 94.4 &  &  &  \\ 
				Fixed & 0.159 &  & 0.150 & 0.150 & 0.138 &  & 92.0 & 91.6 & 90.5 & 0.106 &  & 0.106 & 0.106 & 0.106 &  & 94.5 & 94.5 & 94.6 \\ 
				Mixed(1$\mid$c) & 0.159 &  & 0.150 & 0.150 & 0.138 &  & 92.2 & 91.7 & 90.4 & 0.106 &  & 0.106 & 0.106 & 0.106 &  & 94.4 & 94.3 & 94.4 \\ 
                Mixed(1$\mid$c) Sam. & 0.260 &  & 0.243 & 0.243 & 0.220 &  & 93.7 & 93.4 & 91.4 & 0.201 &  & 0.196 & 0.197 & 0.183 &  & 94.1 & 93.8 & 93.1 \\ 
				Mixed(1+A$\mid$c) & 0.159 &  & 0.150 & 0.150 & 0.138 &  & 92.1 & 91.6 & 90.1 & 0.106 &  & 0.106 & 0.106 & 0.106 &  & 94.2 & 94.3 & 94.5 \\ 
                Mixed(1+A$\mid$c) Sam. & 0.260 &  & 0.243 & 0.243 & 0.221 &  & 93.7 & 93.4 & 91.5 & 0.201 &  & 0.196 & 0.197 & 0.183 &  & 94.1 & 93.8 & 93.3 \\ 
                \multicolumn{19}{l}{\textbf{Adjusted}}\\
				Na\"ive  & 0.158 & 0.069 &  &  &  & 61.4 &  &  &  & 0.090 & 0.093 &  &  &  & 95.3 &  &  &  \\ 
				Fixed  & 0.156 &  & 0.148 & 0.149 & 0.137 &  & 91.6 & 91.9 & 89.8 & 0.094 &  & 0.095 & 0.095 & 0.095 &  & 95.6 & 95.5 & 95.3 \\ 
				Mixed(1$\mid$c)  & 0.156 &  & 0.148 & 0.148 & 0.137 &  & 91.7 & 92.0 & 90.1 & 0.094 &  & 0.095 & 0.095 & 0.095 &  & 95.4 & 95.4 & 95.4 \\ 
                Mixed(1$\mid$c) Sam. & 0.260 &  & 0.242 & 0.242 & 0.219 &  & 93.2 & 93.1 & 91.2 & 0.198 &  & 0.190 & 0.190 & 0.176 &  & 93.7 & 93.7 & 91.9 \\ 
				Mixed(1+A$\mid$c)  & 0.156 &  & 0.148 & 0.149 & 0.137 &  & 92.1 & 91.6 & 90.1 & 0.094 &  & 0.095 & 0.095 & 0.095 &  & 94.2 & 94.3 & 94.5 \\
                Mixed(1+A$\mid$c) Sam. & 0.260 &  & 0.242 & 0.242 & 0.220 &  & 93.2 & 93.1 & 91.1 & 0.198 &  & 0.191 & 0.191 & 0.178 &  & 93.8 & 93.7 & 92.1 \\ 
				\midrule
				&&&&& \multicolumn{9}{c}{\textit{Random-effects variances}: $\sigma^2_{b0}$=0.15, $\sigma^2_{b1}$=0.15, $\sigma^2_{b2}$=0} &&&&&\\
                \multicolumn{19}{l}{\textbf{Unadjusted}}\\
				Na\"ive & 0.264 & 0.076 &  &  &  & 43.5 &  &  &  & 0.211 & 0.106 &  &  &  & 67.4 &  &  &  \\ 
				Fixed & 0.258 &  & 0.240 & 0.240 & 0.218 &  & 93.1 & 93.0 & 91.9 & 0.206 &  & 0.194 & 0.194 & 0.180 &  & 92.3 & 92.4 & 91.0 \\ 
				Mixed(1$\mid$c) & 0.258 &  & 0.240 & 0.240 & 0.218 &  & 93.2 & 93.1 & 91.8 & 0.206 &  & 0.194 & 0.194 & 0.180 &  & 92.3 & 92.3 & 91.4 \\
                Mixed(1$\mid$c) Sam. & 0.263 &  & 0.243 & 0.243 & 0.220 &  & 93.7 & 93.3 & 91.6 & 0.214 &  & 0.196 & 0.196 & 0.183 &  & 92.3 & 93.0 & 92.0 \\ 
				Mixed(1+A$\mid$c) & 0.258 &  & 0.241 & 0.240 & 0.218 &  & 93.1 & 93.2 & 92.0 & 0.206 &  & 0.195 & 0.195 & 0.181 &  & 92.3 & 92.2 & 91.4 \\ 
                Mixed(1+A$\mid$c) Sam. & 0.263 &  & 0.243 & 0.244 & 0.220 &  & 93.7 & 93.3 & 91.8 & 0.214 &  & 0.196 & 0.196 & 0.184 &  & 92.3 & 93.0 & 92.0 \\ 
                \multicolumn{19}{l}{\textbf{Adjusted}}\\
				Na\"ive   & 0.263 & 0.073 &  &  &  & 42.4 &  &  &  & 0.207 & 0.097 &  &  &  & 64.3 &  &  &  \\ 
				Fixed  & 0.256 &  & 0.239 & 0.238 & 0.216 &  & 93.4 & 93.2 & 92.2 & 0.200 &  & 0.187 & 0.187 & 0.173 &  & 92.8 & 92.8 & 91.4 \\ 
				Mixed(1$\mid$c)  & 0.256 &  & 0.239 & 0.238 & 0.216 &  & 93.4 & 93.1 & 91.9 & 0.200 &  & 0.187 & 0.187 & 0.173 &  & 92.9 & 92.9 & 91.4 \\ 
                Mixed(1$\mid$c) Sam. & 0.261 &  & 0.242 & 0.242 & 0.219 &  & 93.9 & 93.7 & 91.1 & 0.208 &  & 0.189 & 0.189 & 0.176 &  & 92.0 & 91.9 & 89.9 \\ 
				Mixed(1+A$\mid$c)  & 0.256 &  & 0.239 & 0.239 & 0.217 &  & 93.1 & 93.2 & 92.0 & 0.200 &  & 0.189 & 0.190 & 0.175 &  & 92.3 & 92.2 & 91.4 \\ 
                Mixed(1+A$\mid$c) Sam. & 0.261 &  & 0.242 & 0.243 & 0.219 &  & 93.9 & 93.7 & 91.1 & 0.208 &  & 0.190 & 0.190 & 0.177 &  & 92.0 & 92.2 & 89.9 \\ 
				\bottomrule		
				&&&&& \multicolumn{9}{c}{\textit{Random-effects variances}: $\sigma^2_{b0}$=0.15, $\sigma^2_{b1}$=0.15, $\sigma^2_{b2}=4\times10^{-6}$} &&&&&\\
                \multicolumn{19}{l}{\textbf{Unadjusted}}\\
				Na\"ive & 0.256 & 0.077 &  &  &  & 45.3 &  &  &  & 0.205 & 0.106 &  &  &  & 70.5 &  &  &  \\ 
				Fixed & 0.248 &  & 0.241 & 0.241 & 0.219 &  & 93.8 & 93.9 & 92.5 & 0.197 &  & 0.190 & 0.190 & 0.177 &  & 92.5 & 92.4 & 91.1 \\ 
				Mixed(1$\mid$c) & 0.248 &  & 0.241 & 0.241 & 0.219 &  & 93.8 & 93.8 & 92.5 & 0.197 &  & 0.190 & 0.190 & 0.177 &  & 92.5 & 92.4 & 91.1 \\ 
                Mixed(1$\mid$c) Sam. & 0.252 &  & 0.240 & 0.239 & 0.218 &  & 93.5 & 93.0 & 92.5 & 0.208 &  & 0.192 & 0.192 & 0.180 &  & 91.5 & 91.6 & 90.7 \\ 
				Mixed(1+A$\mid$c) & 0.247 &  & 0.241 & 0.241 & 0.219 &  & 93.9 & 93.7 & 92.5 & 0.197 &  & 0.191 & 0.191 & 0.178 &  & 92.7 & 92.6 & 91.1 \\ 
                Mixed(1+A$\mid$c) Sam. & 0.252 &  & 0.240 & 0.239 & 0.218 &  & 93.5 & 92.9 & 92.4 & 0.208 &  & 0.193 & 0.192 & 0.180 &  & 91.5 & 91.6 & 90.7 \\ 
                \multicolumn{19}{l}{\textbf{Adjusted}}\\
				Na\"ive  & 0.256 & 0.073 &  &  &  & 43.9 &  &  &  & 0.203 & 0.097 &  &  &  & 66.8 &  &  &  \\ 
				Fixed  & 0.247 &  & 0.239 & 0.239 & 0.217 &  & 93.6 & 93.4 & 92.8 & 0.193 &  & 0.183 & 0.183 & 0.170 &  & 92.0 & 92.0 & 91.3 \\ 
				Mixed(1$\mid$c)  & 0.247 &  & 0.239 & 0.239 & 0.217 &  & 93.7 & 93.5 & 92.8 & 0.193 &  & 0.183 & 0.183 & 0.170 &  & 92.1 & 92.1 & 91.3 \\
                Mixed(1$\mid$c) Sam. & 0.251 &  & 0.239 & 0.238 & 0.217 &  & 94.2 & 94.0 & 92.8 & 0.203 &  & 0.186 & 0.186 & 0.173 &  & 91.5 & 91.3 & 89.9 \\ 
				Mixed(1+A$\mid$c)  & 0.247 &  & 0.240 & 0.240 & 0.218 &  & 93.9 & 93.7 & 92.5 & 0.193 &  & 0.185 & 0.185 & 0.172 &  & 92.7 & 92.6 & 91.1 \\ 
                Mixed(1+A$\mid$c) Sam. & 0.251 &  & 0.240 & 0.239 & 0.218 &  & 94.2 & 93.9 & 92.8 & 0.204 &  & 0.188 & 0.187 & 0.175 &  & 91.6 & 91.3 & 90.0 \\ 
				\bottomrule					
			\end{tabular}%
		}
	\end{center}
	\tiny
	\vspace{0.01cm} 
   \label{table:appe_k_5_nc_100_cont_ch2}
\end{table}

\clearpage
\autoref{table:appe_k_100_nc_5_mis_cont_ch2} reports the Monte Carlo standard deviation, average standard errors, and 95\% confidence interval coverage for the counterfactual mean under treatment and the ATE for continuous outcomes under a misspecified outcome model. The results show that the proposed estimators perform well to this misspecification---specifically, when the fitted model includes only main effects of the covariates even though the true outcome model contains a nonlinear covariate transformation and an interaction term.

\begin{table}[!h]
\caption{Simulation results: Monte Carlo standard deviation, average standard errors and coverage probabilities of 95\% confidence intervals for both counterfactual mean on treatment and ATE based on 1000 simulations. Setting: Continuous outcomes with a misspecified outcome model, $k$=100, $n_c$: Avg=5, Min=1, Max=24. Results are based on weighting centers equally.}
	\begin{center}
		\resizebox{\textwidth}{!}{
			\begin{tabular}{>{\raggedright\arraybackslash}p{3.5cm}cccccccccccccccccc}
				\toprule
				& \multicolumn{9}{c}{\textbf{Counterfactual mean on treatment}} &\multicolumn{9}{c}{\textbf{ATE}} \\
				\cmidrule(lr){2-10} \cmidrule(lr){11-19}
				&& \multicolumn{4}{c}{\textbf{SE}} & \multicolumn{4}{c}{\textbf{Coverage (\%)}} &&\multicolumn{4}{c}{\textbf{SE}}&\multicolumn{4}{c}{\textbf{Coverage (\%)}} \\
				\cmidrule(lr){3-6} \cmidrule(lr){7-10} \cmidrule(lr){12-15} \cmidrule(lr){16-19}
				Method &  SD & Na\"ive & REML & DL & DB & Na\"ive & REML & DL & DB  & SD & Na\"ive & REML & DL & DB & Naive & REML & DL & DB \\ \midrule
				&&&&& \multicolumn{9}{c}{\textit{Random-effects variances}: $\sigma^2_{b0}$=0.15, $\sigma^2_{b1}$=0.15, $\sigma^2_{b2}=0$: $\bar{\rho}_1=0.0875$, $\bar{\rho}=0.0273$} &&&&&\\
                \multicolumn{19}{l}{\textbf{Unadjusted}}\\
				Na\"ive & 0.097 & 0.075 &  &  &  & 86.7 &  &  &  & 0.114 & 0.103 &  &  &  & 92.7 &  &  &  \\ 
                Fixed & 0.100 &  & 0.098 & 0.097 & 0.093 &  & 94.7 & 94.5 & 93.3 & 0.120 &  & 0.111 & 0.109 & 0.106 &  & 93.5 & 93.1 & 92.2 \\ 
                Mixed(1$\mid$c) & 0.101 &  & 0.102 & 0.102 & 0.101 &  & 95.3 & 95.3 & 95.0 & 0.124 &  & 0.119 & 0.119 & 0.119 &  & 94.7 & 94.5 & 94.5\\
                Mixed(1$\mid$c) Sam. & 0.105 &  & 0.102 & 0.102 & 0.102 &  & 94.4 & 94.7 & 94.4 & 0.127 &  & 0.131 & 0.131 & 0.133 &  & 96.2 & 96.5 & 96.5 \\ 
                Mixed(1+A$\mid$c) & 0.101 &  & 0.100 & 0.100 & 0.100 &  & 95.0 & 95.0 & 94.6 & 0.125 &  & 0.121 & 0.121 & 0.121 &  & 94.7 & 94.7 & 94.8 \\ 
                Mixed(1+A$\mid$c) Sam. & 0.105 &  & 0.102 & 0.102 & 0.102 &  & 94.5 & 94.8 & 94.6 & 0.128 &  & 0.131 & 0.131 & 0.133 &  & 96.2 & 96.5 & 96.4 \\ 
                \multicolumn{19}{l}{\textbf{Adjusted}}\\
               Na\"ive  & 0.097 & 0.073 &  &  &  & 85.9 &  &  &  & 0.110 & 0.098 &  &  &  & 91.9 &  &  &  \\ 
               Fixed  & 0.098 &  & 0.097 & 0.096 & 0.092 &  & 94.2 & 94.2 & 92.5 & 0.115 &  & 0.105 & 0.103 & 0.100 &  & 92.5 & 92.0 & 91.2 \\ 
               Mixed(1$\mid$c)  & 0.101 &  & 0.100 & 0.100 & 0.100 &  & 94.7 & 94.9 & 94.4 & 0.120 &  & 0.113 & 0.113 & 0.113 &  & 94.4 & 94.0 & 94.6 \\ 
               Mixed(1$\mid$c) Sam. & 0.103 &  & 0.101 & 0.101 & 0.101 &  & 95.1 & 95.2 & 95.3 & 0.121 &  & 0.125 & 0.125 & 0.127 &  & 95.9 & 96.0 & 95.9 \\ 
               Mixed(1+A$\mid$c)  & 0.100 &  & 0.099 & 0.099 & 0.098 &  & 95.0 & 95.0 & 94.6 & 0.121 &  & 0.115 & 0.115 & 0.115 &  & 94.7 & 94.7 & 94.8 \\ 
            Mixed(1+A$\mid$c) Sam. & 0.103 &  & 0.101 & 0.101 & 0.101 &  & 95.1 & 95.2 & 95.2 & 0.121 &  & 0.125 & 0.125 & 0.127 &  & 96.2 & 96.0 & 96.0 \\
				\midrule												
			\end{tabular}%
		}
	\end{center}
	\tiny
	\vspace{0.01cm} 
  \label{table:appe_k_100_nc_5_mis_cont_ch2}
\end{table}

\autoref{table:appe_k_100_nc_5_bias_MSE_cont_ch2}, \autoref{table:appe_k_50_nc_10_bias_MSE_cont_ch2}, \autoref{table:appe_k_10_nc_50_bias_MSE_cont_ch2}, \autoref{table:appe_k_5_nc_100_bias_MSE_cont_ch2} and \autoref{table:appe_k_100_nc_100_bias_MSE_cont_ch2} show the estimated bias and mean square error for both counterfactual mean on treatment and ATE for simulation Settings 1 through 5, respectively. The results show that all the considered estimators provide unbiased effect estimates.

\clearpage
\begin{table}[!h]
	\caption{Simulation results of estimated bias and mean square error (MSE) for both counterfactual mean on treatment and ATE based on 1000 simulations. Setting: Continuous outcome, $k$=100, $n_c$(Avg=5, Min=1, Max=24), and different values of random-effects variance. Results are based on weighting centers equally.}
	\begin{center}
		\resizebox{\textwidth}{!}{
			\begin{tabular}{ccccccccccccc}
				\toprule
				& \multicolumn{12}{c}{\textbf{Counterfactual mean on treatment}}\\
				\cmidrule(lr){2-13} 
                & \multicolumn{6}{c}{\textbf{Unadjusted}} &\multicolumn{6}{c}{\textbf{Adjusted}}\\
				\cmidrule(lr){2-7} \cmidrule(lr){8-13} 
				Measure &  Na\"ive & Fixed & Mixed(1$\mid$c) & Mixed(1$\mid$c) Sam. & Mixed(1+A$\mid$c) & Mixed(1+A$\mid$c) Sam. & Na\"ive  & Fixed  & Mixed(1$\mid$c) & Mixed(1$\mid$c) Sam. & Mixed(1+A$\mid$c) & Mixed(1+A$\mid$c) Sam. \\
				\midrule
				\multicolumn{12}{c}{\textit{Random-effects variances}: $\sigma^2_{b0}$=0, $\sigma^2_{b1}$=0, $\sigma^2_{b2}$=0}\\
                Bias & 0.0023 & 0.0010 & 0.0031 & -0.0047 & 0.0031 & -0.0047 & 0.0017 & 0.0006 & 0.0026 & -0.0032 & 0.0026 & -0.0032 \\ 
                MSE  & 0.0048 & 0.0070 & 0.0070 & 0.0115 & 0.0076 & 0.0115 & 0.0042 & 0.0062 & 0.0062 & 0.0108 & 0.0067 & 0.0109 \\ 
				\midrule
				\multicolumn{12}{c}{\textit{Random-effects variances}: $\sigma^2_{b0}$=0.15, $\sigma^2_{b1}$=0.15, $\sigma^2_{b2}$=0}\\
                Bias & 0.0012 & -0.0005 & -0.0005 & 0.0025 & -0.0007 & 0.0025 & 0.0005 & -0.0004 & -0.0002 & 0.0035 & -0.0005 & 0.0035 \\ 
                MSE  & 0.0098 & 0.0102  & 0.0102  & 0.0114 & 0.0108  & 0.0114 & 0.0094 & 0.0098  & 0.0098  & 0.0107 & 0.0101  & 0.0107 \\ 
				\midrule
				\multicolumn{12}{c}{\textit{Random-effects variances}: $\sigma^2_{b0}$=0.15, $\sigma^2_{b1}$=0.15, $\sigma^2_{b2}=4\times10^{-6}$} \\
                Bias & -0.0021 & -0.0020 & -0.0012 & 0.0008 & -0.0012 & 0.0008 & -0.0021 & -0.0022 & -0.0013 & -0.0000 & -0.0013 & -0.0000 \\ 
               MSE  & 0.0182 & 0.0159 & 0.0159 & 0.0119 & 0.0162 & 0.0119 & 0.0179 & 0.0152 & 0.0152 & 0.0113 & 0.0156 & 0.0113 \\ 
				\midrule		
                & \multicolumn{12}{c}{\textbf{ATE}}\\
				\cmidrule(lr){2-13} 
                & \multicolumn{6}{c}{\textbf{Unadjusted}} &\multicolumn{6}{c}{\textbf{Adjusted}}\\
				\cmidrule(lr){2-7} \cmidrule(lr){8-13} 
				Measure &  Na\"ive & Fixed & Mixed(1$\mid$c) & Mixed(1$\mid$c) Sam. & Mixed(1+A$\mid$c) & Mixed(1+A$\mid$c) Sam. & Na\"ive  & Fixed  & Mixed(1$\mid$c) & Mixed(1$\mid$c) Sam. & Mixed(1+A$\mid$c) & Mixed(1+A$\mid$c) Sam. \\		
              \midrule
				\multicolumn{12}{c}{\textit{Random-effects variances}: $\sigma^2_{b0}$=0, $\sigma^2_{b1}$=0, $\sigma^2_{b2}$=0}\\
                Bias & -0.0012 & -0.0026 & 0.0017 & -0.0060 & 0.0017 & -0.0060 & -0.0022 & -0.0033 & 0.0007 & -0.0023 & 0.0007 & -0.0024 \\ 
                MSE  & 0.0095  & 0.0129  & 0.0129 & 0.0190  & 0.0147 & 0.0190 & 0.0075  & 0.0105  & 0.0105 & 0.0163  & 0.0120 & 0.0163 \\ 
				\midrule
			   \multicolumn{12}{c}{\textit{Random-effects variances}: $\sigma^2_{b0}$=0.15, $\sigma^2_{b1}$=0.15, $\sigma^2_{b2}$=0}\\
                Bias & -0.0004 & -0.0032 & -0.0030 & 0.0028 & -0.0031 & 0.0029 & -0.0015 & -0.0031 & -0.0025 & 0.0047 & -0.0026 & 0.0047 \\ 
              MSE  & 0.0135 & 0.0155 & 0.0155 & 0.0194 & 0.0169 & 0.0194 & 0.0118 & 0.0133 & 0.0133 & 0.0166 & 0.0141 & 0.0166 \\ 
				\midrule
				\multicolumn{12}{c}{\textit{Random-effects variances}: $\sigma^2_{b0}$=0.15, $\sigma^2_{b1}$=0.15, $\sigma^2_{b2}=4\times10^{-6}$}  \\
                Bias & 0.0003 & 0.0011 & 0.0029 & 0.0028 & 0.0030 & 0.0028 & 0.0004 & 0.0009 & 0.0028 & 0.0019 & 0.0029 & 0.0019 \\ 
                MSE  & 0.0187 & 0.0169 & 0.0169 & 0.0193 & 0.0187 & 0.0193 & 0.0170 & 0.0141 & 0.0141 & 0.0161 & 0.0159 & 0.0161 \\ 
				\midrule
			\end{tabular}
		}
	\end{center}
    	\tiny
	\vspace{0.01cm} 
   \label{table:appe_k_100_nc_5_bias_MSE_cont_ch2}
\end{table}

\clearpage
\begin{table}[!h]
	\caption{Simulation results of estimated bias and mean square error (MSE) for both counterfactual mean on treatment and ATE based on 1000 simulations. Setting: Continuous outcome, $k$=50, $n_c$(Avg=10, Min=2, Max=48), and different values of random-effects variance. Results are based on weighting centers equally.}
	\begin{center}
		\resizebox{\textwidth}{!}{
			\begin{tabular}{ccccccccccccc}
				\toprule
				& \multicolumn{12}{c}{\textbf{Counterfactual mean on treatment}}\\
				\cmidrule(lr){2-13} 
                & \multicolumn{6}{c}{\textbf{Unadjusted}} &\multicolumn{6}{c}{\textbf{Adjusted}}\\
				\cmidrule(lr){2-7} \cmidrule(lr){8-13} 
				Measure &  Na\"ive & Fixed & Mixed(1$\mid$c) & Mixed(1$\mid$c) Sam. & Mixed(1+A$\mid$c) & Mixed(1+A$\mid$c) Sam. & Na\"ive  & Fixed  & Mixed(1$\mid$c) & Mixed(1$\mid$c) Sam. & Mixed(1+A$\mid$c) & Mixed(1+A$\mid$c) Sam. \\
				\midrule
				\multicolumn{12}{c}{\textit{Random-effects variances}: $\sigma^2_{b0}$=0, $\sigma^2_{b1}$=0, $\sigma^2_{b2}$=0}\\
                Bias & -0.0026 & -0.0019 & -0.0029 & -0.0041 & -0.0027 & -0.0042 & -0.0016 & -0.0014 & -0.0018 & -0.0040 & -0.0018 & -0.0040 \\ 
                 MSE  & 0.0046 & 0.0062 & 0.0062 & 0.0123 & 0.0063 & 0.0123 & 0.0043 & 0.0057 & 0.0057 & 0.0117 & 0.0057 & 0.0117 \\ 
				\midrule
				\multicolumn{12}{c}{\textit{Random-effects variances}: $\sigma^2_{b0}$=0.15, $\sigma^2_{b1}$=0.15, $\sigma^2_{b2}$=0}\\
                Bias & -0.0019 & -0.0033 & -0.0048 & -0.0045 & -0.0049 & -0.0045 & -0.0028 & -0.0034 & -0.0047 & -0.0048 & -0.0046 & -0.0048 \\ 
                MSE  & 0.0145 & 0.0134 & 0.0134 & 0.0130 & 0.0137 & 0.0130 & 0.0139 & 0.0128 & 0.0128 & 0.0122 & 0.0131 & 0.0121 \\ 
				\midrule
				\multicolumn{12}{c}{\textit{Random-effects variances}: $\sigma^2_{b0}$=0.15, $\sigma^2_{b1}$=0.15, $\sigma^2_{b2}=4\times10^{-6}$} \\
                Bias & 0.0031 & 0.0048 & 0.0056 & 0.0012 & 0.0057 & 0.0012 & 0.0043 & 0.0052 & 0.0059 & 0.0018 & 0.0059 & 0.0018 \\ 
                MSE  & 0.0330 & 0.0259 & 0.0259 & 0.0131 & 0.0261 & 0.0131 & 0.0326 & 0.0253 & 0.0253 & 0.0123 & 0.0254 & 0.0124 \\ 
				\midrule	
                & \multicolumn{12}{c}{\textbf{Counterfactual mean on treatment}}\\
				\cmidrule(lr){2-13} 
                & \multicolumn{6}{c}{\textbf{Unadjusted}} &\multicolumn{6}{c}{\textbf{Adjusted}}\\
				\cmidrule(lr){2-7} \cmidrule(lr){8-13} 
				Measure &  Na\"ive & Fixed & Mixed(1$\mid$c) & Mixed(1$\mid$c) Sam. & Mixed(1+A$\mid$c) & Mixed(1+A$\mid$c) Sam. & Na\"ive  & Fixed  & Mixed(1$\mid$c) & Mixed(1$\mid$c) Sam. & Mixed(1+A$\mid$c) & Mixed(1+A$\mid$c) Sam. \\
				\midrule
				\multicolumn{12}{c}{\textit{Random-effects variances}: $\sigma^2_{b0}$=0, $\sigma^2_{b1}$=0, $\sigma^2_{b2}$=0}\\
                Bias & -0.0008 & -0.0023 & -0.0043 & -0.0013 & -0.0043 & -0.0014 & 0.0013 & -0.0011 & -0.0018 & -0.0012 & -0.0019 & -0.0013 \\ 
                 MSE  & 0.0095 & 0.0125 & 0.0125 & 0.0180 & 0.0125 & 0.0180 & 0.0078 & 0.0103 & 0.0103 & 0.0153 & 0.0102 & 0.0154 \\ 
				\midrule
				\multicolumn{12}{c}{\textit{Random-effects variances}: $\sigma^2_{b0}$=0.15, $\sigma^2_{b1}$=0.15, $\sigma^2_{b2}$=0} \\
                Bias & 0.0025 & -0.0010 & -0.0040 & -0.0018 & -0.0040 & -0.0019 & 0.0008 & -0.0017 & -0.0042 & -0.0021 & -0.0041 & -0.0021 \\ 
                MSE  & 0.0159 & 0.0164 & 0.0164 & 0.0172 & 0.0168 & 0.0172 & 0.0137 & 0.0138 & 0.0138 & 0.0145 & 0.0144 & 0.0145 \\ 
				\midrule
				\multicolumn{12}{c}{\textit{Random-effects variances}: $\sigma^2_{b0}$=0.15, $\sigma^2_{b1}$=0.15, $\sigma^2_{b2}=4\times10^{-6}$}  \\
                Bias & 0.0015 & 0.0018 & 0.0037 & -0.0058 & 0.0040 & -0.0058 & 0.0036 & 0.0023 & 0.0038 & -0.0045 & 0.0041 & -0.0044 \\ 
                 MSE  & 0.0198 & 0.0182 & 0.0182 & 0.0179 & 0.0184 & 0.0180 & 0.0182 & 0.0154 & 0.0154 & 0.0152 & 0.0155 & 0.0153 \\ 
				\midrule			
			\end{tabular}
		}
	\end{center}
    	\tiny
	\vspace{0.01cm} 
       \label{table:appe_k_50_nc_10_bias_MSE_cont_ch2}
\end{table}

\clearpage
\begin{table}[!h]
	\caption{Simulation results of estimated bias and mean square error (MSE) for both counterfactual mean on treatment and ATE based on 1000 simulations. Setting: Continuous outcome, $k$=10, $n_c$(Avg=50, Min=25, Max=80), and different values of random-effects variance. Results are based on weighting centers equally.}
	\begin{center}
		\resizebox{\textwidth}{!}{
			\begin{tabular}{ccccccccccccc}
				\toprule
				& \multicolumn{12}{c}{\textbf{Counterfactual mean on treatment}}\\
				\cmidrule(lr){2-13} 
                & \multicolumn{6}{c}{\textbf{Unadjusted}} &\multicolumn{6}{c}{\textbf{Adjusted}}\\
				\cmidrule(lr){2-7} \cmidrule(lr){8-13} 
				Measure &  Na\"ive & Fixed & Mixed(1$\mid$c) & Mixed(1$\mid$c) Sam. & Mixed(1+A$\mid$c) & Mixed(1+A$\mid$c) Sam. & Na\"ive  & Fixed  & Mixed(1$\mid$c) & Mixed(1$\mid$c) Sam. & Mixed(1+A$\mid$c) & Mixed(1+A$\mid$c) Sam. \\
				\midrule
				\multicolumn{12}{c}{\textit{Random-effects variances}: $\sigma^2_{b0}$=0, $\sigma^2_{b1}$=0, $\sigma^2_{b2}$=0} \\
                Bias & -0.0061 & -0.0067 & -0.0065 & -0.0002 & -0.0064 & -0.0002 & -0.0052 & -0.0060 & -0.0058 & -0.0010 & -0.0058 & -0.0010 \\ 
                MSE  & 0.0051 & 0.0056 & 0.0056 & 0.0369 & 0.0056 & 0.0369 & 0.0047 & 0.0052 & 0.0052 & 0.0361 & 0.0051 & 0.0361 \\ 
				\midrule
				\multicolumn{12}{c}{\textit{Random-effects variances}: $\sigma^2_{b0}$=0.15, $\sigma^2_{b1}$=0.15, $\sigma^2_{b2}$=0}\\
                Bias & 0.0031 & 0.0015 & 0.0014 & -0.0060 & 0.0016 & -0.0059 & 0.0034 & 0.0017 & 0.0016 & -0.0057 & 0.0018 & -0.0057 \\ 
                MSE  & 0.0429 & 0.0392 & 0.0392 & 0.0355 & 0.0392 & 0.0355 & 0.0420 & 0.0381 & 0.0381 & 0.0351 & 0.0381 & 0.0351 \\ 
				\midrule
				\multicolumn{12}{c}{\textit{Random-effects variances}: $\sigma^2_{b0}$=0.15, $\sigma^2_{b1}$=0.15, $\sigma^2_{b2}=4\times10^{-6}$} \\
                Bias & 0.0091 & 0.0080 & 0.0078 & -0.0016 & 0.0076 & -0.0017 & 0.0101 & 0.0089 & 0.0088 & -0.0018 & 0.0086 & -0.0018 \\ 
                MSE  & 0.0370 & 0.0340 & 0.0340 & 0.0365 & 0.0339 & 0.0365 & 0.0362 & 0.0332 & 0.0332 & 0.0359 & 0.0332 & 0.0359 \\ 
				\midrule
                & \multicolumn{12}{c}{\textbf{ATE}}\\
				\cmidrule(lr){2-13} 
                & \multicolumn{6}{c}{\textbf{Unadjusted}} &\multicolumn{6}{c}{\textbf{Adjusted}}\\
				\cmidrule(lr){2-7} \cmidrule(lr){8-13} 
				Measure &  Na\"ive & Fixed & Mixed(1$\mid$c) & Mixed(1$\mid$c) Sam. & Mixed(1+A$\mid$c) & Mixed(1+A$\mid$c) Sam. & Na\"ive  & Fixed  & Mixed(1$\mid$c) & Mixed(1$\mid$c) Sam. & Mixed(1+A$\mid$c) & Mixed(1+A$\mid$c) Sam. \\
				\midrule
				\multicolumn{12}{c}{\textit{Random-effects variances}: $\sigma^2_{b0}$=0, $\sigma^2_{b1}$=0, $\sigma^2_{b2}$=0}\\
                Bias & -0.0045 & -0.0059 & -0.0054 & -0.0005 & -0.0054 & -0.0005 & -0.0032 & -0.0048 & -0.0044 & -0.0014 & -0.0044 & -0.0015 \\ 
                MSE  & 0.0101 & 0.0113 & 0.0113 & 0.0286 & 0.0111 & 0.0286 & 0.0085 & 0.0094 & 0.0094 & 0.0258 & 0.0093 & 0.0258 \\ 
				\midrule
				\multicolumn{12}{c}{\textit{Random-effects variances}: $\sigma^2_{b0}$=0.15, $\sigma^2_{b1}$=0.15, $\sigma^2_{b2}$=0}\\
                Bias & -0.0036 & -0.0048 & -0.0049 & -0.0055 & -0.0047 & -0.0054 & -0.0028 & -0.0041 & -0.0043 & -0.0048 & -0.0042 & -0.0049 \\ 
                MSE  & 0.0296 & 0.0273 & 0.0273 & 0.0267 & 0.0274 & 0.0268 & 0.0271 & 0.0242 & 0.0242 & 0.0250 & 0.0243 & 0.0250 \\ 
				\midrule
				\multicolumn{12}{c}{\textit{Random-effects variances}: $\sigma^2_{b0}$=0.15, $\sigma^2_{b1}$=0.15, $\sigma^2_{b2}=4\times10^{-6}$}  \\
                Bias & -0.0034 & -0.0035 & -0.0038 & 0.0014 & -0.0040 & 0.0013 & -0.0016 & -0.0019 & -0.0022 & 0.0006 & -0.0023 & 0.0006 \\ 
                MSE  & 0.0280 & 0.0269 & 0.0269 & 0.0265 & 0.0268 & 0.0265 & 0.0257 & 0.0246 & 0.0246 & 0.0242 & 0.0246 & 0.0241 \\ 
				\midrule
			\end{tabular}
		}
	\end{center}
    	\tiny
	\vspace{0.01cm} 
       \label{table:appe_k_10_nc_50_bias_MSE_cont_ch2}
\end{table}

\clearpage
\begin{table}[!h]
	\caption{Simulation results of estimated bias and mean square error (MSE) for both counterfactual mean on treatment and ATE based on 1000 simulations. Setting: Continuous outcome, $k$=5, $n_c$(Avg=100, Min=50, Max=150), and different values of random-effects variance. Results are based on weighting centers equally.}
	\begin{center}
		\resizebox{\textwidth}{!}{
			\begin{tabular}{ccccccccccccc}
				\toprule
				& \multicolumn{12}{c}{\textbf{Counterfactual mean on treatment}}\\
				\cmidrule(lr){2-13} 
                & \multicolumn{6}{c}{\textbf{Unadjusted}} &\multicolumn{6}{c}{\textbf{Adjusted}}\\
				\cmidrule(lr){2-7} \cmidrule(lr){8-13} 
				Measure &  Na\"ive & Fixed & Mixed(1$\mid$c) & Mixed(1$\mid$c) Sam. & Mixed(1+A$\mid$c) & Mixed(1+A$\mid$c) Sam. & Na\"ive  & Fixed  & Mixed(1$\mid$c) & Mixed(1$\mid$c) Sam. & Mixed(1+A$\mid$c) & Mixed(1+A$\mid$c) Sam. \\
				\midrule
			    \multicolumn{12}{c}{\textit{Random-effects variances}: $\sigma^2_{b0}$=0, $\sigma^2_{b1}$=0, $\sigma^2_{b2}$=0} \\
                Bias & -0.0013 & -0.0015 & -0.0015 & -0.0001 & -0.0015 & -0.0001 & -0.0000 & 0.0003 & 0.0003 & -0.0013 & 0.0004 & -0.0013 \\ 
                MSE  & 0.0050 & 0.0056 & 0.0056 & 0.0643 & 0.0057 & 0.0643 & 0.0044 & 0.0050 & 0.0050 & 0.0635 & 0.0050 & 0.0635 \\ 
				\midrule
				\multicolumn{12}{c}{\textit{Random-effects variances}: $\sigma^2_{b0}$=0.15, $\sigma^2_{b1}$=0.15, $\sigma^2_{b2}$=0} \\
                 Bias & -0.0046 & -0.0098 & -0.0097 & -0.0009 & -0.0098 & -0.0009 & -0.0047 & -0.0095 & -0.0095 & -0.0021 & -0.0097 & -0.0020 \\ 
                 MSE  & 0.0694 & 0.0664 & 0.0664 & 0.0693 & 0.0664 & 0.0693 & 0.0692 & 0.0658 & 0.0658 & 0.0683 & 0.0658 & 0.0683 \\ 
				\midrule
				\multicolumn{12}{c}{\textit{Random-effects variances}: $\sigma^2_{b0}$=0.15, $\sigma^2_{b1}$=0.15, $\sigma^2_{b2}=4\times10^{-6}$} \\
                Bias & 0.0104 & 0.0125 & 0.0126 & 0.0059 & 0.0125 & 0.0058 & 0.0096 & 0.0119 & 0.0119 & 0.0045 & 0.0118 & 0.0045 \\ 
                 MSE  & 0.0655 & 0.0614 & 0.0614 & 0.0636 & 0.0613 & 0.0636 & 0.0653 & 0.0612 & 0.0612 & 0.0630 & 0.0612 & 0.0630 \\ 
				\midrule	
           & \multicolumn{12}{c}{\textbf{ATE}}\\
				\cmidrule(lr){2-13} 
                & \multicolumn{6}{c}{\textbf{Unadjusted}} &\multicolumn{6}{c}{\textbf{Adjusted}}\\
				\cmidrule(lr){2-7} \cmidrule(lr){8-13} 
				Measure &  Na\"ive & Fixed & Mixed(1$\mid$c) & Mixed(1$\mid$c) Sam. & Mixed(1+A$\mid$c) & Mixed(1+A$\mid$c) Sam. & Na\"ive  & Fixed  & Mixed(1$\mid$c) & Mixed(1$\mid$c) Sam. & Mixed(1+A$\mid$c) & Mixed(1+A$\mid$c) Sam. \\
            \midrule
           \multicolumn{12}{c}{\textit{Random-effects variances}: $\sigma^2_{b0}$=0, $\sigma^2_{b1}$=0, $\sigma^2_{b2}$=0} \\
                Bias & 0.0011 & 0.0002 & 0.0002 & -0.0008 & 0.0002 & -0.0008 & 0.0037 & 0.0040 & 0.0041 & -0.0033 & 0.0042 & -0.0033 \\ 
                MSE  & 0.0105 & 0.0119 & 0.0119 & 0.0427 & 0.0119 & 0.0427 & 0.0081 & 0.0091 & 0.0091 & 0.0398 & 0.0092 & 0.0399 \\ 
				\midrule
				\multicolumn{12}{c}{\textit{Random-effects variances}: $\sigma^2_{b0}$=0.15, $\sigma^2_{b1}$=0.15, $\sigma^2_{b2}$=0}\\
                Bias & -0.0063 & -0.0096 & -0.0095 & 0.0107 & -0.0096 & 0.0107 & -0.0069 & -0.0097 & -0.0097 & 0.0083 & -0.0098 & 0.0084 \\ 
                MSE  & 0.0445 & 0.0426 & 0.0426 & 0.0457 & 0.0427 & 0.0458 & 0.0430 & 0.0401 & 0.0401 & 0.0432 & 0.0402 & 0.0432 \\
				\midrule
				\multicolumn{12}{c}{\textit{Random-effects variances}: $\sigma^2_{b0}$=0.15, $\sigma^2_{b1}$=0.15, $\sigma^2_{b2}=4\times10^{-6}$}  \\
                Bias & 0.0057 & 0.0070 & 0.0070 & -0.0001 & 0.0071 & -0.0001 & 0.0038 & 0.0052 & 0.0052 & -0.0031 & 0.0052 & -0.0031 \\ 
               MSE  & 0.0422 & 0.0389 & 0.0389 & 0.0431 & 0.0390 & 0.0432 & 0.0410 & 0.0373 & 0.0373 & 0.0414 & 0.0374 & 0.0414 \\ 
                \midrule
			\end{tabular}
		}
	\end{center}
    	\tiny
	\vspace{0.01cm} 
       \label{table:appe_k_5_nc_100_bias_MSE_cont_ch2}
\end{table}

\clearpage
\begin{table}[!h]
    \caption{Simulation results of estimated bias and mean square error (MSE) for both counterfactual mean on treatment and ATE based on 1000 simulations. Setting: Continuous outcome, $k=100$, $n_c$ (Avg=100, Min=50, Max=145), and higher values of random-effects variance. Results are based on weighting centers equally.}
    \begin{center}
		\resizebox{\textwidth}{!}{
			\begin{tabular}{ccccccccccccc}
				\toprule
				& \multicolumn{12}{c}{\textbf{Counterfactual mean on treatment}}\\
				\cmidrule(lr){2-13} 
                & \multicolumn{6}{c}{\textbf{Unadjusted}} &\multicolumn{6}{c}{\textbf{Adjusted}}\\
				\cmidrule(lr){2-7} \cmidrule(lr){8-13} 
				Measure &  Na\"ive & Fixed & Mixed(1$\mid$c) & Mixed(1$\mid$c) Sam. & Mixed(1+A$\mid$c) & Mixed(1+A$\mid$c) Sam. & Na\"ive  & Fixed  & Mixed(1$\mid$c) & Mixed(1$\mid$c) Sam. & Mixed(1+A$\mid$c) & Mixed(1+A$\mid$c) Sam. \\
            \midrule
         \multicolumn{9}{c}{\textit{Random-effects variances}: $\sigma^2_{b0}$=0.15, $\sigma^2_{b1}$=0.15, $\sigma^2_{b2}=4\times10^{-6}$} \\
        Bias & 0.0015 & 0.0012 & 0.0012 & 0.0005 & 0.0013 & 0.0005 & 0.0016 & 0.0014 & 0.0014 & 0.0004 & 0.0015 & 0.0004 \\ 
        MSE  & 0.0036 & 0.0032 & 0.0032 & 0.0034 & 0.0032 & 0.0034 & 0.0036 & 0.0032 & 0.0032 & 0.0034 & 0.0032 & 0.0034 \\ 
            \midrule
            & \multicolumn{12}{c}{\textbf{ATE}}\\
				\cmidrule(lr){2-13} 
                & \multicolumn{6}{c}{\textbf{Unadjusted}} &\multicolumn{6}{c}{\textbf{Adjusted}}\\
				\cmidrule(lr){2-7} \cmidrule(lr){8-13} 
				Measure &  Na\"ive & Fixed & Mixed(1$\mid$c) & Mixed(1$\mid$c) Sam. & Mixed(1+A$\mid$c) & Mixed(1+A$\mid$c) Sam. & Na\"ive  & Fixed  & Mixed(1$\mid$c) & Mixed(1$\mid$c) Sam. & Mixed(1+A$\mid$c) & Mixed(1+A$\mid$c) Sam. \\
            \midrule
            \multicolumn{12}{c}{\textit{Random-effects variances}: $\sigma^2_{b0}$=0.15, $\sigma^2_{b1}$=0.15, $\sigma^2_{b2}=4\times10^{-6}$}  \\
        Bias & -0.0004 & -0.0002 & -0.0002 & 0.0008 & -0.0002 & 0.0008 & -0.0002 & 0.0002 & 0.0002 & 0.0006 & 0.0002 & 0.0006 \\ 
        MSE  & 0.0022 & 0.0021 & 0.0021 & 0.0021 & 0.0021 & 0.0021 & 0.0021 & 0.0020 & 0.0020 & 0.0020 & 0.0020 & 0.0020 \\ 
            \bottomrule
        \end{tabular}
        }
    \end{center}
    \vspace{0.1cm}
    \footnotesize
       \label{table:appe_k_100_nc_100_bias_MSE_cont_ch2}
\end{table}

\clearpage
\autoref{table:appe_k_100_nc_5_mis_bias_cont_ch2} reports the estimated bias and mean squared error for the counterfactual mean under treatment and the ATE under outcome model misspecification with continuous outcomes. All estimators yield unbiased effect estimates in this setting.  In addition, \autoref{table:appe_k_5_nc_100_cont_inf_center_size_ch2} presents the Monte Carlo standard deviation, average standard errors, and 95\% confidence interval coverage in setting where treatment effects and center sizes are not independent.

\begin{table}[!h]
	\caption{Simulation results of estimated bias and mean square error (MSE) for both counterfactual mean on treatment and ATE based on 1000 simulations. Setting: Continuous outcomes with a misspecified outcome model, $k$=100, $n_c$(Avg=5, Min=1, Max=24). Results are based on weighting centers equally.}
	\begin{center}
		\resizebox{\textwidth}{!}{
			\begin{tabular}{ccccccccccccc}
				\toprule
				& \multicolumn{12}{c}{\textbf{Counterfactual mean on treatment}}\\
				\cmidrule(lr){2-13} 
                & \multicolumn{6}{c}{\textbf{Unadjusted}} &\multicolumn{6}{c}{\textbf{Adjusted}}\\
				\cmidrule(lr){2-7} \cmidrule(lr){8-13} 
				Measure &  Na\"ive & Fixed & Mixed(1$\mid$c) & Mixed(1$\mid$c) Sam. & Mixed(1+A$\mid$c) & Mixed(1+A$\mid$c) Sam. & Na\"ive  & Fixed  & Mixed(1$\mid$c) & Mixed(1$\mid$c) Sam. & Mixed(1+A$\mid$c) & Mixed(1+A$\mid$c) Sam. \\
				\midrule
				\multicolumn{12}{c}{\textit{Random-effects variances}: $\sigma^2_{b0}$=0.15, $\sigma^2_{b1}$=0.15, $\sigma^2_{b2}$=0}\\
                Bias & 0.0068 & 0.0050 & 0.0051 & 0.0026 & 0.0054 & 0.0025 & 0.0059 & 0.0044 & 0.0042 & 0.0025 & 0.0044 & 0.0025 \\ 
                MSE  & 0.0094 & 0.0100 & 0.0100 & 0.0110 & 0.0101 & 0.0110 & 0.0094 & 0.0097 & 0.0097 & 0.0106 & 0.0100 & 0.0106 \\ 
				\midrule
               & \multicolumn{12}{c}{\textbf{ATE}}\\
				\cmidrule(lr){2-13} 
                & \multicolumn{6}{c}{\textbf{Unadjusted}} &\multicolumn{6}{c}{\textbf{Adjusted}}\\
				\cmidrule(lr){2-7} \cmidrule(lr){8-13} 
				Measure &  Na\"ive & Fixed & Mixed(1$\mid$c) & Mixed(1$\mid$c) Sam. & Mixed(1+A$\mid$c) & Mixed(1+A$\mid$c) Sam. & Na\"ive  & Fixed  & Mixed(1$\mid$c) & Mixed(1$\mid$c) Sam. & Mixed(1+A$\mid$c) & Mixed(1+A$\mid$c) Sam. \\
				\midrule
				\multicolumn{12}{c}{\textit{Random-effects variances}: $\sigma^2_{b0}$=0.15, $\sigma^2_{b1}$=0.15, $\sigma^2_{b2}$=0}\\
                Bias & 0.0083 & 0.0087 & 0.0089 & 0.0007 & 0.0092 & 0.0006 & 0.0068 & 0.0078 & 0.0073 & 0.0001 & 0.0075 & 0.0000 \\ 
                MSE  & 0.0130 & 0.0145 & 0.0145 & 0.0162 & 0.0158 & 0.0163 & 0.0122 & 0.0133 & 0.0133 & 0.0147 & 0.0146 & 0.0147 \\ 
				\midrule
			\end{tabular}
		}
	\end{center}
    	\tiny
	\vspace{0.01cm} 
     \label{table:appe_k_100_nc_5_mis_bias_cont_ch2}
\end{table}

\clearpage
\begin{table}[!h]
	\caption{Simulation results: Monte Carlo standard deviation, average standard errors and coverage probabilities of 95\% confidence intervals for both counterfactual mean on treatment and ATE based on 1000 simulations. Setting: Continuous outcome, $k$=5, $n_c$(Avg=100, Min=50, Max=150), and treatment effects and center sizes are not independent. Results are based on weighting centers equally.}
	\begin{center}
		\resizebox{\textwidth}{!}{
			\begin{tabular}{>{\raggedright\arraybackslash}p{3.5cm}cccccccccccccccccc}
				\toprule
				& \multicolumn{9}{c}{\textbf{Counterfactual mean on treatment}} &\multicolumn{9}{c}{\textbf{ATE}} \\
				\cmidrule(lr){2-10} \cmidrule(lr){11-19}
				&& \multicolumn{4}{c}{\textbf{SE}} & \multicolumn{4}{c}{\textbf{Coverage (\%)}} &&\multicolumn{4}{c}{\textbf{SE}}&\multicolumn{4}{c}{\textbf{Coverage (\%)}} \\
				\cmidrule(lr){3-6} \cmidrule(lr){7-10} \cmidrule(lr){12-15} \cmidrule(lr){16-19}
				Method &  SD & Na\"ive & REML & DL & DB & Na\"ive & REML & DL & DB  & SD & Na\"ive & REML & DL & DB & Naive & REML & DL & DB \\ \midrule
				&&&&& \multicolumn{9}{c}{\textit{Random-effects variances}: $\sigma^2_{b0}$=0.15, $\sigma^2_{b1}$=0.15, $\sigma^2_{b2}$=0} &&&&&\\
               \multicolumn{19}{l}{\textbf{Unadjusted}}\\
                Na\"ive & 0.263 & 0.082 &  &  &  & 37.2 &  &  &  & 0.207 & 0.112 &  &  &  & 61.0 &  &  &  \\ 
                Fixed & 0.249 &  & 0.329 & 0.324 & 0.298 &  & 98.2 & 98.0 & 97.8 & 0.199 &  & 0.241 & 0.239 & 0.223 &  & 96.0 & 96.1 & 95.7 \\ 
                Mixed(1$\mid$c) & 0.249 &  & 0.329 & 0.324 & 0.298 &  & 98.3 & 98.2 & 97.6 & 0.199 &  & 0.241 & 0.239 & 0.223 &  & 96.1 & 96.3 & 95.7 \\
                Mixed(1$\mid$c) Sam. & 0.256 &  & 0.328 & 0.323 & 0.297 &  & 97.2 & 97.1 & 96.1 & 0.207 &  & 0.245 & 0.244 & 0.228 &  & 94.6 & 94.6 & 93.7 \\ 
                Mixed(1+A$\mid$c) & 0.248 &  & 0.330 & 0.325 & 0.298 &  & 98.2 & 98.1 & 97.6 & 0.199 &  & 0.243 & 0.241 & 0.224 &  & 96.2 & 96.3 & 95.8 \\ 
                Mixed(1+A$\mid$c) Sam. & 0.256 &  & 0.328 & 0.323 & 0.297 &  & 97.2 & 97.1 & 96.1 & 0.206 &  & 0.245 & 0.244 & 0.228 &  & 94.6 & 94.8 & 93.6 \\ 
                \multicolumn{19}{l}{\textbf{Adjusted}}\\
                Na\"ive  & 0.262 & 0.079 &  &  &  & 36.1 &  &  &  & 0.202 & 0.102 &  &  &  & 59.0 &  &  &  \\ 
                Fixed  & 0.247 &  & 0.328 & 0.323 & 0.296 &  & 98.1 & 98.2 & 97.3 & 0.193 &  & 0.235 & 0.233 & 0.216 &  & 96.5 & 96.5 & 95.6 \\ 
                Mixed(1$\mid$c)  & 0.247 &  & 0.328 & 0.323 & 0.296 &  & 97.9 & 98.2 & 97.3 & 0.193 &  & 0.235 & 0.233 & 0.216 &  & 96.5 & 96.4 & 95.7 \\
                 Mixed(1$\mid$c) Sam. & 0.255 &  & 0.327 & 0.321 & 0.295 &  & 97.5 & 97.4 & 96.4 & 0.202 &  & 0.240 & 0.238 & 0.222 &  & 95.4 & 95.4 & 93.8 \\ 
                Mixed(1+A$\mid$c) & 0.247 &  & 0.329 & 0.324 & 0.297 &  & 98.2 & 98.1 & 97.6 & 0.193 &  & 0.238 & 0.236 & 0.219 &  & 96.2 & 96.3 & 95.8 \\ 
                Mixed(1+A$\mid$c) Sam. & 0.255 &  & 0.328 & 0.322 & 0.296 &  & 97.6 & 97.4 & 96.4 & 0.202 &  & 0.242 & 0.240 & 0.224 &  & 95.4 & 95.3 & 94.2 \\ 
				\bottomrule					
			\end{tabular}%
		}
	\end{center}
	\tiny
	\vspace{0.1cm} 
            \label{table:appe_k_5_nc_100_cont_inf_center_size_ch2}
\end{table}

\autoref{table:appe_k_50_nc_10_bin_ch2}, \autoref{table:appe_k_10_nc_50_bin_ch2} and \autoref{table:appe_k_5_nc_100_bin_ch2} present the Monte Carlo standard deviation, average standard errors, and 95\% confidence interval coverage for the counterfactual mean under treatment and the ATE for binary outcomes for 50, 10 and 5 centers, respectively. As for continuous outcomes, the results demonstrate that greater between-center heterogeneity and a smaller number of centers lead to undercoverage across the estimators considered. Further, \autoref{table:appe_k_100_nc_5_mis_bin_ch2} shows the results when the outcome model is misspecified for binary outcomes.

\clearpage
\begin{table}[!h]
	\caption{Simulation results: Monte Carlo standard deviation, average standard errors and coverage probabilities of 95\% confidence intervals for both counterfactual mean on treatment and ATE based on 1000 simulations. Setting: Binary outcome, $k$=50, $n_c$(Avg=10, Min=2, Max=48), and different values of random-effects variance. Results are based on weighting centers equally.}
	\begin{center}
		\resizebox{\textwidth}{!}{
			\begin{tabular}{>{\raggedright\arraybackslash}p{3.5cm}cccccccccccccccccc}
				\toprule
				& \multicolumn{9}{c}{\textbf{Counterfactual mean on treatment}} &\multicolumn{9}{c}{\textbf{ATE}} \\
				\cmidrule(lr){2-10} \cmidrule(lr){11-19}
				&& \multicolumn{4}{c}{\textbf{SE}} & \multicolumn{4}{c}{\textbf{Coverage (\%)}} &&\multicolumn{4}{c}{\textbf{SE}}&\multicolumn{4}{c}{\textbf{Coverage (\%)}} \\
				\cmidrule(lr){3-6} \cmidrule(lr){7-10} \cmidrule(lr){12-15} \cmidrule(lr){16-19}
				Method &  SD & Na\"ive & REML & DL & DB & Na\"ive & REML & DL & DB  & SD & Na\"ive & REML & DL & DB & Naive & REML & DL & DB \\ \midrule
				&&&&& \multicolumn{9}{c}{\textit{Random-effects variances}: $\sigma^2_{b0}$=0, $\sigma^2_{b1}$=0, $\sigma^2_{b2}$=0} &&&&&\\
                \multicolumn{19}{l}{\textbf{Unadjusted}}\\
				Na\"ive & 0.031 & 0.032 &  &  &  & 95.1 &  &  &  & 0.044 & 0.044 &  &  &  & 94.8 &  &  &  \\ 
				Fixed & 0.036 &  & 0.037 & 0.037 & 0.037 &  & 95.0 & 95.1 & 95.1 & 0.050 &  & 0.050 & 0.050 & 0.050 &  & 95.4 & 95.4 & 95.2 \\ 
				Mixed(1$\mid$c) & 0.036 &  & 0.037 & 0.037 & 0.038 &  & 94.9 & 95.1 & 95.7 & 0.050 &  & 0.052 & 0.053 & 0.053 &  & 95.6 & 95.7 & 95.8 \\
                Mixed(1$\mid$c) Sam. & 0.035 &  & 0.037 & 0.037 & 0.038 &  & 96.9 & 96.9 & 97.2 & 0.053 &  & 0.053 & 0.053 & 0.053 &  & 95.1 & 95.4 & 95.5 \\ 
				Mixed(1+A$\mid$c) & 0.036 &  & 0.037 & 0.037 & 0.038 &  & 95.0 & 95.2 & 95.6 & 0.050 &  & 0.052 & 0.053 & 0.053 &  & 95.6 & 95.6 & 95.7 \\ 
                Mixed(1+A$\mid$c) Sam. & 0.035 &  & 0.037 & 0.037 & 0.038 &  & 96.9 & 96.9 & 97.2 & 0.053 &  & 0.053 & 0.053 & 0.053 &  & 95.1 & 95.4 & 95.5 \\ 
                \multicolumn{19}{l}{\textbf{Adjusted}}\\
				Na\"ive   & 0.031 & 0.030 &  &  &  & 94.7 &  &  &  & 0.041 & 0.041 &  &  &  & 94.6 &  &  &  \\ 
				Fixed   & 0.035 &  & 0.035 & 0.035 & 0.035 &  & 94.9 & 94.9 & 95.0 & 0.047 &  & 0.047 & 0.046 & 0.046 &  & 94.9 & 94.5 & 94.4 \\ 
				Mixed(1$\mid$c)  & 0.035 &  & 0.036 & 0.036 & 0.037 &  & 95.1 & 95.1 & 95.1 & 0.047 &  & 0.049 & 0.049 & 0.049 &  & 94.9 & 95.1 & 95.4 \\ 
                Mixed(1$\mid$c) Sam. & 0.034 &  & 0.036 & 0.036 & 0.037 &  & 96.6 & 96.5 & 97.1 & 0.048 &  & 0.049 & 0.049 & 0.050 &  & 95.0 & 95.1 & 95.5 \\ 
				Mixed(1+A$\mid$c)  & 0.035 &  & 0.036 & 0.036 & 0.037 &  & 95.0 & 95.2 & 95.6 & 0.047 &  & 0.049 & 0.049 & 0.049 &  & 95.6 & 95.6 & 95.7 \\
                Mixed(1+A$\mid$c) Sam. & 0.034 &  & 0.036 & 0.036 & 0.037 &  & 96.6 & 96.5 & 97.0 & 0.048 &  & 0.049 & 0.049 & 0.050 &  & 95.1 & 95.2 & 95.5\\ 
				\midrule
				&&&&& \multicolumn{9}{c}{\textit{Random-effects variances}: $\sigma^2_{b0}$=0.5, $\sigma^2_{b1}$=0, $\sigma^2_{b2}$=0} &&&&&\\
                \multicolumn{19}{l}{\textbf{Unadjusted}}\\
				Na\"ive & 0.039 & 0.031 &  &  &  & 89.4 &  &  &  & 0.044 & 0.044 &  &  &  & 95.7 &  &  &  \\ 
				Fixed & 0.040 &  & 0.040 & 0.040 & 0.040 &  & 94.5 & 94.3 & 94.0 & 0.049 &  & 0.048 & 0.048 & 0.048 &  & 94.9 & 94.8 & 94.9 \\ 
				Mixed(1$\mid$c) & 0.041 &  & 0.041 & 0.041 & 0.041 &  & 94.8 & 94.8 & 95.1 & 0.051 &  & 0.050 & 0.050 & 0.050 &  & 95.1 & 95.2 & 95.4 \\ 
                Mixed(1$\mid$c) Sam. & 0.041 &  & 0.041 & 0.041 & 0.041 &  & 95.1 & 95.0 & 94.6 & 0.051 &  & 0.053 & 0.053 & 0.053 &  & 96.0 & 95.9 & 96.1 \\ 
				Mixed(1+A$\mid$c) & 0.041 &  & 0.041 & 0.041 & 0.041 &  & 94.7 & 94.8 & 95.0 & 0.051 &  & 0.050 & 0.050 & 0.051 &  & 95.2 & 95.2 & 95.4 \\
                Mixed(1+A$\mid$c) Sam. & 0.041 &  & 0.041 & 0.041 & 0.041 &  & 94.9 & 95.1 & 94.5 & 0.051 &  & 0.053 & 0.053 & 0.053 &  & 95.9 & 95.9 & 96.1 \\ 
                \multicolumn{19}{l}{\textbf{Adjusted}}\\
				Na\"ive  & 0.039 & 0.031 &  &  &  & 87.6 &  &  &  & 0.042 & 0.042 &  &  &  & 94.5 &  &  &  \\ 
				Fixed  & 0.039 &  & 0.039 & 0.039 & 0.039 &  & 94.7 & 94.6 & 94.3 & 0.046 &  & 0.045 & 0.045 & 0.045 &  & 95.1 & 95.2 & 94.8 \\ 
				Mixed(1$\mid$c)  & 0.040 &  & 0.040 & 0.040 & 0.040 &  & 95.0 & 94.9 & 95.0 & 0.048 &  & 0.046 & 0.047 & 0.047 &  & 94.7 & 94.8 & 95.1 \\ 
                Mixed(1$\mid$c) Sam. & 0.041 &  & 0.040 & 0.040 & 0.040 &  & 94.5 & 94.3 & 94.3 & 0.048 &  & 0.049 & 0.050 & 0.050 &  & 96.5 & 96.6 & 96.6 \\ 
				Mixed(1+A$\mid$c)  & 0.040 &  & 0.040 & 0.040 & 0.040 &  & 94.7 & 94.8 & 95.0 & 0.048 &  & 0.047 & 0.047 & 0.047 &  & 95.2 & 95.2 & 95.4 \\ 
                Mixed(1+A$\mid$c) Sam. & 0.041 &  & 0.040 & 0.040 & 0.040 &  & 94.8 & 94.5 & 94.1 & 0.048 &  & 0.050 & 0.050 & 0.050 &  & 96.6 & 96.7 & 96.5 \\ 
				\bottomrule
				&&&&& \multicolumn{9}{c}{\textit{Random-effects variances}: $\sigma^2_{b0}$=0.5, $\sigma^2_{b1}$=0.25, $\sigma^2_{b2}$=0.25} &&&&&\\
                \multicolumn{19}{l}{\textbf{Unadjusted}}\\
				Na\"ive & 0.041 & 0.031 &  &  &  & 86.6 &  &  &  & 0.047 & 0.044 &  &  &  & 93.3 &  &  &  \\ 
				Fixed & 0.041 &  & 0.042 & 0.042 & 0.041 &  & 95.2 & 95.3 & 95.1 & 0.050 &  & 0.049 & 0.049 & 0.048 &  & 94.3 & 94.1 & 93.5 \\ 
				Mixed(1$\mid$c) & 0.042 &  & 0.042 & 0.043 & 0.042 &  & 95.1 & 95.3 & 95.4 & 0.052 &  & 0.051 & 0.051 & 0.051 &  & 94.5 & 94.3 & 94.6 \\ 
                Mixed(1$\mid$c) Sam. & 0.043 &  & 0.043 & 0.043 & 0.042 &  & 93.7 & 93.7 & 93.9 & 0.053 &  & 0.053 & 0.054 & 0.054 &  & 95.0 & 95.1 & 95.3 \\ 
				Mixed(1+A$\mid$c) & 0.042 &  & 0.042 & 0.042 & 0.042 &  & 95.2 & 95.5 & 95.4 & 0.052 &  & 0.051 & 0.051 & 0.051 &  & 94.5 & 94.3 & 94.4 \\ 
                Mixed(1+A$\mid$c) Sam. & 0.043 &  & 0.043 & 0.043 & 0.042 &  & 93.7 & 93.6 & 93.9 & 0.053 &  & 0.053 & 0.054 & 0.054 &  & 95.1 & 95.3 & 95.5 \\ 
                \multicolumn{19}{l}{\textbf{Adjusted}}\\
				Na\"ive  & 0.040 & 0.031 &  &  &  & 87.2 &  &  &  & 0.045 & 0.042 &  &  &  & 92.8 &  &  &  \\ 
				Fixed  & 0.040 &  & 0.041 & 0.041 & 0.041 &  & 95.0 & 95.1 & 94.7 & 0.047 &  & 0.046 & 0.045 & 0.045 &  & 93.8 & 93.8 & 93.6 \\ 
				Mixed(1$\mid$c)  & 0.040 &  & 0.042 & 0.042 & 0.042 &  & 95.5 & 95.6 & 95.1 & 0.049 &  & 0.047 & 0.047 & 0.048 &  & 93.9 & 93.9 & 93.7 \\ 
                Mixed(1$\mid$c) Sam. & 0.042 &  & 0.042 & 0.042 & 0.042 &  & 94.4 & 94.3 & 93.8 & 0.050 &  & 0.051 & 0.051 & 0.051 &  & 95.5 & 95.2 & 95.3 \\ 
				Mixed(1+A$\mid$c)  & 0.040 &  & 0.042 & 0.042 & 0.041 &  & 95.2 & 95.5 & 95.4 & 0.049 &  & 0.048 & 0.048 & 0.048 &  & 94.5 & 94.3 & 94.4 \\ 
                Mixed(1+A$\mid$c) Sam. & 0.042 &  & 0.042 & 0.042 & 0.042 &  & 94.3 & 94.1 & 93.8 & 0.050 &  & 0.051 & 0.051 & 0.051 &  & 95.3 & 95.2 & 95.2 \\ 
				\bottomrule
				&&&&& \multicolumn{9}{c}{\textit{Random-effects variances}: $\sigma^2_{b0}$=0.75, $\sigma^2_{b1}$=0.5, $\sigma^2_{b2}$=0.5} &&&&&\\
                \multicolumn{19}{l}{\textbf{Unadjusted}}\\
				Na\"ive & 0.047 & 0.031 &  &  &  & 80.2 &  &  &  & 0.048 & 0.044 &  &  &  & 92.5 &  &  &  \\ 
				Fixed & 0.044 &  & 0.045 & 0.045 & 0.044 &  & 95.4 & 95.3 & 94.8 & 0.050 &  & 0.049 & 0.049 & 0.048 &  & 95.1 & 94.9 & 94.3 \\ 
				Mixed(1$\mid$c) & 0.044 &  & 0.045 & 0.045 & 0.045 &  & 95.4 & 95.0 & 94.9 & 0.050 &  & 0.050 & 0.050 & 0.050 &  & 94.9 & 95.0 & 94.7 \\ 
                Mixed(1$\mid$c) Sam. & 0.047 &  & 0.045 & 0.045 & 0.045 &  & 93.3 & 93.5 & 93.3 & 0.055 &  & 0.054 & 0.054 & 0.055 &  & 93.2 & 93.5 & 93.6 \\ 
				Mixed(1+A$\mid$c) & 0.044 &  & 0.045 & 0.045 & 0.045 &  & 95.3 & 95.0 & 94.9 & 0.050 &  & 0.051 & 0.051 & 0.051 &  & 94.9 & 94.9 & 95.0 \\ 
                 Mixed(1+A$\mid$c) Sam. & 0.047 &  & 0.045 & 0.045 & 0.045 &  & 93.4 & 93.4 & 93.3 & 0.055 &  & 0.054 & 0.054 & 0.055 &  & 93.2 & 93.4 & 93.6 \\ 
                \multicolumn{19}{l}{\textbf{Adjusted}}\\
				Na\"ive  & 0.047 & 0.031 &  &  &  & 80.3 &  &  &  & 0.046 & 0.042 &  &  &  & 92.7 &  &  &  \\ 
				Fixed  & 0.044 &  & 0.044 & 0.044 & 0.043 &  & 94.6 & 94.5 & 94.0 & 0.048 &  & 0.046 & 0.046 & 0.045 &  & 93.8 & 93.4 & 92.8 \\ 
				Mixed(1$\mid$c)  & 0.044 &  & 0.044 & 0.044 & 0.044 &  & 95.4 & 95.3 & 94.9 & 0.048 &  & 0.047 & 0.047 & 0.047 &  & 93.9 & 93.9 & 93.8 \\ 
                Mixed(1$\mid$c) Sam. & 0.047 &  & 0.044 & 0.044 & 0.044 &  & 93.3 & 93.4 & 92.7 & 0.053 &  & 0.052 & 0.052 & 0.052 &  & 94.1 & 94.2 & 94.2 \\ 
				Mixed(1+A$\mid$c)  & 0.044 &  & 0.044 & 0.044 & 0.044 &  & 95.3 & 95.0 & 94.9 & 0.048 &  & 0.048 & 0.048 & 0.048 &  & 94.9 & 94.9 & 95.0 \\ 
                Mixed(1+A$\mid$c) Sam. & 0.047 &  & 0.044 & 0.044 & 0.044 &  & 93.4 & 93.3 & 92.8 & 0.053 &  & 0.052 & 0.052 & 0.052 &  & 94.2 & 94.3 & 94.1 \\ 
				\bottomrule					  
			\end{tabular}%
		}
	\end{center}
	\tiny
	\vspace{0.01cm} 
    \label{table:appe_k_50_nc_10_bin_ch2}
\end{table}

\clearpage
\begin{table}[!h]
	\caption{Simulation results: Monte Carlo standard deviation, average standard errors and coverage probabilities of 95\% confidence intervals for both counterfactual mean on treatment and ATE based on 1000 simulations. Setting: Binary outcome, $k$=10, $n_c$(Avg=50, Min=25, Max=80), and different values of random-effects variance. Results are based on weighting centers equally.}
	\begin{center}
		\resizebox{\textwidth}{!}{
			\begin{tabular}{>{\raggedright\arraybackslash}p{3.5cm}cccccccccccccccccc}
				\toprule
				& \multicolumn{9}{c}{\textbf{Counterfactual mean on treatment}} &\multicolumn{9}{c}{\textbf{ATE}} \\
				\cmidrule(lr){2-10} \cmidrule(lr){11-19}
				&& \multicolumn{4}{c}{\textbf{SE}} & \multicolumn{4}{c}{\textbf{Coverage (\%)}} &&\multicolumn{4}{c}{\textbf{SE}}&\multicolumn{4}{c}{\textbf{Coverage (\%)}} \\
				\cmidrule(lr){3-6} \cmidrule(lr){7-10} \cmidrule(lr){12-15} \cmidrule(lr){16-19}
				Method &  SD & Na\"ive & REML & DL & DB & Na\"ive & REML & DL & DB  & SD & Na\"ive & REML & DL & DB & Naive & REML & DL & DB \\ \midrule
				&&&&& \multicolumn{9}{c}{\textit{Random-effects variances}: $\sigma^2_{b0}$=0, $\sigma^2_{b1}$=0, $\sigma^2_{b2}$=0} &&&&&\\
                \multicolumn{19}{l}{\textbf{Unadjusted}}\\
				Na\"ive & 0.033 & 0.032 &  &  &  & 94.4 &  &  &  & 0.046 & 0.044 &  &  &  & 93.8 &  &  &  \\ 
				Fixed & 0.034 &  & 0.035 & 0.035 & 0.035 &  & 95.3 & 95.5 & 95.8 & 0.049 &  & 0.049 & 0.050 & 0.050 &  & 95.2 & 95.2 & 95.1 \\ 
				Mixed(1$\mid$c) & 0.034 &  & 0.035 & 0.035 & 0.035 &  & 95.6 & 95.8 & 95.9 & 0.049 &  & 0.050 & 0.050 & 0.050 &  & 95.4 & 95.5 & 95.4 \\ 
                Mixed(1$\mid$c) Sam. & 0.032 &  & 0.035 & 0.035 & 0.035 &  & 96.2 & 96.4 & 96.5 & 0.045 &  & 0.050 & 0.050 & 0.050 &  & 97.3 & 97.3 & 97.3 \\ 
				Mixed(1+A$\mid$c) & 0.034 &  & 0.035 & 0.035 & 0.035 &  & 95.6 & 95.8 & 95.9 & 0.049 &  & 0.050 & 0.050 & 0.050 &  & 95.4 & 95.4 & 95.3 \\ 
                 Mixed(1+A$\mid$c) Sam. & 0.032 &  & 0.035 & 0.035 & 0.035 &  & 96.2 & 96.4 & 96.5 & 0.045 &  & 0.050 & 0.050 & 0.050 &  & 97.3 & 97.3 & 97.3 \\ 
                \multicolumn{19}{l}{\textbf{Adjusted}}\\
				Na\"ive & 0.032 & 0.031 &  &  &  & 94.8 &  &  &  & 0.043 & 0.041 &  &  &  & 94.7 &  &  &  \\ 
				Fixed  & 0.033 &  & 0.034 & 0.034 & 0.034 &  & 95.8 & 96.0 & 96.1 & 0.045 &  & 0.046 & 0.046 & 0.046 &  & 95.1 & 95.0 & 95.6 \\ 
				Mixed(1$\mid$c)  & 0.033 &  & 0.034 & 0.034 & 0.034 &  & 95.9 & 96.0 & 96.3 & 0.045 &  & 0.046 & 0.047 & 0.047 &  & 95.7 & 95.7 & 95.7 \\ 
                  Mixed(1$\mid$c) Sam. & 0.032 &  & 0.034 & 0.034 & 0.034 &  & 96.6 & 96.7 & 96.6 & 0.042 &  & 0.046 & 0.046 & 0.046 &  & 97.3 & 97.4 & 97.0 \\
				Mixed(1+A$\mid$c)  & 0.033 &  & 0.034 & 0.034 & 0.034 &  & 95.6 & 95.8 & 95.9 & 0.045 &  & 0.046 & 0.047 & 0.047 &  & 95.4 & 95.4 & 95.3 \\
                  Mixed(1+A$\mid$c) Sam. & 0.032 &  & 0.034 & 0.034 & 0.034 &  & 96.6 & 96.7 & 96.6 & 0.042 &  & 0.046 & 0.046 & 0.046 &  & 97.3 & 97.4 & 97.1 \\ 
				\midrule
				&&&&& \multicolumn{9}{c}{\textit{Random-effects variances}: $\sigma^2_{b0}$=0.5, $\sigma^2_{b1}$=0, $\sigma^2_{b2}$=0} &&&&&\\
               \multicolumn{19}{l}{\textbf{Unadjusted}}\\
				Na\"ive & 0.056 & 0.031 &  &  &  & 74.0 &  &  &  & 0.042 & 0.044 &  &  &  & 95.7 &  &  &  \\ 
				Fixed & 0.055 &  & 0.054 & 0.054 & 0.052 &  & 93.9 & 93.7 & 92.9 & 0.043 &  & 0.044 & 0.044 & 0.044 &  & 96.0 & 96.0 & 96.1 \\ 
				Mixed(1$\mid$c) & 0.055 &  & 0.054 & 0.054 & 0.052 &  & 93.7 & 93.7 & 92.8 & 0.043 &  & 0.044 & 0.044 & 0.044 &  & 96.2 & 96.1 & 96.1 \\ 
                Mixed(1$\mid$c) Sam. & 0.055 &  & 0.054 & 0.054 & 0.053 &  & 94.0 & 93.8 & 93.5 & 0.046 &  & 0.050 & 0.050 & 0.050 &  & 95.8 & 96.0 & 96.2 \\ 
				Mixed(1+A$\mid$c) & 0.055 &  & 0.054 & 0.054 & 0.052 &  & 93.7 & 93.7 & 92.8 & 0.043 &  & 0.044 & 0.044 & 0.044 &  & 96.1 & 96.1 & 96.1 \\ 
                Mixed(1+A$\mid$c) Sam. & 0.055 &  & 0.054 & 0.054 & 0.053 &  & 94.1 & 93.9 & 93.5 & 0.046 &  & 0.050 & 0.050 & 0.050 &  & 95.7 & 95.9 & 96.1 \\
                \multicolumn{19}{l}{\textbf{Adjusted}}\\
				Na\"ive  & 0.055 & 0.031 &  &  &  & 72.1 &  &  &  & 0.039 & 0.042 &  &  &  & 95.6 &  &  &  \\ 
				Fixed & 0.054 &  & 0.053 & 0.053 & 0.051 &  & 93.8 & 93.7 & 93.0 & 0.039 &  & 0.041 & 0.041 & 0.041 &  & 96.0 & 95.9 & 95.9 \\ 
				Mixed(1$\mid$c)  & 0.054 &  & 0.053 & 0.053 & 0.051 &  & 93.7 & 93.7 & 93.1 & 0.039 &  & 0.041 & 0.041 & 0.041 &  & 95.7 & 95.8 & 95.8 \\ 
                Mixed(1$\mid$c) Sam. & 0.054 &  & 0.054 & 0.054 & 0.052 &  & 93.9 & 93.9 & 93.2 & 0.042 &  & 0.047 & 0.047 & 0.047 &  & 96.1 & 96.1 & 96.2 \\
				Mixed(1+A$\mid$c)  & 0.054 &  & 0.053 & 0.053 & 0.051 &  & 93.7 & 93.7 & 92.8 & 0.039 &  & 0.041 & 0.041 & 0.041 &  & 96.1 & 96.1 & 96.1 \\ 
                Mixed(1+A$\mid$c) Sam. & 0.054 &  & 0.054 & 0.054 & 0.052 &  & 93.8 & 93.9 & 93.3 & 0.042 &  & 0.047 & 0.047 & 0.047 &  & 96.2 & 96.2 & 96.3 \\ 
				\bottomrule	
				&&&&& \multicolumn{9}{c}{\textit{Random-effects variances}: $\sigma^2_{b0}$=0.5, $\sigma^2_{b1}$=0.25, $\sigma^2_{b2}$=0.25}  &&&&&\\
                \multicolumn{19}{l}{\textbf{Unadjusted}}\\
				Na\"ive & 0.065 & 0.031 &  &  &  & 63.4 &  &  &  & 0.054 & 0.044 &  &  &  & 89.3 &  &  &  \\ 
				Fixed & 0.063 &  & 0.060 & 0.060 & 0.058 &  & 91.9 & 92.4 & 90.6 & 0.054 &  & 0.053 & 0.053 & 0.053 &  & 92.8 & 92.8 & 92.9 \\ 
				Mixed(1$\mid$c) & 0.063 &  & 0.060 & 0.060 & 0.058 &  & 92.0 & 92.1 & 90.4 & 0.054 &  & 0.054 & 0.054 & 0.053 &  & 93.4 & 93.5 & 93.4 \\ 
                Mixed(1$\mid$c) Sam. & 0.063 &  & 0.060 & 0.060 & 0.058 &  & 93.6 & 93.7 & 92.6 & 0.054 &  & 0.055 & 0.055 & 0.055 &  & 94.5 & 94.6 & 94.5\\ 
				Mixed(1+A$\mid$c) & 0.063 &  & 0.060 & 0.060 & 0.058 &  & 92.3 & 92.5 & 90.7 & 0.054 &  & 0.054 & 0.054 & 0.053 &  & 93.3 & 93.2 & 93.5 \\ 
                Mixed(1+A$\mid$c) Sam. & 0.063 &  & 0.060 & 0.060 & 0.058 &  & 93.6 & 93.7 & 92.5 & 0.054 &  & 0.055 & 0.055 & 0.055 &  & 94.6 & 94.5 & 94.5 \\ 
                \multicolumn{19}{l}{\textbf{Adjusted}}\\
				Na\"ive  & 0.065 & 0.031 &  &  &  & 62.4 &  &  &  & 0.052 & 0.042 &  &  &  & 87.7 &  &  &  \\ 
				Fixed  & 0.063 &  & 0.060 & 0.060 & 0.057 &  & 91.5 & 91.8 & 90.2 & 0.051 &  & 0.051 & 0.051 & 0.050 &  & 93.5 & 93.4 & 92.8 \\ 
				Mixed(1$\mid$c)  & 0.063 &  & 0.060 & 0.060 & 0.057 &  & 91.4 & 91.7 & 90.1 & 0.051 &  & 0.051 & 0.051 & 0.050 &  & 93.5 & 93.4 & 92.7 \\ 
                Mixed(1$\mid$c) Sam. & 0.063 &  & 0.060 & 0.060 & 0.058 &  & 92.8 & 92.8 & 92.1 & 0.053 &  & 0.053 & 0.053 & 0.052 &  & 94.8 & 94.9 & 94.3 \\ 
				Mixed(1+A$\mid$c)  & 0.063 &  & 0.060 & 0.060 & 0.058 &  & 92.3 & 92.5 & 90.7 & 0.051 &  & 0.051 & 0.051 & 0.050 &  & 93.3 & 93.2 & 93.5 \\ 
                Mixed(1+A$\mid$c) Sam. & 0.063 &  & 0.060 & 0.060 & 0.058 &  & 92.8 & 92.8 & 92.1 & 0.053 &  & 0.053 & 0.053 & 0.052 &  & 94.9 & 95.0 & 94.4 \\ 
				\bottomrule
				&&&&& \multicolumn{9}{c}{\textit{Random-effects variances}: $\sigma^2_{b0}$=0.75, $\sigma^2_{b1}$=0.5, $\sigma^2_{b2}$=0.5}  &&&&&\\
                \multicolumn{19}{l}{\textbf{Unadjusted}}\\
				Na\"ive & 0.076 & 0.031 &  &  &  & 57.3 &  &  &  & 0.064 & 0.044 &  &  &  & 83.7 &  &  &  \\ 
				Fixed & 0.073 &  & 0.071 & 0.071 & 0.068 &  & 93.0 & 93.1 & 91.8 & 0.062 &  & 0.058 & 0.058 & 0.057 &  & 94.0 & 94.1 & 92.9 \\ 
				Mixed(1$\mid$c) & 0.073 &  & 0.071 & 0.071 & 0.068 &  & 92.9 & 92.9 & 91.9 & 0.062 &  & 0.058 & 0.058 & 0.057 &  & 94.0 & 94.1 & 92.9 \\ 
                Mixed(1$\mid$c) Sam. & 0.074 &  & 0.071 & 0.071 & 0.068 &  & 92.9 & 92.6 & 91.6 & 0.061 &  & 0.060 & 0.060 & 0.059 &  & 92.9 & 93.5 & 93.3 \\ 
				Mixed(1+A$\mid$c) & 0.073 &  & 0.071 & 0.071 & 0.068 &  & 93.1 & 93.1 & 91.7 & 0.062 &  & 0.059 & 0.059 & 0.057 &  & 94.4 & 94.3 & 93.1 \\ 
                Mixed(1+A$\mid$c) Sam. & 0.074 &  & 0.071 & 0.071 & 0.068 &  & 93.0 & 92.6 & 91.6 & 0.061 &  & 0.060 & 0.060 & 0.059 &  & 92.9 & 93.4 & 93.3 \\
                \multicolumn{19}{l}{\textbf{Adjusted}}\\
				Na\"ive   & 0.076 & 0.031 &  &  &  & 57.0 &  &  &  & 0.061 & 0.042 &  &  &  & 82.7 &  &  &  \\ 
				Fixed  & 0.072 &  & 0.070 & 0.070 & 0.067 &  & 93.3 & 93.4 & 92.3 & 0.059 &  & 0.056 & 0.056 & 0.054 &  & 93.4 & 93.4 & 92.9 \\ 
				Mixed(1$\mid$c)  & 0.072 &  & 0.070 & 0.070 & 0.067 &  & 93.4 & 93.5 & 92.3 & 0.059 &  & 0.056 & 0.056 & 0.055 &  & 93.6 & 93.6 & 92.9 \\ 
                Mixed(1$\mid$c) Sam. & 0.074 &  & 0.071 & 0.071 & 0.068 &  & 93.1 & 93.1 & 91.8 & 0.060 &  & 0.058 & 0.058 & 0.057 &  & 93.6 & 93.8 & 93.4 \\ 
				Mixed(1+A$\mid$c)  & 0.072 &  & 0.071 & 0.071 & 0.068 &  & 93.1 & 93.1 & 91.7 & 0.059 &  & 0.057 & 0.057 & 0.055 &  & 94.4 & 94.3 & 93.1 \\
                Mixed(1+A$\mid$c) Sam. & 0.074 &  & 0.071 & 0.071 & 0.068 &  & 93.0 & 93.2 & 91.7 & 0.060 &  & 0.059 & 0.059 & 0.057 &  & 93.6 & 93.7 & 93.4 \\ 
				\bottomrule					
			\end{tabular}%
		}
	\end{center}
	\tiny
	\vspace{0.01cm} 
        \label{table:appe_k_10_nc_50_bin_ch2}
\end{table}

\clearpage
\begin{table}[!h]
	\caption{Simulation results: Monte Carlo standard deviation, average standard errors and coverage probabilities of 95\% confidence intervals for both counterfactual mean on treatment and ATE based on 1000 simulations. Setting: Binary outcome, $k$=5, $n_c$(Avg=100, Min=50, Max=150), and different values of random-effects variance. Results are based on weighting centers equally.}
	\begin{center}
		\resizebox{\textwidth}{!}{
			\begin{tabular}{>{\raggedright\arraybackslash}p{3.5cm}cccccccccccccccccc}
				\toprule
				& \multicolumn{9}{c}{\textbf{Counterfactual mean on treatment}} &\multicolumn{9}{c}{\textbf{ATE}} \\
				\cmidrule(lr){2-10} \cmidrule(lr){11-19}
				&& \multicolumn{4}{c}{\textbf{SE}} & \multicolumn{4}{c}{\textbf{Coverage (\%)}} &&\multicolumn{4}{c}{\textbf{SE}}&\multicolumn{4}{c}{\textbf{Coverage (\%)}} \\
				\cmidrule(lr){3-6} \cmidrule(lr){7-10} \cmidrule(lr){12-15} \cmidrule(lr){16-19}
				Method &  SD & Na\"ive & REML & DL & DB & Na\"ive & REML & DL & DB  & SD & Na\"ive & REML & DL & DB & Naive & REML & DL & DB \\ \midrule
				&&&&& \multicolumn{9}{c}{\textit{Random-effects variances}: $\sigma^2_{b0}$=0, $\sigma^2_{b1}$=0, $\sigma^2_{b2}$=0} &&&&&\\
                \multicolumn{19}{l}{\textbf{Unadjusted}}\\
				Na\"ive & 0.032 & 0.032 &  &  &  & 94.9 &  &  &  & 0.045 & 0.045 &  &  &  & 94.6 &  &  &  \\ 
				Fixed & 0.035 &  & 0.034 & 0.034 & 0.034 &  & 95.1 & 95.3 & 95.3 & 0.047 &  & 0.047 & 0.047 & 0.047 &  & 95.2 & 95.1 & 95.0 \\ 
				Mixed(1$\mid$c) & 0.035 &  & 0.034 & 0.034 & 0.034 &  & 95.5 & 95.6 & 95.4 & 0.047 &  & 0.047 & 0.047 & 0.047 &  & 95.4 & 95.5 & 95.6 \\ 
                Mixed(1$\mid$c) Sam. & 0.034 &  & 0.037 & 0.037 & 0.037 &  & 96.6 & 96.6 & 96.7 & 0.048 &  & 0.052 & 0.052 & 0.052 &  & 96.6 & 96.4 & 96.3 \\ 
				Mixed(1+A$\mid$c) & 0.035 &  & 0.034 & 0.034 & 0.034 &  & 95.5 & 95.6 & 95.4 & 0.047 &  & 0.047 & 0.047 & 0.047 &  & 95.4 & 95.5 & 95.4 \\ 
                Mixed(1+A$\mid$c) Sam. & 0.034 &  & 0.037 & 0.037 & 0.037 &  & 96.6 & 96.6 & 96.7 & 0.048 &  & 0.052 & 0.052 & 0.052 &  & 96.6 & 96.4 & 96.3 \\ 
               \multicolumn{19}{l}{\textbf{Adjusted}}\\
				Na\"ive   & 0.032 & 0.031 &  &  &  & 94.8 &  &  &  & 0.042 & 0.041 &  &  &  & 94.4 &  &  &  \\ 
				Fixed  & 0.034 &  & 0.033 & 0.033 & 0.033 &  & 95.4 & 95.4 & 95.4 & 0.045 &  & 0.044 & 0.044 & 0.044 &  & 94.7 & 94.7 & 94.6 \\ 
				Mixed(1$\mid$c)  & 0.034 &  & 0.033 & 0.033 & 0.033 &  & 95.7 & 95.8 & 95.7 & 0.045 &  & 0.044 & 0.044 & 0.044 &  & 94.8 & 94.8 & 94.9\\ 
                Mixed(1$\mid$c) Sam. & 0.033 &  & 0.036 & 0.036 & 0.036 &  & 96.1 & 96.3 & 96.3 & 0.045 &  & 0.048 & 0.048 & 0.048 &  & 96.4 & 96.5 & 96.6 \\ 
				Mixed(1+A$\mid$c) & 0.034 &  & 0.033 & 0.033 & 0.033 &  & 95.5 & 95.6 & 95.4 & 0.045 &  & 0.044 & 0.044 & 0.044 &  & 95.4 & 95.5 & 95.4 \\ 
                Mixed(1+A$\mid$c) Sam. & 0.033 &  & 0.036 & 0.036 & 0.036 &  & 96.1 & 96.3 & 96.3 & 0.045 &  & 0.049 & 0.049 & 0.048 &  & 96.4 & 96.5 & 96.6 \\ 
				\midrule        
				&&&&& \multicolumn{9}{c}{\textit{Random-effects variances}: $\sigma^2_{b0}$=0.5, $\sigma^2_{b1}$=0, $\sigma^2_{b2}$=0} &&&&&\\
                \multicolumn{19}{l}{\textbf{Unadjusted}}\\
				Na\"ive & 0.076 & 0.031 &  &  &  & 58.3 &  &  &  & 0.044 & 0.044 &  &  &  & 95.4 &  &  &  \\ 
				Fixed & 0.073 &  & 0.068 & 0.067 & 0.063 &  & 90.7 & 90.8 & 90.4 & 0.045 &  & 0.045 & 0.045 & 0.045 &  & 96.2 & 96.3 & 96.5 \\ 
				Mixed(1$\mid$c) & 0.073 &  & 0.068 & 0.067 & 0.063 &  & 90.7 & 90.7 & 90.9 & 0.045 &  & 0.045 & 0.045 & 0.045 &  & 96.2 & 96.3 & 96.1 \\ 
                Mixed(1$\mid$c) Sam. & 0.072 &  & 0.067 & 0.067 & 0.062 &  & 91.5 & 91.4 & 88.9 & 0.048 &  & 0.052 & 0.052 & 0.051 &  & 96.8 & 97.1 & 96.9 \\
				Mixed(1+A$\mid$c) & 0.073 &  & 0.068 & 0.067 & 0.063 &  & 90.7 & 90.8 & 90.9 & 0.046 &  & 0.045 & 0.045 & 0.045 &  & 96.1 & 96.2 & 96.3 \\ 
                  Mixed(1+A$\mid$c) Sam. & 0.072 &  & 0.067 & 0.067 & 0.062 &  & 91.5 & 91.4 & 89.0 & 0.048 &  & 0.052 & 0.052 & 0.051 &  & 96.9 & 97.0 & 96.8 \\ 
                \multicolumn{19}{l}{\textbf{Adjusted}}\\
				Na\"ive   & 0.076 & 0.030 &  &  &  & 57.3 &  &  &  & 0.042 & 0.041 &  &  &  & 94.4 &  &  &  \\ 
				Fixed  & 0.073 &  & 0.067 & 0.067 & 0.062 &  & 91.6 & 91.5 & 90.7 & 0.043 &  & 0.042 & 0.042 & 0.042 &  & 95.5 & 95.5 & 95.5 \\ 
				Mixed(1$\mid$c)  & 0.073 &  & 0.067 & 0.067 & 0.062 &  & 91.2 & 91.5 & 90.4 & 0.043 &  & 0.042 & 0.042 & 0.042 &  & 95.5 & 95.5 & 95.5 \\ 
                Mixed(1$\mid$c) Sam. & 0.071 &  & 0.067 & 0.067 & 0.062 &  & 91.4 & 91.4 & 89.6 & 0.045 &  & 0.049 & 0.049 & 0.048 &  & 97.1 & 97.4 & 97.1 \\
				Mixed(1+A$\mid$c)  & 0.073 &  & 0.067 & 0.067 & 0.062 &  & 90.7 & 90.8 & 90.9 & 0.043 &  & 0.042 & 0.042 & 0.042 &  & 96.1 & 96.2 & 96.3 \\ 
                Mixed(1+A$\mid$c) Sam. & 0.071 &  & 0.067 & 0.067 & 0.062 &  & 91.5 & 91.4 & 89.6 & 0.045 &  & 0.049 & 0.049 & 0.048 &  & 97.0 & 97.4 & 97.0 \\ 
				\midrule
				&&&&& \multicolumn{9}{c}{\textit{Random-effects variances}: $\sigma^2_{b0}$=0.5, $\sigma^2_{b1}$=0.25, $\sigma^2_{b2}$=0.25} &&&&&\\
               \multicolumn{19}{l}{\textbf{Unadjusted}}\\
				Na\"ive & 0.085 & 0.031 &  &  &  & 50.2 &  &  &  & 0.064 & 0.044 &  &  &  & 82.4 &  &  &  \\ 
				Fixed & 0.083 &  & 0.077 & 0.078 & 0.071 &  & 91.3 & 91.3 & 89.1 & 0.064 &  & 0.060 & 0.060 & 0.058 &  & 91.0 & 91.3 & 90.6 \\ 
				Mixed(1$\mid$c) & 0.083 &  & 0.077 & 0.078 & 0.071 &  & 91.4 & 91.5 & 88.9 & 0.064 &  & 0.060 & 0.060 & 0.058 &  & 90.9 & 90.8 & 90.6 \\ 
                Mixed(1$\mid$c) Sam. & 0.084 &  & 0.080 & 0.080 & 0.073 &  & 91.9 & 92.0 & 89.6 & 0.065 &  & 0.063 & 0.063 & 0.061 &  & 93.8 & 93.8 & 93.0 \\ 
				Mixed(1+A$\mid$c) & 0.083 &  & 0.078 & 0.078 & 0.071 &  & 91.3 & 91.5 & 89.0 & 0.064 &  & 0.061 & 0.060 & 0.058 &  & 90.8 & 90.8 & 90.6 \\
                Mixed(1+A$\mid$c) Sam. & 0.084 &  & 0.080 & 0.080 & 0.073 &  & 92.0 & 92.0 & 89.7 & 0.065 &  & 0.063 & 0.063 & 0.061 &  & 93.9 & 94.0 & 93.1 \\ 
                \multicolumn{19}{l}{\textbf{Adjusted}}\\
				Na\"ive & 0.084 & 0.030 &  &  &  & 49.4 &  &  &  & 0.062 & 0.042 &  &  &  & 80.8 &  &  &  \\ 
				Fixed  & 0.082 &  & 0.077 & 0.077 & 0.070 &  & 91.2 & 91.5 & 89.8 & 0.062 &  & 0.058 & 0.058 & 0.055 &  & 90.3 & 90.4 & 90.2 \\ 
				Mixed(1$\mid$c)  & 0.082 &  & 0.077 & 0.077 & 0.070 &  & 91.2 & 91.2 & 89.7 & 0.062 &  & 0.058 & 0.058 & 0.055 &  & 90.5 & 90.2 & 90.1 \\ 
                Mixed(1$\mid$c) Sam. & 0.083 &  & 0.080 & 0.080 & 0.073 &  & 92.4 & 92.4 & 90.3 & 0.062 &  & 0.061 & 0.061 & 0.058 &  & 93.1 & 92.9 & 92.8 \\ 
				Mixed(1+A$\mid$c)  & 0.082 &  & 0.077 & 0.077 & 0.071 &  & 91.3 & 91.5 & 89.0 & 0.062 &  & 0.058 & 0.058 & 0.056 &  & 90.8 & 90.8 & 90.6 \\
                Mixed(1+A$\mid$c) Sam. & 0.083 &  & 0.080 & 0.080 & 0.073 &  & 92.4 & 92.6 & 90.3 & 0.062 &  & 0.061 & 0.061 & 0.058 &  & 93.1 & 92.9 & 92.8 \\ 
				\midrule
				&&&&& \multicolumn{9}{c}{\textit{Random-effects variances}: $\sigma^2_{b0}$=0.75, $\sigma^2_{b1}$=0.5, $\sigma^2_{b2}$=0.5} &&&&&\\
                \multicolumn{19}{l}{\textbf{Unadjusted}}\\
				Na\"ive & 0.104 & 0.031 &  &  &  & 43.0 &  &  &  & 0.073 & 0.044 &  &  &  & 76.8 &  &  &  \\ 
				Fixed & 0.100 &  & 0.094 & 0.094 & 0.085 &  & 91.7 & 91.9 & 89.9 & 0.072 &  & 0.071 & 0.071 & 0.067 &  & 93.7 & 93.4 & 92.4 \\ 
				Mixed(1$\mid$c) & 0.100 &  & 0.094 & 0.094 & 0.085 &  & 91.6 & 91.9 & 89.8 & 0.072 &  & 0.071 & 0.071 & 0.067 &  & 94.0 & 93.6 & 92.6 \\
                Mixed(1$\mid$c) Sam. & 0.096 &  & 0.093 & 0.093 & 0.084 &  & 92.7 & 92.4 & 90.5 & 0.073 &  & 0.071 & 0.071 & 0.067 &  & 92.5 & 92.9& 91.9 \\ 
				Mixed(1+A$\mid$c) & 0.099 &  & 0.094 & 0.094 & 0.085 &  & 91.8 & 92.0 & 89.8 & 0.072 &  & 0.072 & 0.071 & 0.067 &  & 94.0 & 93.8 & 92.6 \\
                Mixed(1+A$\mid$c) Sam. & 0.096 &  & 0.093 & 0.093 & 0.084 &  & 92.7 & 92.4 & 90.4 & 0.073 &  & 0.071 & 0.071 & 0.067 &  & 92.5 & 92.9 & 91.9 \\ 
                \multicolumn{19}{l}{\textbf{Adjusted}}\\
				Na\"ive  & 0.104 & 0.030 &  &  &  & 41.7 &  &  &  & 0.072 & 0.042 &  &  &  & 73.0 &  &  &  \\ 
				Fixed   & 0.099 &  & 0.093 & 0.093 & 0.084 &  & 92.1 & 92.3 & 89.9 & 0.070 &  & 0.069 & 0.069 & 0.065 &  & 92.3 & 92.4 & 91.8 \\ 
				Mixed(1$\mid$c)  & 0.099 &  & 0.093 & 0.093 & 0.084 &  & 91.8 & 92.2 & 90.0 & 0.070 &  & 0.069 & 0.069 & 0.065 &  & 92.5 & 92.4 & 91.7 \\
                Mixed(1$\mid$c) Sam. & 0.096 &  & 0.092 & 0.092 & 0.084 &  & 92.3 & 92.1 & 90.9 & 0.071 &  & 0.069 & 0.069 & 0.065 &  & 92.3 & 92.4 & 91.9 \\ 
				Mixed(1+A$\mid$c)  & 0.099 &  & 0.093 & 0.093 & 0.085 &  & 91.8 & 92.0 & 89.8 & 0.070 &  & 0.069 & 0.069 & 0.065 &  & 94.0 & 93.8 & 92.6 \\
                Mixed(1+A$\mid$c) Sam. & 0.096 &  & 0.093 & 0.092 & 0.084 &  & 92.4 & 92.1 & 90.9 & 0.071 &  & 0.070 & 0.070 & 0.066 &  & 92.4 & 92.5 & 91.8 \\ 
				\midrule											
			\end{tabular}%
		}
	\end{center}
	\tiny
	\vspace{0.01cm} 
        \label{table:appe_k_5_nc_100_bin_ch2}
\end{table}

\clearpage
\begin{table}[!h]
\caption{Simulation results: Monte Carlo standard deviation, average standard errors and coverage probabilities of 95\% confidence intervals for both counterfactual mean on treatment and ATE based on 1000 simulations. Setting: Binary outcome with a misspecified outcome model, $k$=100, $n_c$: Avg=5, Min=1, Max=24. Results are based on weighting centers equally.}
	\begin{center}
		\resizebox{\textwidth}{!}{
			\begin{tabular}{>{\raggedright\arraybackslash}p{3.5cm}cccccccccccccccccc}
				\toprule
				& \multicolumn{9}{c}{\textbf{Counterfactual mean on treatment}} &\multicolumn{9}{c}{\textbf{ATE}} \\
				\cmidrule(lr){2-10} \cmidrule(lr){11-19}
				&& \multicolumn{4}{c}{\textbf{SE}} & \multicolumn{4}{c}{\textbf{Coverage (\%)}} &&\multicolumn{4}{c}{\textbf{SE}}&\multicolumn{4}{c}{\textbf{Coverage (\%)}} \\
				\cmidrule(lr){3-6} \cmidrule(lr){7-10} \cmidrule(lr){12-15} \cmidrule(lr){16-19}
				Method &  SD & Na\"ive & REML & DL & DB & Na\"ive & REML & DL & DB  & SD & Na\"ive & REML & DL & DB & Naive & REML & DL & DB \\ \midrule
				&&&&& \multicolumn{9}{c}{\textit{Random-effects variances}: $\sigma^2_{b0}$=0.5, $\sigma^2_{b1}$=0.5, $\sigma^2_{b2}=0$: $\bar{\rho}_1=0.0983$, $\bar{\rho}=0.0221$} &&&&&\\
                \multicolumn{19}{l}{\textbf{Unadjusted}}\\
				Na\"ive & 0.038 & 0.031 &  &  &  & 89.0 &  &  &  & 0.047 & 0.044 &  &  &  & 93.4 &  &  &  \\ 
                Fixed & 0.039 &  & 0.040 & 0.040 & 0.038 &  & 95.5 & 95.3 & 94.3 & 0.047 &  & 0.048 & 0.047 & 0.047 &  & 95.8 & 95.6 & 95.0 \\ 
                Mixed(1$\mid$c) & 0.041 &  & 0.042 & 0.042 & 0.041 &  & 95.0 & 94.9 & 94.7 & 0.055 &  & 0.053 & 0.053 & 0.052 &  & 94.3 & 94.4 & 94.4 \\ 
                Mixed(1$\mid$c) Sam. & 0.041 &  & 0.042 & 0.042 & 0.042 &  & 95.4 & 95.5 & 94.7 & 0.054 &  & 0.056 & 0.056 & 0.056 &  & 95.0 & 95.0 & 94.9 \\ 
                Mixed(1+A$\mid$c) & 0.041 &  & 0.041 & 0.041 & 0.041 &  & 95.0 & 95.0 & 94.7 & 0.055 &  & 0.053 & 0.053 & 0.053 &  & 94.3 & 94.5 & 94.4 \\ 
                Mixed(1+A$\mid$c) Sam. & 0.041 &  & 0.042 & 0.042 & 0.042 &  & 95.4 & 95.5 & 94.7 & 0.055 &  & 0.056 & 0.056 & 0.056 &  & 94.9 & 95.0 & 95.0 \\ 
                \multicolumn{19}{l}{\textbf{Adjusted}}\\
                Na\"ive  & 0.037 & 0.031 &  &  &  & 89.8 &  &  &  & 0.045 & 0.042 &  &  &  & 93.4 &  &  &  \\ 
                Fixed  & 0.039 &  & 0.039 & 0.039 & 0.037 &  & 95.6 & 95.3 & 94.0 & 0.044 &  & 0.045 & 0.044 & 0.043 &  & 95.2 & 94.5 & 94.2 \\ 
                Mixed(1$\mid$c)  & 0.041 &  & 0.041 & 0.041 & 0.041 &  & 94.5 & 94.3 & 94.6 & 0.051 &  & 0.049 & 0.049 & 0.049 &  & 95.0 & 95.0 & 94.8 \\ 
                Mixed(1$\mid$c) Sam. & 0.041 &  & 0.041 & 0.041 & 0.041 &  & 94.7 & 94.7 & 95.0 & 0.052 &  & 0.053 & 0.053 & 0.053 &  & 95.8 & 95.9 & 96.2 \\ 
                Mixed(1+A$\mid$c)  & 0.040 &  & 0.040 & 0.040 & 0.040 &  & 95.0 & 95.0 & 94.7 & 0.051 &  & 0.050 & 0.050 & 0.050 &  & 94.3 & 94.5 & 94.4 \\ 
                 Mixed(1+A$\mid$c) Sam. & 0.041 &  & 0.041 & 0.041 & 0.041 &  & 94.8 & 94.7 & 95.0 & 0.052 &  & 0.053 & 0.053 & 0.053 &  & 95.8 & 95.8 & 96.1 \\ 
				\midrule												
			\end{tabular}%
		}
	\end{center}
	\tiny
	\vspace{0.01cm} 
  \label{table:appe_k_100_nc_5_mis_bin_ch2}
\end{table}

\autoref{table:appe_k_100_nc_5_bias_MSE_bin_ch2}, \autoref{table:appe_k_50_nc_10_bias_MSE_bin_ch2}, \autoref{table:appe_k_10_nc_50_bias_MSE_bin_ch2}, \autoref{table:appe_k_5_nc_100_bias_MSE_bin_ch2} and \autoref{table:appe_k_100_nc_100_bias_MSE_bin_ch2} show the estimated bias and mean square error for both counterfactual mean on treatment and ATE for simulation Settings 1 through 5 with binary outcomes, respectively. Fixed-effects estimators lead to biased estimates of both the counterfactual mean on treatment and the ATE for Setting 1 (many small centers). Further, \autoref{table:appe_k_100_nc_5_mis_bias_MSE_bin_ch2} presents the estimated bias and mean square error when the outcome model is misspecified under Setting 1.

\clearpage
\begin{table}[!h]
	\caption{Simulation results of estimated bias and mean square error (MSE) for both counterfactual mean on treatment and ATE based on 1000 simulations. Setting: Binary outcome, $k$=100, $n_c$(Avg=5, Min=1, Max=24), and different values of random-effects variance. Results are based on weighting centers equally.}
	\begin{center}
		\resizebox{\textwidth}{!}{
			\begin{tabular}{ccccccccccccc}
				\toprule
				& \multicolumn{12}{c}{\textbf{Counterfactual mean on treatment}}\\
				\cmidrule(lr){2-13} 
                & \multicolumn{6}{c}{\textbf{Unadjusted}} &\multicolumn{6}{c}{\textbf{Adjusted}}\\
				\cmidrule(lr){2-7} \cmidrule(lr){8-13} 
				Measure &  Na\"ive & Fixed & Mixed(1$\mid$c) & Mixed(1$\mid$c) Sam. & Mixed(1+A$\mid$c) & Mixed(1+A$\mid$c) Sam. & Na\"ive  & Fixed  & Mixed(1$\mid$c) & Mixed(1$\mid$c) Sam. & Mixed(1+A$\mid$c) & Mixed(1+A$\mid$c) Sam. \\
				\midrule
				\multicolumn{12}{c}{\textit{Random-effects variances}: $\sigma^2_{b0}$=0, $\sigma^2_{b1}$=0, $\sigma^2_{b2}$=0}\\
                Bias & -0.0006 & -0.0048 & -0.0012 & -0.0009 & -0.0012 & -0.0008 & -0.0009 & -0.0049 & -0.0017 & -0.0014 & -0.0017 & -0.0014 \\ 
                MSE  & 0.0009 & 0.0013 & 0.0013 & 0.0015 & 0.0015 & 0.0015 & 0.0009 & 0.0013 & 0.0013 & 0.0014 & 0.0014 & 0.0014 \\ 
				\midrule
				\multicolumn{12}{c}{\textit{Random-effects variances}: $\sigma^2_{b0}$=0.5, $\sigma^2_{b1}$=0, $\sigma^2_{b2}$=0}\\
                Bias & -0.0001 & -0.0038 & 0.0001 & 0.0009 & 0.0000 & 0.0009 & -0.0002 & -0.0040 & -0.0001 & 0.0009 & -0.0002 & 0.0009 \\ 
                 MSE  & 0.0012 & 0.0014 & 0.0014 & 0.0017 & 0.0016 & 0.0017 & 0.0012 & 0.0014 & 0.0014 & 0.0016 & 0.0015 & 0.0016 \\ 
				\midrule
				\multicolumn{12}{c}{\textit{Random-effects variances}: $\sigma^2_{b0}$=0.75, $\sigma^2_{b1}$=0.5, $\sigma^2_{b2}$=0.5} \\
                Bias & -0.0002 & -0.0039 & -0.0000 & 0.0019 & 0.0000 & 0.0018 & -0.0000 & -0.0040 & -0.0003 & 0.0015 & -0.0002 & 0.0015 \\ 
                MSE  & 0.0016 & 0.0017 & 0.0017 & 0.0018 & 0.0019 & 0.0017 & 0.0016 & 0.0016 & 0.0016 & 0.0017 & 0.0019 & 0.0017 \\ 
				\midrule
               & \multicolumn{12}{c}{\textbf{ATE}}\\
				\cmidrule(lr){2-13} 
                & \multicolumn{6}{c}{\textbf{Unadjusted}} &\multicolumn{6}{c}{\textbf{Adjusted}}\\
				\cmidrule(lr){2-7} \cmidrule(lr){8-13} 
				Measure &  Na\"ive & Fixed & Mixed(1$\mid$c) & Mixed(1$\mid$c) Sam. & Mixed(1+A$\mid$c) & Mixed(1+A$\mid$c) Sam. & Na\"ive  & Fixed  & Mixed(1$\mid$c) & Mixed(1$\mid$c) Sam. & Mixed(1+A$\mid$c) & Mixed(1+A$\mid$c) Sam. \\
				\midrule
				\multicolumn{12}{c}{\textit{Random-effects variances}: $\sigma^2_{b0}$=0, $\sigma^2_{b1}$=0, $\sigma^2_{b2}$=0}\\
                Bias & -0.0006 & -0.0086 & -0.0013 & 0.0010 & -0.0012 & 0.0010 & -0.0009 & -0.0087 & -0.0021 & 0.0001 & -0.0021 & 0.0001 \\ 
                 MSE  & 0.0019 & 0.0023 & 0.0023 & 0.0029 & 0.0029 & 0.0029 & 0.0017 & 0.0021 & 0.0021 & 0.0025 & 0.0027 & 0.0025 \\ 
				\midrule
				\multicolumn{12}{c}{\textit{Random-effects variances}: $\sigma^2_{b0}$=0.5, $\sigma^2_{b1}$=0, $\sigma^2_{b2}$=0}\\
                Bias & -0.0001 & -0.0073 & 0.0003 & -0.0013 & 0.0003 & -0.0013  & -0.0002 & -0.0074 & 0.0000 & -0.0011 & -0.0000 & -0.0011 \\ 
                MSE  & 0.0019 & 0.0020 & 0.0020 & 0.0030 & 0.0027 & 0.0030 & 0.0017 & 0.0018 & 0.0018 & 0.0026 & 0.0024 & 0.0026 \\ 
				\midrule
				\multicolumn{12}{c}{\textit{Random-effects variances}: $\sigma^2_{b0}$=0.75, $\sigma^2_{b1}$=0.5, $\sigma^2_{b2}$=0.5}\\
                Bias & -0.0017 & -0.0082 & -0.0004 & 0.0013 & -0.0003 & 0.0013 & -0.0015 & -0.0083 & -0.0009 & 0.0007 & -0.0009 & 0.0007 \\ 
                MSE  & 0.0022 & 0.0022 & 0.0022 & 0.0030 & 0.0029 & 0.0030 & 0.0019 & 0.0019 & 0.0019 & 0.0027 & 0.0025 & 0.0027 \\ 
				\midrule
			\end{tabular}
		}
	\end{center}
    	\tiny
	\vspace{0.01cm} 
    \label{table:appe_k_100_nc_5_bias_MSE_bin_ch2}
\end{table}

\clearpage
\begin{table}[!h]
	\caption{Simulation results of estimated bias and mean square error (MSE) for both counterfactual mean on treatment and ATE based on 1000 simulations. Setting: Binary outcome, $k$=50, $n_c$(Avg=10, Min=2, Max=48), and different values of random-effects variance. Results are based on weighting centers equally.}
	\begin{center}
		\resizebox{\textwidth}{!}{
			\begin{tabular}{ccccccccccccc}
				\toprule
				& \multicolumn{12}{c}{\textbf{Counterfactual mean on treatment}}\\
				\cmidrule(lr){2-13} 
                & \multicolumn{6}{c}{\textbf{Unadjusted}} &\multicolumn{6}{c}{\textbf{Adjusted}}\\
				\cmidrule(lr){2-7} \cmidrule(lr){8-13} 
				Measure &  Na\"ive & Fixed & Mixed(1$\mid$c) & Mixed(1$\mid$c) Sam. & Mixed(1+A$\mid$c) & Mixed(1+A$\mid$c) Sam. & Na\"ive  & Fixed  & Mixed(1$\mid$c) & Mixed(1$\mid$c) Sam. & Mixed(1+A$\mid$c) & Mixed(1+A$\mid$c) Sam. \\
				\midrule
				\multicolumn{12}{c}{\textit{Random-effects variances}: $\sigma^2_{b0}$=0, $\sigma^2_{b1}$=0, $\sigma^2_{b2}$=0}\\
                Bias & 0.0004 & -0.0002 & 0.0005 & 0.0008 & 0.0005 & 0.0008 & 0.0005 & -0.0002 & 0.0004 & 0.0006 & 0.0004 & 0.0006 \\ 
                MSE  & 0.0010 & 0.0013 & 0.0013 & 0.0012 & 0.0013 & 0.0012 & 0.0009 & 0.0012 & 0.0012 & 0.0012 & 0.0013 & 0.0012 \\ 
				\midrule
				\multicolumn{12}{c}{\textit{Random-effects variances}: $\sigma^2_{b0}$=0.5, $\sigma^2_{b1}$=0, $\sigma^2_{b2}$=0} \\
                Bias & 0.0017 & 0.0011 & 0.0013 & -0.0018 & 0.0013 & -0.0018 & 0.0019 & 0.0012 & 0.0014 & -0.0018 & 0.0014 & -0.0018 \\ 
                MSE  & 0.0016 & 0.0016 & 0.0016 & 0.0017 & 0.0017 & 0.0017 & 0.0015 & 0.0015 & 0.0015 & 0.0017 & 0.0016 & 0.0017 \\ 
				\midrule
				\multicolumn{12}{c}{\textit{Random-effects variances}: $\sigma^2_{b0}$=0.75, $\sigma^2_{b1}$=0.5, $\sigma^2_{b2}$=0.5} \\
                Bias & -0.0000 & -0.0002 & 0.0001 & 0.0001 & 0.0000 & 0.0001 & 0.0001 & -0.0002 & 0.0003 & 0.0001 & 0.0002 & 0.0001 \\ 
                MSE  & 0.0022 & 0.0020 & 0.0020 & 0.0022 & 0.0020 & 0.0022 & 0.0022 & 0.0020 & 0.0020 & 0.0022 & 0.0020 & 0.0022 \\ 
				\midrule
                & \multicolumn{12}{c}{\textbf{ATE}}\\
				\cmidrule(lr){2-13} 
                & \multicolumn{6}{c}{\textbf{Unadjusted}} &\multicolumn{6}{c}{\textbf{Adjusted}}\\
				\cmidrule(lr){2-7} \cmidrule(lr){8-13} 
				Measure &  Na\"ive & Fixed & Mixed(1$\mid$c) & Mixed(1$\mid$c) Sam. & Mixed(1+A$\mid$c) & Mixed(1+A$\mid$c) Sam. & Na\"ive  & Fixed  & Mixed(1$\mid$c) & Mixed(1$\mid$c) Sam. & Mixed(1+A$\mid$c) & Mixed(1+A$\mid$c) Sam. \\
				\midrule
				\multicolumn{12}{c}{\textit{Random-effects variances}: $\sigma^2_{b0}$=0, $\sigma^2_{b1}$=0, $\sigma^2_{b2}$=0}\\
                Bias & -0.0007 & -0.0020 & -0.0006 & 0.0005 & -0.0005 & 0.0004 & -0.0006 & -0.0018 & -0.0006 & 0.0001 & -0.0007 & 0.0001 \\ 
                MSE  & 0.0020 & 0.0025 & 0.0025 & 0.0028 & 0.0025 & 0.0028 & 0.0017 & 0.0022 & 0.0022 & 0.0023 & 0.0022 & 0.0023 \\ 
				\midrule
				\multicolumn{12}{c}{\textit{Random-effects variances}: $\sigma^2_{b0}$=0.5, $\sigma^2_{b1}$=0, $\sigma^2_{b2}$=0}\\
                Bias & -0.0005 & -0.0008 & -0.0003 & -0.0013 & -0.0003 & -0.0013 & -0.0000 & -0.0007 & -0.0002 & -0.0011 & -0.0001 & -0.0011 \\ 
                 MSE  & 0.0019 & 0.0024 & 0.0024 & 0.0026 & 0.0026 & 0.0026 & 0.0018 & 0.0021 & 0.0021 & 0.0023 & 0.0023 & 0.0023 \\ 
				\midrule
				\multicolumn{12}{c}{\textit{Random-effects variances}: $\sigma^2_{b0}$=0.75, $\sigma^2_{b1}$=0.5, $\sigma^2_{b2}$=0.5}\\
                Bias & -0.0008 & -0.0014 & -0.0007 & 0.0013 & -0.0006 & 0.0013 & -0.0006 & -0.0015 & -0.0002 & 0.0012 & -0.0002 & 0.0011 \\ 
                 MSE  & 0.0023 & 0.0025 & 0.0025 & 0.0031 & 0.0025 & 0.0031 & 0.0021 & 0.0023 & 0.0023 & 0.0028 & 0.0023 & 0.0028 \\ 
				\midrule	
			\end{tabular}
		}
	\end{center}
    	\tiny
	\vspace{0.01cm} 
        \label{table:appe_k_50_nc_10_bias_MSE_bin_ch2}
\end{table}

\clearpage
\begin{table}[!h]
	\caption{Simulation results of estimated bias and mean square error (MSE) for both counterfactual mean on treatment and ATE based on 1000 simulations. Setting: Binary outcome, $k$=10, $n_c$(Avg=50, Min=25, Max=80), and different values of random-effects variance. Results are based on weighting centers equally.}
	\begin{center}
		\resizebox{\textwidth}{!}{
			\begin{tabular}{ccccccccccccc}
				\toprule
				& \multicolumn{12}{c}{\textbf{Counterfactual mean on treatment}}\\
				\cmidrule(lr){2-13} 
                & \multicolumn{6}{c}{\textbf{Unadjusted}} &\multicolumn{6}{c}{\textbf{Adjusted}}\\
				\cmidrule(lr){2-7} \cmidrule(lr){8-13} 
				Measure &  Na\"ive & Fixed & Mixed(1$\mid$c) & Mixed(1$\mid$c) Sam. & Mixed(1+A$\mid$c) & Mixed(1+A$\mid$c) Sam. & Na\"ive  & Fixed  & Mixed(1$\mid$c) & Mixed(1$\mid$c) Sam. & Mixed(1+A$\mid$c) & Mixed(1+A$\mid$c) Sam. \\
				\midrule
				\multicolumn{12}{c}{\textit{Random-effects variances}: $\sigma^2_{b0}$=0, $\sigma^2_{b1}$=0, $\sigma^2_{b2}$=0}\\
                Bias & 0.0012 & 0.0009 & 0.0009 & -0.0005 & 0.0009 & -0.0005 & 0.0010 & 0.0006 & 0.0006 & -0.0005 & 0.0006 & -0.0005 \\ 
                MSE  & 0.0011 & 0.0012 & 0.0012 & 0.0011 & 0.0012 & 0.0011 & 0.0010 & 0.0011 & 0.0011 & 0.0010 & 0.0011 & 0.0010 \\ 
				\midrule
				\multicolumn{12}{c}{\textit{Random-effects variances}: $\sigma^2_{b0}$=0.5, $\sigma^2_{b1}$=0, $\sigma^2_{b2}$=0}\\
                Bias & -0.0005 & 0.0000 & 0.0000 & 0.0010 & 0.0000 & 0.0010 & -0.0008 & -0.0003 & -0.0004 & 0.0006 & -0.0003 & 0.0005 \\ 
                MSE  & 0.0031 & 0.0030 & 0.0030 & 0.0030 & 0.0030 & 0.0030 & 0.0030 & 0.0029 & 0.0029 & 0.0029 & 0.0029 & 0.0029 \\ 
				\midrule
				\multicolumn{12}{c}{\textit{Random-effects variances}: $\sigma^2_{b0}$=0.75, $\sigma^2_{b1}$=0.5, $\sigma^2_{b2}$=0.5}\\
                Bias & -0.0030 & -0.0021 & -0.0021 & -0.0002 & -0.0021 & -0.0002 & -0.0025 & -0.0015 & -0.0015 & -0.0004 & -0.0016 & -0.0004 \\ 
                 MSE  & 0.0058 & 0.0053 & 0.0053 & 0.0055 & 0.0053 & 0.0055 & 0.0057 & 0.0052 & 0.0052 & 0.0055 & 0.0053 & 0.0055 \\ 
				\midrule	
               & \multicolumn{12}{c}{\textbf{ATE}}\\
				\cmidrule(lr){2-13} 
                & \multicolumn{6}{c}{\textbf{Unadjusted}} &\multicolumn{6}{c}{\textbf{Adjusted}}\\
				\cmidrule(lr){2-7} \cmidrule(lr){8-13} 
				Measure &  Na\"ive & Fixed & Mixed(1$\mid$c) & Mixed(1$\mid$c) Sam. & Mixed(1+A$\mid$c) & Mixed(1+A$\mid$c) Sam. & Na\"ive  & Fixed  & Mixed(1$\mid$c) & Mixed(1$\mid$c) Sam. & Mixed(1+A$\mid$c) & Mixed(1+A$\mid$c) Sam. \\
				\midrule
				\multicolumn{12}{c}{\textit{Random-effects variances}: $\sigma^2_{b0}$=0, $\sigma^2_{b1}$=0, $\sigma^2_{b2}$=0}\\
                Bias & 0.0014 & 0.0012 & 0.0012 & 0.0003 & 0.0012 & 0.0003 & 0.0011 & 0.0008 & 0.0008 & 0.0002 & 0.0008 & 0.0002 \\ 
               MSE  & 0.0022 & 0.0024 & 0.0024 & 0.0021 & 0.0024 & 0.0021 & 0.0018 & 0.0020 & 0.0020 & 0.0018 & 0.0020 & 0.0018 \\ 
				\midrule
				\multicolumn{12}{c}{\textit{Random-effects variances}: $\sigma^2_{b0}$=0.5, $\sigma^2_{b1}$=0, $\sigma^2_{b2}$=0}\\
                Bias & -0.0003 & -0.0001 & -0.0002 & -0.0003 & -0.0001 & -0.0003 & -0.0009 & -0.0007 & -0.0008 & -0.0008 & -0.0008 & -0.0008 \\ 
                 MSE  & 0.0018 & 0.0018 & 0.0018 & 0.0021 & 0.0018 & 0.0021 & 0.0016 & 0.0015 & 0.0015 & 0.0018 & 0.0015 & 0.0018 \\ 
				\midrule
				\multicolumn{12}{c}{\textit{Random-effects variances}: $\sigma^2_{b0}$=0.75, $\sigma^2_{b1}$=0.5, $\sigma^2_{b2}$=0.5}\\
                Bias & -0.0026 & -0.0018 & -0.0018 & -0.0014 & -0.0018 & -0.0014 & -0.0018 & -0.0008 & -0.0009 & -0.0018 & -0.0008 & -0.0018 \\ 
                 MSE  & 0.0040 & 0.0038 & 0.0038 & 0.0038 & 0.0038 & 0.0038 & 0.0038 & 0.0035 & 0.0035 & 0.0036 & 0.0035 & 0.0036 \\ 
				\midrule			
			\end{tabular}
		}
	\end{center}
    	\tiny
	\vspace{0.01cm} 
        \label{table:appe_k_10_nc_50_bias_MSE_bin_ch2}
\end{table}

\clearpage
\begin{table}[!h]
	\caption{Simulation results of estimated bias and mean square error (MSE) for both counterfactual mean on treatment and ATE based on 1000 simulations. Setting: Binary outcome, $k$=5, $n_c$(Avg=100, Min=50, Max=150), and different values of random-effects variance. Results are based on weighting centers equally.}
	\begin{center}
		\resizebox{\textwidth}{!}{
			\begin{tabular}{ccccccccccccc}
				\toprule
				& \multicolumn{12}{c}{\textbf{Counterfactual mean on treatment}}\\
				\cmidrule(lr){2-13} 
                & \multicolumn{6}{c}{\textbf{Unadjusted}} &\multicolumn{6}{c}{\textbf{Adjusted}}\\
				\cmidrule(lr){2-7} \cmidrule(lr){8-13} 
				Measure &  Na\"ive & Fixed & Mixed(1$\mid$c) & Mixed(1$\mid$c) Sam. & Mixed(1+A$\mid$c) & Mixed(1+A$\mid$c) Sam. & Na\"ive  & Fixed  & Mixed(1$\mid$c) & Mixed(1$\mid$c) Sam. & Mixed(1+A$\mid$c) & Mixed(1+A$\mid$c) Sam. \\
				\midrule
				\multicolumn{12}{c}{\textit{Random-effects variances}: $\sigma^2_{b0}$=0, $\sigma^2_{b1}$=0, $\sigma^2_{b2}$=0}\\
                Bias & 0.0007 & 0.0004 & 0.0003 & 0.0014 & 0.0004 & 0.0014 & 0.0010 & 0.0007 & 0.0007 & 0.0016 & 0.0007 & 0.0016 \\ 
               MSE  & 0.0010 & 0.0012 & 0.0012 & 0.0011 & 0.0012 & 0.0011 & 0.0010 & 0.0012 & 0.0012 & 0.0011 & 0.0012 & 0.0011 \\ 
				\midrule
				\multicolumn{12}{c}{\textit{Random-effects variances}: $\sigma^2_{b0}$=0.5, $\sigma^2_{b1}$=0, $\sigma^2_{b2}$=0}\\
                Bias & 0.0034 & 0.0019 & 0.0019 & 0.0023 & 0.0020 & 0.0023 & 0.0037 & 0.0023 & 0.0023 & 0.0021 & 0.0023 & 0.0021 \\ 
                 MSE  & 0.0058 & 0.0054 & 0.0054 & 0.0052 & 0.0054 & 0.0052 & 0.0058 & 0.0053 & 0.0053 & 0.0051 & 0.0053 & 0.0051 \\ 
				\midrule
				\multicolumn{12}{c}{\textit{Random-effects variances}: $\sigma^2_{b0}$=0.75, $\sigma^2_{b1}$=0.5, $\sigma^2_{b2}$=0.5}\\
                Bias & 0.0021 & 0.0011 & 0.0011 & 0.0092 & 0.0011 & 0.0092 & 0.0023 & 0.0014 & 0.0014 & 0.0089 & 0.0014 & 0.0089 \\ 
               MSE  & 0.0109 & 0.0099 & 0.0099 & 0.0093 & 0.0099 & 0.0093 & 0.0108 & 0.0098 & 0.0098 & 0.0092 & 0.0098 & 0.0092 \\ 
				\midrule		
               & \multicolumn{12}{c}{\textbf{ATE}}\\
				\cmidrule(lr){2-13} 
                & \multicolumn{6}{c}{\textbf{Unadjusted}} &\multicolumn{6}{c}{\textbf{Adjusted}}\\
				\cmidrule(lr){2-7} \cmidrule(lr){8-13} 
				Measure &  Na\"ive & Fixed & Mixed(1$\mid$c) & Mixed(1$\mid$c) Sam. & Mixed(1+A$\mid$c) & Mixed(1+A$\mid$c) Sam. & Na\"ive  & Fixed  & Mixed(1$\mid$c) & Mixed(1$\mid$c) Sam. & Mixed(1+A$\mid$c) & Mixed(1+A$\mid$c) Sam. \\
				\midrule
				\multicolumn{12}{c}{\textit{Random-effects variances}: $\sigma^2_{b0}$=0, $\sigma^2_{b1}$=0, $\sigma^2_{b2}$=0}\\
                Bias & -0.0010 & -0.0013 & -0.0014 & 0.0019 & -0.0014 & 0.0019 & -0.0006 & -0.0008 & -0.0008 & 0.0023 & -0.0008 & 0.0023 \\ 
                 MSE  & 0.0020 & 0.0022 & 0.0022 & 0.0023 & 0.0022 & 0.0023 & 0.0018 & 0.0020 & 0.0020 & 0.0020 & 0.0020 & 0.0020 \\ 
				\midrule
				\multicolumn{12}{c}{\textit{Random-effects variances}: $\sigma^2_{b0}$=0.5, $\sigma^2_{b1}$=0, $\sigma^2_{b2}$=0}\\
                Bias & 0.0006 & 0.0007 & 0.0007 & 0.0007 & 0.0007 & 0.0007 & 0.0012 & 0.0014 & 0.0014 & 0.0005 & 0.0013 & 0.0005 \\ 
                 MSE  & 0.0019 & 0.0021 & 0.0021 & 0.0023 & 0.0021 & 0.0023 & 0.0017 & 0.0018 & 0.0018 & 0.0020 & 0.0018 & 0.0020 \\ 
				\midrule
				\multicolumn{12}{c}{\textit{Random-effects variances}: $\sigma^2_{b0}$=0.75, $\sigma^2_{b1}$=0.5, $\sigma^2_{b2}$=0.5}\\
                Bias & 0.0003 & -0.0006 & -0.0007 & 0.0037 & -0.0007 & 0.0037 & 0.0008 & -0.0000 & -0.0001 & 0.0032 & -0.0001 & 0.0032 \\ 
               MSE  & 0.0053 & 0.0051 & 0.0051 & 0.0053 & 0.0051 & 0.0053 & 0.0051 & 0.0049 & 0.0049 & 0.0050 & 0.0049 & 0.0050 \\ 
				\midrule		
			\end{tabular}
		}
	\end{center}
    	\tiny
	\vspace{0.01cm} 
        \label{table:appe_k_5_nc_100_bias_MSE_bin_ch2}
\end{table}

\clearpage
\begin{table}[!h]
	\caption{Simulation results of estimated bias and mean square error (MSE) for both counterfactual mean on treatment and ATE based on 1000 simulations. Setting: Binary outcome, $k$=100, $n_c$(Avg=100, Min=50, Max=145), and only at higher values of random-effects variance. Results are based on weighting centers equally.}
	\begin{center}
		\resizebox{\textwidth}{!}{
			\begin{tabular}{ccccccccccccc}
				\toprule
				& \multicolumn{12}{c}{\textbf{Counterfactual mean on treatment}}\\
				\cmidrule(lr){2-13} 
                & \multicolumn{6}{c}{\textbf{Unadjusted}} &\multicolumn{6}{c}{\textbf{Adjusted}}\\
				\cmidrule(lr){2-7} \cmidrule(lr){8-13} 
				Measure &  Na\"ive & Fixed & Mixed(1$\mid$c) & Mixed(1$\mid$c) Sam. & Mixed(1+A$\mid$c) & Mixed(1+A$\mid$c) Sam. & Na\"ive  & Fixed  & Mixed(1$\mid$c) & Mixed(1$\mid$c) Sam. & Mixed(1+A$\mid$c) & Mixed(1+A$\mid$c) Sam. \\
				\midrule
				\multicolumn{12}{c}{\textit{Random-effects variances}: $\sigma^2_{b0}$=0.75, $\sigma^2_{b1}$=0.5, $\sigma^2_{b2}$=0.5}\\
                Bias & 0.0010 & 0.0008 & 0.0008 & 0.0004 & 0.0008 & 0.0003 & 0.0010 & 0.0007 & 0.0007 & 0.0003 & 0.0007 & 0.0003 \\ 
                MSE  & 0.0005 & 0.0004 & 0.0004 & 0.0005 & 0.0004 & 0.0005 & 0.0005 & 0.0004 & 0.0004 & 0.0005 & 0.0004 & 0.0005 \\ 
				\midrule
                & \multicolumn{12}{c}{\textbf{Counterfactual mean on treatment}}\\
				\cmidrule(lr){2-13} 
                & \multicolumn{6}{c}{\textbf{Unadjusted}} &\multicolumn{6}{c}{\textbf{Adjusted}}\\
				\cmidrule(lr){2-7} \cmidrule(lr){8-13} 
				Measure &  Na\"ive & Fixed & Mixed(1$\mid$c) & Mixed(1$\mid$c) Sam. & Mixed(1+A$\mid$c) & Mixed(1+A$\mid$c) Sam. & Na\"ive  & Fixed  & Mixed(1$\mid$c) & Mixed(1$\mid$c) Sam. & Mixed(1+A$\mid$c) & Mixed(1+A$\mid$c) Sam. \\
				\midrule
				\multicolumn{12}{c}{\textit{Random-effects variances}: $\sigma^2_{b0}$=0.75, $\sigma^2_{b1}$=0.5, $\sigma^2_{b2}$=0.5}\\
                Bias & 0.0010 & 0.0007 & 0.0007 & -0.0008 & 0.0007 & -0.0008 & 0.0008 & 0.0005 & 0.0005 & -0.0009 & 0.0005 & -0.0009 \\ 
                MSE  & 0.0003 & 0.0002 & 0.0002 & 0.0003 & 0.0002 & 0.0003 & 0.0003 & 0.0002 & 0.0002 & 0.0002 & 0.0002 & 0.0002 \\ 
				\midrule
			\end{tabular}
		}
	\end{center}
    	\tiny
	\vspace{0.01cm} 
     \label{table:appe_k_100_nc_100_bias_MSE_bin_ch2}
\end{table}

\clearpage
\begin{table}[!h]
	\caption{Simulation results of estimated bias and mean square error (MSE) for both counterfactual mean on treatment and ATE based on 1000 simulations. Setting: Binary outcome with a misspecified outcome model, $k$=100, $n_c$(Avg=5, Min=1, Max=24). Results are based on weighting centers equally.}
	\begin{center}
		\resizebox{\textwidth}{!}{
			\begin{tabular}{ccccccccccccc}
				\toprule
				& \multicolumn{12}{c}{\textbf{Counterfactual mean on treatment}}\\
				\cmidrule(lr){2-13} 
                & \multicolumn{6}{c}{\textbf{Unadjusted}} &\multicolumn{6}{c}{\textbf{Adjusted}}\\
				\cmidrule(lr){2-7} \cmidrule(lr){8-13} 
				Measure &  Na\"ive & Fixed & Mixed(1$\mid$c) & Mixed(1$\mid$c) Sam. & Mixed(1+A$\mid$c) & Mixed(1+A$\mid$c) Sam. & Na\"ive  & Fixed  & Mixed(1$\mid$c) & Mixed(1$\mid$c) Sam. & Mixed(1+A$\mid$c) & Mixed(1+A$\mid$c) Sam. \\
				\midrule
				\multicolumn{12}{c}{\textit{Random-effects variances}: $\sigma^2_{b0}$=0.5, $\sigma^2_{b1}$=0.5, $\sigma^2_{b2}$=0}\\
                Bias & 0.0018 & -0.0016 & 0.0012 & 0.0005 & 0.0013 & 0.0005 & 0.0018 & -0.0017 & 0.0012 & 0.0007 & 0.0014 & 0.0007 \\ 
                MSE  & 0.0014 & 0.0015 & 0.0015 & 0.0017 & 0.0017 & 0.0017 & 0.0014 & 0.0015 & 0.0015 & 0.0016 & 0.0016 & 0.0016 \\ 
				\midrule
                & \multicolumn{12}{c}{\textbf{ATE}}\\
				\cmidrule(lr){2-13} 
                & \multicolumn{6}{c}{\textbf{Unadjusted}} &\multicolumn{6}{c}{\textbf{Adjusted}}\\
				\cmidrule(lr){2-7} \cmidrule(lr){8-13} 
				Measure &  Na\"ive & Fixed & Mixed(1$\mid$c) & Mixed(1$\mid$c) Sam. & Mixed(1+A$\mid$c) & Mixed(1+A$\mid$c) Sam. & Na\"ive  & Fixed  & Mixed(1$\mid$c) & Mixed(1$\mid$c) Sam. & Mixed(1+A$\mid$c) & Mixed(1+A$\mid$c) Sam. \\ 
				\midrule
				\multicolumn{12}{c}{\textit{Random-effects variances}: $\sigma^2_{b0}$=0.5, $\sigma^2_{b1}$=0.5, $\sigma^2_{b2}$=0}\\
                Bias & 0.0009 & -0.0052 & 0.0008 & -0.0003 & 0.0009 & -0.0003 & 0.0011 & -0.0052 & 0.0009 & 0.0000 & 0.0011 & 0.0001 \\ 
               MSE  & 0.0022 & 0.0022 & 0.0022 & 0.0030 & 0.0030 & 0.0030 & 0.0020 & 0.0020 & 0.0020 & 0.0027 & 0.0026 & 0.0027 \\ 
				\midrule
			\end{tabular}
		}
	\end{center}
    	\tiny
	\vspace{0.01cm} 
     \label{table:appe_k_100_nc_5_mis_bias_MSE_bin_ch2}
\end{table}

\autoref{table:appe_k_100_nc_5_bin_weighing_by_nc_ch2} presents Monte Carlo standard deviation, average standard errors and coverage probabilities of 95\% confidence intervals for both counterfactual mean on treatment and ATE based on weighting patients equally under Setting 1 with binary outcomes.

\clearpage
\begin{table}
	\caption{Simulation results: Monte Carlo standard deviation, average standard errors and coverage probabilities of 95\% confidence intervals for both counterfactual mean on treatment and ATE based on 1000 simulations. Setting: Continuous outcome, $k$=100, $n_c$: Avg=5, Min=1, Max=24, and different values of random-effects variance. Results are based on weighting patients equally.}
	\begin{center}
		\resizebox{\textwidth}{!}{
			\begin{tabular}{>{\raggedright\arraybackslash}p{3.5cm}cccccccccccccccccc}
				\toprule
				& \multicolumn{9}{c}{\textbf{Counterfactual mean on treatment}} &\multicolumn{9}{c}{\textbf{ATE}} \\
				\cmidrule(lr){2-10} \cmidrule(lr){11-19}
				&& \multicolumn{4}{c}{\textbf{SE}} & \multicolumn{4}{c}{\textbf{Coverage (\%)}} &&\multicolumn{4}{c}{\textbf{SE}}& \multicolumn{4}{c}{\textbf{Coverage (\%)}} \\
				\cmidrule(lr){3-6} \cmidrule(lr){7-10} \cmidrule(lr){12-15} \cmidrule(lr){16-19}
				Method &  SD & Na\"ive & REML & DL & DB & Na\"ive & REML & DL & DB  & SD & Na\"ive & REML & DL & DB & Naive & REML & DL & DB \\
				\midrule
				&&&& \multicolumn{9}{c}{\textit{Random-effects variances}: $\sigma^2_{b0}$=0, $\sigma^2_{b1}$=0, $\sigma^2_{b2}$=0: $\bar{\rho}_1=0.0008$, $\bar{\rho}=-0.0002$} &&&&&&\\
 \multicolumn{19}{l}{\textbf{Unadjusted}}\\
 Na\"ive & 0.072 & 0.070 &  &  &  & 94.5 &  &  &  & 0.102 & 0.099 &  &  &  & 94.8 &  &  &  \\ 
  Fixed & 0.076 &  & 0.069 & 0.068 & 0.067 &  & 92.1 & 92.0 & 91.4 & 0.113 &  & 0.091 & 0.090 & 0.089 &  & 89.4 & 88.8 & 88.1 \\ 
  Mixed(1$\mid$c) & 0.072 &  & 0.073 & 0.073 & 0.076 &  & 95.8 & 95.4 & 96.1 & 0.102 &  & 0.102 & 0.103 & 0.106 &  & 95.4 & 95.3 & 96.0 \\ 
  Mixed(1$\mid$c) Sam. & 0.074 &  & 0.072 & 0.072 & 0.076 &  & 94.1 & 94.2 & 94.9 & 0.103 &  & 0.101 & 0.102 & 0.105 &  & 94.6 & 94.6 & 94.6 \\ 
  Mixed(1+A$\mid$c) & 0.072 &  & 0.073 & 0.073 & 0.076 &  & 95.6 & 95.3 & 96.1 & 0.102 &  & 0.102 & 0.103 & 0.106 &  & 95.4 & 95.4 & 96.0 \\ 
  Mixed(1+A$\mid$c) Sam. & 0.074 &  & 0.072 & 0.072 & 0.076 &  & 94.1 & 94.1 & 94.9 & 0.103 &  & 0.102 & 0.102 & 0.105 &  & 94.7 & 94.7 & 94.7 \\ 
  \multicolumn{19}{l}{\textbf{Adjusted}}\\
  Na\"ive & 0.068 & 0.067 &  &  &  & 93.7 &  &  &  & 0.092 & 0.089 &  &  &  & 94.2 &  &  &  \\ 
  Fixed & 0.071 &  & 0.066 & 0.065 & 0.065 &  & 92.9 & 92.8 & 92.3 & 0.102 &  & 0.082 & 0.080 & 0.079 &  & 88.9 & 88.1 & 87.9 \\ 
  Mixed(1$\mid$c) & 0.068 &  & 0.069 & 0.069 & 0.072 &  & 94.8 & 95.0 & 95.8 & 0.092 &  & 0.092 & 0.092 & 0.095 &  & 95.0 & 95.3 & 95.7 \\ 
  Mixed(1$\mid$c) Sam. & 0.069 &  & 0.069 & 0.069 & 0.072 &  & 94.9 & 95.0 & 95.5 & 0.091 &  & 0.091 & 0.092 & 0.094 &  & 95.3 & 95.2 & 95.5 \\
  Mixed(1+A$\mid$c) & 0.068 &  & 0.069 & 0.069 & 0.072 &  & 94.8 & 94.9 & 95.8 & 0.092 &  & 0.092 & 0.092 & 0.095 &  & 95.0 & 95.2 & 95.7 \\
  Mixed(1+A$\mid$c) Sam. & 0.069 &  & 0.069 & 0.069 & 0.072 &  & 94.9 & 95.0 & 95.6 & 0.091 &  & 0.091 & 0.092 & 0.094 &  & 95.4 & 95.2 & 95.6 \\ 
\bottomrule						
\multicolumn{19}{c}{\textit{Random-effects variances}: $\sigma^2_{b0}$=0.15, $\sigma^2_{b1}$=0, $\sigma^2_{b2}$=0: $\bar{\rho}_1=0.0603$, $\bar{\rho}= -0.0007$}\\
 \multicolumn{19}{l}{\textbf{Unadjusted}}\\
Na\"ive & 0.086 & 0.074 &  &  &  & 90.8 &  &  &  & 0.106 & 0.105 &  &  &  & 95.1 &  &  &  \\ 
  Fixed & 0.089 &  & 0.080 & 0.079 & 0.073 &  & 92.4 & 91.4 & 88.5 & 0.114 &  & 0.092 & 0.090 & 0.089 &  & 87.9 & 87.2 & 87.2 \\ 
   Mixed(1$\mid$c) & 0.086 &  & 0.086 & 0.086 & 0.087 &  & 94.4 & 94.6 & 93.7 & 0.104 &  & 0.109 & 0.109 & 0.113 &  & 95.6 & 95.9 & 96.7 \\ 
   Mixed(1$\mid$c) Sam. & 0.087 &  & 0.085 & 0.085 & 0.086 &  & 94.2 & 94.1 & 92.9 & 0.105 &  & 0.107 & 0.108 & 0.111 &  & 95.1 & 95.5 & 95.5 \\ 
  Mixed(1+A$\mid$c) & 0.086 &  & 0.086 & 0.086 & 0.087 &  & 94.4 & 94.7 & 94.0 & 0.104 &  & 0.109 & 0.109 & 0.113 &  & 95.8 & 96.0 & 96.7 \\ 
  Mixed(1+A$\mid$c) Sam. & 0.087 &  & 0.085 & 0.085 & 0.086 &  & 94.2 & 94.1 & 93.0 & 0.105 &  & 0.107 & 0.108 & 0.111 &  & 95.0 & 95.4 & 95.5 \\ 
   \multicolumn{19}{l}{\textbf{Adjusted}}\\
  Na\"ive & 0.083 & 0.071 &  &  &  & 90.8 &  &  &  & 0.097 & 0.095 &  &  &  & 94.8 &  &  &  \\ 
  Fixed & 0.084 &  & 0.078 & 0.076 & 0.072 &  & 92.3 & 91.9 & 90.1 & 0.102 &  & 0.082 & 0.080 & 0.079 &  & 88.8 & 88.0 & 87.3 \\ 
   Mixed(1$\mid$c) & 0.082 &  & 0.082 & 0.083 & 0.083 &  & 94.5 & 94.6 & 94.3 & 0.095 &  & 0.099 & 0.099 & 0.102 &  & 95.7 & 95.8 & 96.5 \\ 
   Mixed(1$\mid$c) Sam. & 0.083 &  & 0.082 & 0.082 & 0.082 &  & 94.3 & 94.5 & 93.4 & 0.094 &  & 0.098 & 0.098 & 0.102 &  & 95.7 & 95.7 & 95.9 \\ 
  Mixed(1+A$\mid$c) & 0.082 &  & 0.082 & 0.083 & 0.083 &  & 94.6 & 94.8 & 94.4 & 0.095 &  & 0.099 & 0.099 & 0.102 &  & 95.8 & 95.9 & 96.4 \\  
  Mixed(1+A$\mid$c) Sam. & 0.083 &  & 0.082 & 0.082 & 0.083 &  & 94.3 & 94.4 & 93.4 & 0.094 &  & 0.098 & 0.098 & 0.102 &  & 95.6 & 95.7 & 96.0 \\
				\bottomrule				
\multicolumn{19}{c}{\textit{Random-effects variances}: $\sigma^2_{b0}$=0.15, $\sigma^2_{b1}$=0.15, $\sigma^2_{b2}$=0: $\bar{\rho}_1=0.1155$, $\bar{\rho}= 0.0331$}\\
 \multicolumn{19}{l}{\textbf{Unadjusted}}\\
Na\"ive & 0.096 & 0.078 &  &  &  & 89.5 &  &  &  & 0.114 & 0.108 &  &  &  & 93.2 &  &  &  \\ 
  Fixed & 0.096 &  & 0.090 & 0.089 & 0.081 &  & 92.9 & 92.7 & 89.9 & 0.116 &  & 0.097 & 0.093 & 0.089 &  & 89.7 & 87.9 & 86.3 \\ 
  Mixed(1$\mid$c) & 0.095 &  & 0.098 & 0.097 & 0.097 &  & 94.8 & 94.9 & 94.3 & 0.110 &  & 0.116 & 0.117 & 0.118 &  & 96.2 & 96.0 & 96.0 \\ 
  Mixed(1$\mid$c) Sam. & 0.097 &  & 0.097 & 0.097 & 0.096 &  & 95.1 & 94.9 & 93.5 & 0.112 &  & 0.115 & 0.115 & 0.118 &  & 95.1 & 95.3 & 95.7 \\ 
  Mixed(1+A$\mid$c) & 0.094 &  & 0.098 & 0.097 & 0.097 &  & 95.0 & 95.0 & 94.7 & 0.110 &  & 0.116 & 0.117 & 0.118 &  & 96.2 & 96.0 & 96.3 \\ 
  Mixed(1+A$\mid$c) Sam. & 0.096 &  & 0.097 & 0.097 & 0.096 &  & 95.1 & 94.9 & 93.4 & 0.112 &  & 0.115 & 0.115 & 0.118 &  & 95.3 & 95.3 & 95.7 \\ 
  \multicolumn{19}{l}{\textbf{Adjusted}}\\
  Na\"ive & 0.095 & 0.075 &  &  &  & 88.4 &  &  &  & 0.108 & 0.098 &  &  &  & 92.4 &  &  &  \\ 
  Fixed & 0.094 &  & 0.088 & 0.087 & 0.080 &  & 92.9 & 92.5 & 90.6 & 0.111 &  & 0.088 & 0.084 & 0.080 &  & 88.0 & 86.6 & 84.4 \\ 
  Mixed(1$\mid$c) & 0.093 &  & 0.095 & 0.095 & 0.094 &  & 95.6 & 95.6 & 94.6 & 0.104 &  & 0.107 & 0.107 & 0.109 &  & 95.4 & 95.5 & 95.7 \\ 
  Mixed(1$\mid$c) Sam. & 0.096 &  & 0.095 & 0.095 & 0.093 &  & 95.2 & 94.8 & 93.8 & 0.106 &  & 0.107 & 0.107 & 0.108 &  & 94.4 & 94.3 & 94.6 \\ 
  Mixed(1+A$\mid$c) & 0.093 &  & 0.095 & 0.095 & 0.094 &  & 95.7 & 95.5 & 94.4 & 0.104 &  & 0.107 & 0.107 & 0.109 &  & 95.8 & 95.9 & 95.9 \\
  Mixed(1+A$\mid$c) Sam. & 0.096 &  & 0.095 & 0.095 & 0.094 &  & 95.1 & 94.8 & 93.7 & 0.106 &  & 0.107 & 0.107 & 0.108 &  & 94.5 & 94.3 & 94.6 \\
				\bottomrule	
\multicolumn{19}{c}{\textit{Random-effects variances}: $\sigma^2_{b0}$=0.15, $\sigma^2_{b1}$=0.15, $\sigma^2_{b2}=4\times10^{-6}$: $\bar{\rho}_1=0.2190$, $\bar{\rho}=0.0280$}\\
 \multicolumn{19}{l}{\textbf{Unadjusted}}\\
Na\"ive & 0.132 & 0.091 &  &  &  & 81.8 &  &  &  & 0.137 & 0.127 &  &  &  & 93.6 &  &  &  \\ 
  Fixed & 0.129 &  & 0.123 & 0.123 & 0.116 &  & 94.2 & 94.1 & 91.7 & 0.126 &  & 0.099 & 0.095 & 0.091 &  & 88.9 & 87.2 & 85.5 \\ 
  Mixed(1$\mid$c) & 0.128 &  & 0.134 & 0.133 & 0.132 &  & 95.3 & 95.4 & 94.9 & 0.121 &  & 0.136 & 0.136 & 0.139 &  & 96.4 & 96.6 & 96.8 \\ 
  Mixed(1$\mid$c) Sam. & 0.132 &  & 0.131 & 0.131 & 0.130 &  & 95.1 & 95.3 & 93.8 & 0.135 &  & 0.133 & 0.133 & 0.136 &  & 94.3 & 94.4 & 95.0 \\ 
  Mixed(1+A$\mid$c) & 0.128 &  & 0.134 & 0.133 & 0.132 &  & 95.5 & 95.5 & 95.0 & 0.121 &  & 0.136 & 0.136 & 0.139 &  & 96.6 & 96.6 & 96.9 \\ 
  Mixed(1+A$\mid$c) Sam. & 0.132 &  & 0.131 & 0.131 & 0.130 &  & 95.1 & 95.1 & 93.5 & 0.135 &  & 0.133 & 0.133 & 0.136 &  & 94.4 & 94.6 & 95.0 \\ 
  \multicolumn{19}{l}{\textbf{Adjusted}}\\
  Na\"ive & 0.131 & 0.088 &  &  &  & 81.0 &  &  &  & 0.130 & 0.119 &  &  &  & 92.9 &  &  &  \\ 
  Fixed & 0.128 &  & 0.121 & 0.121 & 0.115 &  & 93.3 & 93.4 & 91.4 & 0.116 &  & 0.090 & 0.086 & 0.082 &  & 87.1 & 85.3 & 83.1 \\ 
  Mixed(1$\mid$c) & 0.127 &  & 0.132 & 0.131 & 0.130 &  & 95.5 & 95.4 & 94.8 & 0.112 &  & 0.128 & 0.128 & 0.131 &  & 96.6 & 96.6 & 96.9 \\ 
   Mixed(1$\mid$c) Sam. & 0.131 &  & 0.130 & 0.129 & 0.128 &  & 94.9 & 95.2 & 94.3 & 0.129 &  & 0.125 & 0.126 & 0.129 &  & 93.6 & 93.9 & 94.2 \\ 
  Mixed(1+A$\mid$c) & 0.126 &  & 0.132 & 0.131 & 0.130 &  & 95.4 & 95.4 & 94.8 & 0.112 &  & 0.128 & 0.128 & 0.131 &  & 96.6 & 96.8 & 96.8 \\
  Mixed(1+A$\mid$c) Sam. & 0.131 &  & 0.130 & 0.129 & 0.128 &  & 95.0 & 95.2 & 94.3 & 0.129 &  & 0.125 & 0.126 & 0.129 &  & 93.7 & 93.9 & 94.2 \\ 
				\bottomrule	
			\end{tabular}%
		}
	\end{center}
	\tiny
	\vspace{0.01cm} 
	$\bar{\rho}_1$ and $\bar{\rho}$ represent the mean values of intra-class correlation in the influence function of the counterfactual mean and the treatment effect, respectively, across 1000 simulations.
  \label{table:appe_k_100_nc_5_cont_weighing_by_nc_ch2}
\end{table}

\clearpage
\begin{table}
	\caption{Simulation results: Monte Carlo standard deviation, average standard errors and coverage probabilities of 95\% confidence intervals for both counterfactual mean on treatment and ATE based on 1000 simulations. Setting: Binary outcome, $k$=100, $n_c$: Avg=5, Min=1, Max=24, and at different values of random-effects variance. Results are based on weighting patients equally.}
	\begin{center}
		\resizebox{\textwidth}{!}{
			\begin{tabular}{>{\raggedright\arraybackslash}p{3.5cm}cccccccccccccccccc}
				\toprule
				& \multicolumn{9}{c}{\textbf{Counterfactual mean on treatment}} &\multicolumn{9}{c}{\textbf{ATE}} \\
				\cmidrule(lr){2-10} \cmidrule(lr){11-19}
				&& \multicolumn{4}{c}{\textbf{SE}} & \multicolumn{4}{c}{\textbf{Coverage (\%)}} &&\multicolumn{4}{c}{\textbf{SE}}&\multicolumn{4}{c}{\textbf{Coverage (\%)}} \\
				\cmidrule(lr){3-6} \cmidrule(lr){7-10} \cmidrule(lr){12-15} \cmidrule(lr){16-19}
				Method &  SD & Na\"ive & REML & DL & DB & Na\"ive & REML & DL & DB  & SD & Na\"ive & REML & DL & DB & Naive & REML & DL & DB \\ \midrule
				&&&&& \multicolumn{9}{c}{\textit{Random-effects variances}: $\sigma^2_{b0}$=0, $\sigma^2_{b1}$=0, $\sigma^2_{b2}$=0: $\bar{\rho_1}=0.0007$, $\bar{\rho}=0.0000$} &&&&&\\
            \multicolumn{19}{l}{\textbf{Unadjusted}}\\
Na\"ive & 0.032 & 0.031 &  &  &  & 95.0 &  &  &  & 0.045 & 0.044 &  &  &  & 93.9 &  &  &  \\ 
  Fixed & 0.033 &  & 0.031 & 0.030 & 0.030 &  & 92.8 & 92.6 & 91.9 & 0.048 &  & 0.041 & 0.040 & 0.039 &  & 91.2 & 90.3 & 90.1 \\ 
  Mixed(1$\mid$c) & 0.032 &  & 0.033 & 0.033 & 0.033 &  & 95.8 & 95.6 & 95.8 & 0.045 &  & 0.046 & 0.046 & 0.046 &  & 94.9 & 94.9 & 95.2 \\ 
  Mixed(1$\mid$c) Sam. & 0.032 &  & 0.032 & 0.032 & 0.033 &  & 94.2 & 94.3 & 94.5 & 0.045 &  & 0.046 & 0.046 & 0.046 &  & 95.5 & 95.5 & 95.3 \\ 
  Mixed(1+A$\mid$c) & 0.032 &  & 0.033 & 0.033 & 0.033 &  & 95.7 & 95.8 & 95.9 & 0.045 &  & 0.046 & 0.046 & 0.046 &  & 94.9 & 94.9 & 95.2 \\ 
  Mixed(1+A$\mid$c) Sam. & 0.032 &  & 0.032 & 0.032 & 0.033 &  & 94.1 & 94.3 & 94.5 & 0.045 &  & 0.046 & 0.046 & 0.046 &  & 95.6 & 95.5 & 95.4 \\
  \multicolumn{19}{l}{\textbf{Adjusted}}\\
  Na\"ive  & 0.031 & 0.030 &  &  &  & 94.8 &  &  &  & 0.043 & 0.041 &  &  &  & 93.8 &  &  &  \\ 
  Fixed  & 0.032 &  & 0.030 & 0.030 & 0.029 &  & 93.1 & 92.9 & 92.1 & 0.046 &  & 0.038 & 0.037 & 0.036 &  & 90.0 & 88.8 & 88.3 \\ 
  Mixed(1$\mid$c) & 0.031 &  & 0.031 & 0.032 & 0.032 &  & 95.5 & 95.5 & 95.6 & 0.043 &  & 0.042 & 0.043 & 0.043 &  & 94.7 & 94.9 & 95.2 \\ 
  Mixed(1$\mid$c) Sam. & 0.031 &  & 0.031 & 0.031 & 0.032 &  & 94.3 & 94.5 & 95.0 & 0.042 &  & 0.042 & 0.042 & 0.043 &  & 96.0 & 96.1 & 96.2 \\ 
  Mixed(1+A$\mid$c)  & 0.031 &  & 0.031 & 0.032 & 0.032 &  & 95.7 & 95.8 & 95.9 & 0.043 &  & 0.043 & 0.043 & 0.043 &  & 94.9 & 94.9 & 95.2 \\ 
  Mixed(1+A$\mid$c) Sam. & 0.031 &  & 0.031 & 0.031 & 0.032 &  & 94.7 & 94.6 & 95.1 & 0.042 &  & 0.042 & 0.043 & 0.043 &  & 96.0 & 96.1 & 96.2 \\ 
\bottomrule
&&&&& \multicolumn{9}{c}{\textit{Random-effects variances}: $\sigma^2_{b0}$=0.5, $\sigma^2_{b1}$=0, $\sigma^2_{b2}$=0: $\bar{\rho}_1=0.04228$, $\bar{\rho}= -0.0010$} &&&&&\\
\multicolumn{19}{l}{\textbf{Unadjusted}}\\
Na\"ive & 0.036 & 0.031 &  &  &  & 92.5 &  &  &  & 0.045 & 0.044 &  &  &  & 94.6 &  &  &  \\ 
  Fixed & 0.036 &  & 0.033 & 0.033 & 0.030 &  & 92.5 & 92.0 & 89.6 & 0.046 &  & 0.039 & 0.038 & 0.038 &  & 91.3 & 90.3 & 89.9 \\ 
  Mixed(1$\mid$c) & 0.036 &  & 0.035 & 0.035 & 0.035 &  & 94.9 & 94.6 & 94.3 & 0.045 &  & 0.046 & 0.046 & 0.046 &  & 96.1 & 96.1 & 96.1 \\ 
  Mixed(1$\mid$c) Sam. & 0.036 &  & 0.035 & 0.035 & 0.035 &  & 94.2 & 94.2 & 94.1 & 0.045 &  & 0.045 & 0.045 & 0.046 &  & 96.0 & 96.2 & 96.4 \\ 
  Mixed(1+A$\mid$c) & 0.036 &  & 0.035 & 0.035 & 0.035 &  & 94.7 & 94.5 & 94.5 & 0.045 &  & 0.046 & 0.046 & 0.046 &  & 96.2 & 96.1 & 96.1 \\ 
  Mixed(1+A$\mid$c) Sam. & 0.036 &  & 0.035 & 0.035 & 0.035 &  & 94.2 & 94.3 & 94.1 & 0.045 &  & 0.045 & 0.045 & 0.046 &  & 96.0 & 96.2 & 96.4 \\
  \multicolumn{19}{l}{\textbf{Adjusted}}\\
  Na\"ive  & 0.035 & 0.031 &  &  &  & 92.6 &  &  &  & 0.042 & 0.042 &  &  &  & 94.2 &  &  &  \\ 
  Fixed  & 0.035 &  & 0.032 & 0.032 & 0.030 &  & 93.0 & 92.0 & 89.5 & 0.043 &  & 0.036 & 0.035 & 0.035 &  & 89.3 & 88.5 & 88.2 \\ 
  Mixed(1$\mid$c)  & 0.035 &  & 0.034 & 0.034 & 0.035 &  & 94.9 & 94.9 & 94.9 & 0.042 &  & 0.043 & 0.043 & 0.044 &  & 95.0 & 95.1 & 95.4 \\ 
  Mixed(1$\mid$c) Sam. & 0.035 &  & 0.034 & 0.034 & 0.034 &  & 94.8 & 94.5 & 94.4 & 0.043 &  & 0.043 & 0.043 & 0.044 &  & 94.9 & 95.0 & 95.7 \\ 
  Mixed(1+A$\mid$c)  & 0.035 &  & 0.034 & 0.034 & 0.035 &  & 94.7 & 94.5 & 94.5 & 0.042 &  & 0.043 & 0.043 & 0.044 &  & 96.2 & 96.1 & 96.1\\ 
  Mixed(1+A$\mid$c) Sam. & 0.035 &  & 0.034 & 0.034 & 0.034 &  & 94.7 & 94.3 & 94.4 & 0.043 &  & 0.043 & 0.043 & 0.044 &  & 95.0 & 95.0 & 95.6 \\ 
				\bottomrule
&&&&& \multicolumn{9}{c}{\textit{Random-effects variances}: $\sigma^2_{b0}$=0.5, $\sigma^2_{b1}$=0.5, $\sigma^2_{b2}$=0: $\bar{\rho}_1=0.0796$, $\bar{\rho}=0.0214$} &&&&&\\
\multicolumn{19}{l}{\textbf{Unadjusted}}\\
  Na\"ive & 0.037 & 0.031 &  &  &  & 90.7 &  &  &  & 0.048 & 0.044 &  &  &  & 92.8 &  &  &  \\ 
  Fixed & 0.037 &  & 0.035 & 0.034 & 0.032 &  & 92.2 & 91.8 & 89.4 & 0.050 &  & 0.040 & 0.039 & 0.037 &  & 89.6 & 88.5 & 87.0 \\ 
  Mixed(1$\mid$c) & 0.037 &  & 0.038 & 0.038 & 0.037 &  & 95.2 & 95.2 & 94.8 & 0.048 &  & 0.047 & 0.047 & 0.048 &  & 94.4 & 94.5 & 94.7 \\ 
  Mixed(1$\mid$c) Sam. & 0.037 &  & 0.037 & 0.037 & 0.037 &  & 94.1 & 94.2 & 94.1 & 0.047 &  & 0.047 & 0.047 & 0.047 &  & 94.5 & 94.7 & 95.0 \\ 
  Mixed(1+A$\mid$c) & 0.037 &  & 0.038 & 0.038 & 0.037 &  & 95.1 & 95.2 & 94.8 & 0.048 &  & 0.047 & 0.047 & 0.048 &  & 94.6 & 94.6 & 94.3 \\ 
  Mixed(1+A$\mid$c) Sam. & 0.037 &  & 0.037 & 0.037 & 0.037 &  & 94.2 & 94.2 & 94.1 & 0.047 &  & 0.047 & 0.047 & 0.047 &  & 94.5 & 94.7 & 95.0 \\ 
 \multicolumn{19}{l}{\textbf{Adjusted}}\\
  Na\"ive  & 0.036 & 0.031 &  &  &  & 90.7 &  &  &  & 0.046 & 0.042 &  &  &  & 92.1 &  &  &  \\ 
  Fixed  & 0.037 &  & 0.034 & 0.034 & 0.031 &  & 92.2 & 91.6 & 89.6 & 0.048 &  & 0.037 & 0.036 & 0.034 &  & 87.9 & 86.5 & 85.3 \\ 
  Mixed(1$\mid$c)  & 0.036 &  & 0.037 & 0.037 & 0.037 &  & 94.9 & 94.7 & 94.7 & 0.046 &  & 0.045 & 0.045 & 0.045 &  & 94.0 & 93.9 & 94.3 \\ 
  Mixed(1$\mid$c) Sam. & 0.036 &  & 0.037 & 0.037 & 0.037 &  & 94.6 & 94.8 & 94.8 & 0.044 &  & 0.044 & 0.045 & 0.045 &  & 94.8 & 95.0 & 95.0 \\ 
  Mixed(1+A$\mid$c)  & 0.036 &  & 0.037 & 0.037 & 0.037 &  & 95.1 & 95.2 & 94.8 & 0.046 &  & 0.045 & 0.045 & 0.045 &  & 94.6 & 94.6 & 94.3 \\ 
  Mixed(1+A$\mid$c) Sam. & 0.036 &  & 0.037 & 0.037 & 0.037 &  & 94.6 & 94.9 & 94.9 & 0.044 &  & 0.044 & 0.045 & 0.045 &  & 95.0 & 95.0 & 95.2 \\ 
\bottomrule		
&&&&& \multicolumn{9}{c}{\textit{Random-effects variances}: $\sigma^2_{b0}$=0.75, $\sigma^2_{b1}$=0.5, $\sigma^2_{b2}$=0.5: $\bar{\rho}_1=0.0974$, $\bar{\rho}=0.0195$} &&&&&\\
\multicolumn{19}{l}{\textbf{Unadjusted}}\\
  Na\"ive & 0.039 & 0.031 &  &  &  & 88.0 &  &  &  & 0.047 & 0.044 &  &  &  & 93.5 &  &  &  \\ 
  Fixed & 0.039 &  & 0.036 & 0.035 & 0.032 &  & 93.1 & 92.6 & 89.6 & 0.047 &  & 0.039 & 0.038 & 0.037 &  & 89.9 & 89.0 & 88.2 \\ 
  Mixed(1$\mid$c) & 0.039 &  & 0.039 & 0.039 & 0.038 &  & 94.9 & 95.1 & 94.9 & 0.046 &  & 0.047 & 0.047 & 0.048 &  & 95.5 & 95.5 & 95.5 \\ 
  Mixed(1$\mid$c) Sam. & 0.038 &  & 0.038 & 0.038 & 0.038 &  & 94.0 & 94.0 & 93.4 & 0.046 &  & 0.047 & 0.047 & 0.047 &  & 95.4 & 94.9 & 95.2 \\ 
  Mixed(1+A$\mid$c) & 0.039 &  & 0.039 & 0.039 & 0.038 &  & 94.8 & 95.0 & 95.0 & 0.046 &  & 0.047 & 0.047 & 0.048 &  & 95.4 & 95.5 & 95.6 \\ 
  Mixed(1+A$\mid$c) Sam. & 0.038 &  & 0.038 & 0.038 & 0.038 &  & 94.1 & 94.0 & 93.4 & 0.046 &  & 0.047 & 0.047 & 0.047 &  & 95.5 & 95.1 & 95.4 \\ 
 \multicolumn{19}{l}{\textbf{Adjusted}}\\
  Na\"ive & 0.039 & 0.031 &  &  &  & 87.7 &  &  &  & 0.045 & 0.042 &  &  &  & 92.9 &  &  &  \\ 
  Fixed  & 0.039 &  & 0.035 & 0.035 & 0.032 &  & 91.9 & 91.0 & 89.0 & 0.046 &  & 0.037 & 0.035 & 0.034 &  & 88.7 & 87.7 & 86.6 \\ 
  Mixed(1$\mid$c)  & 0.038 &  & 0.038 & 0.038 & 0.038 &  & 94.2 & 93.8 & 93.8 & 0.044 &  & 0.045 & 0.045 & 0.045 &  & 95.2 & 95.1 & 95.5 \\
  Mixed(1$\mid$c) Sam. & 0.038 &  & 0.038 & 0.038 & 0.038 &  & 93.7 & 93.8 & 93.8 & 0.045 &  & 0.044 & 0.045 & 0.045 &  & 94.2 & 94.4 & 94.7 \\ 
  Mixed(1+A$\mid$c)  & 0.039 &  & 0.038 & 0.038 & 0.038 &  & 94.8 & 95.0 & 95.0 & 0.044 &  & 0.045 & 0.045 & 0.045 &  & 95.4 & 95.5 & 95.6 \\ 
  Mixed(1+A$\mid$c) Sam. & 0.038 &  & 0.038 & 0.038 & 0.038 &  & 93.7 & 93.7 & 93.8 & 0.045 &  & 0.045 & 0.045 & 0.045 &  & 94.1 & 94.5 & 94.7 \\
				\bottomrule	
			\end{tabular}%
		}
	\end{center}
	\tiny
	\vspace{0.01cm} 
   \label{table:appe_k_100_nc_5_bin_weighing_by_nc_ch2}
\end{table}

\clearpage
\subsection{Additional data application results}\label{appe_data_anlysis_results_ch2}
\begin{figure}[!h]
    \centering
    \includegraphics[width=1\textwidth]{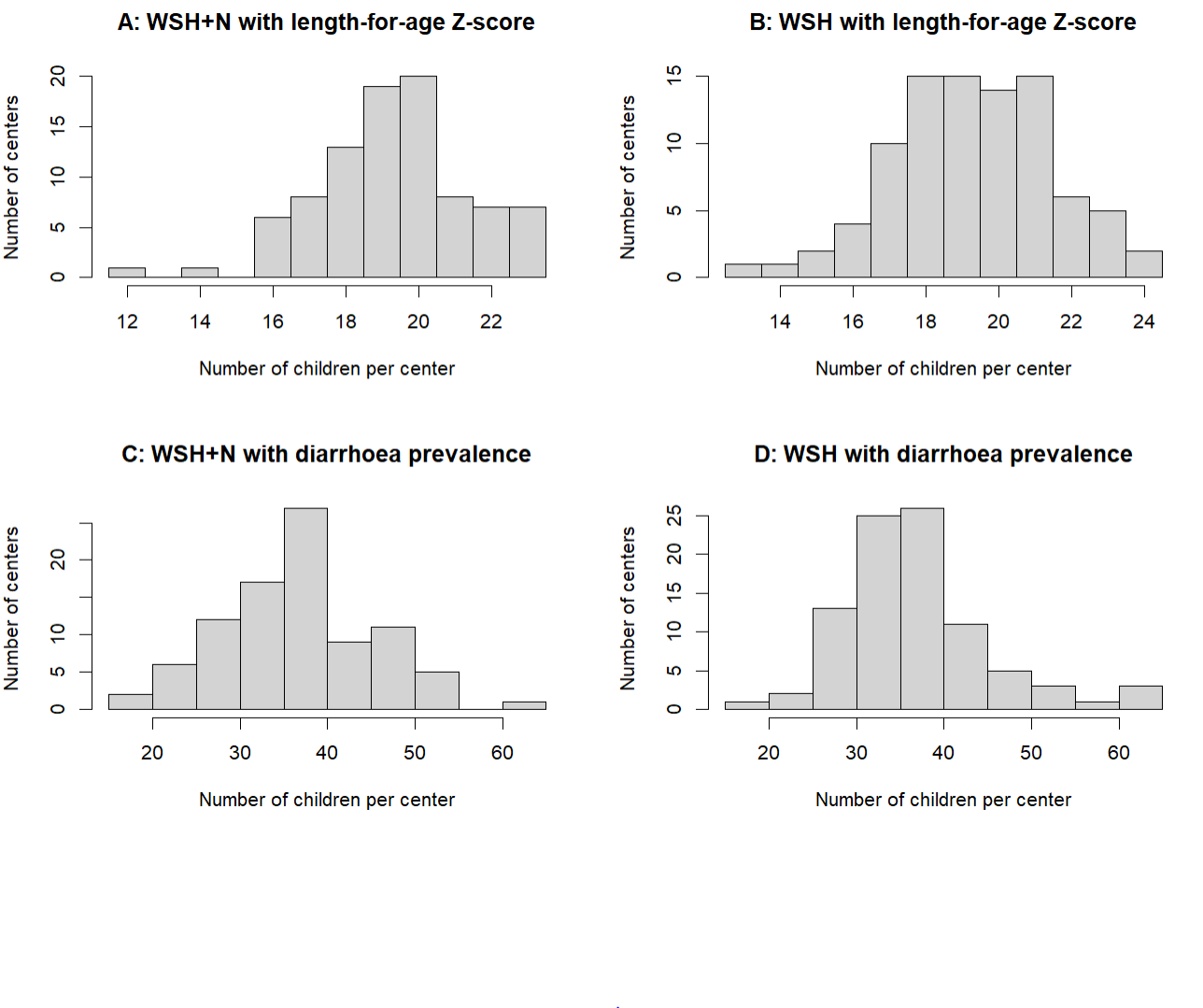}
    \caption{Number of children per center in each intervention group for each outcome. A: WSH intervention with length-for-age Z-score outcome; B: WSH+N intervention with length-for-age Z-score outcome; C: WSH intervention with diarrhoea prevalence outcome, and D: WSH+N intervention with diarrhoea prevalence outcome.}
       \label{figure:appe_DA_description_ch2}
\end{figure}

\autoref{figure:appe_DA_description_ch2} shows the distribution of the number of children per center across the intervention groups and outcomes considered in the data analysis, with all panels indicating substantial variation in center size.

\clearpage
\autoref{table:appe_WSH_N_CM_ch2} shows the counterfactual mean on treatment estimates and 95\% confidence intervals for the WSH+N intervention compared to the control group on length-for-age Z-score and the diarrhoea prevalence in the WASH Benefits Bangladesh study. 
\begin{table}[!h]
	\caption{The counterfactual mean on treatment estimates and 95\% confidence intervals for the WSH+N intervention compared to the control group on length-for-age Z-score (continuous outcome) and the diarrhoea prevalence (binary outcome) in the WASH Benefits Bangladesh study. Results are based on weighting centers equally.}
	\begin{center}
		\resizebox{\textwidth}{!}{
	\begin{tabular}{>{\raggedright\arraybackslash}p{3.5cm}ccccc}
				\toprule
				& \multicolumn{2}{c}{\textbf{Length-for-age Z-score (continuous outcome)}} &\multicolumn{2}{c}{\textbf{Diarrhoea prevalence (binary outcome)}} \\
				\cmidrule(lr){2-3} \cmidrule(lr){4-5}
				Method & Estimate & Confidence interval &  Estimate &  Confidence interval\\
				\midrule
				\multicolumn{5}{c}{\textit{The analysis is based on data from the WSH+N intervention and control arms}.}\\
                \multicolumn{5}{l}{\textbf{Unadjusted}}\\
				Na\"ive & -1.673 & (-1.755, -1.591) &  0.038 & (0.030, 0.046) \\ 
				Fixed & -1.660 & (-1.748, -1.572)  & 0.036 & (0.027, 0.045) \\ 
				Mixed(1$\mid$c:cl) & -1.665 & (-1.748, -1.582) & 0.037 & (0.028, 0.045) \\ 
                Mixed(1$\mid$c:cl) Sam. & -1.668 & (-1.774, -1.563) & 0.036 & (0.026, 0.046)  \\ 
				Mixed(1+A$\mid$c:cl) & -1.666 & (-1.749, -1.583) & 0.036 & (0.027, 0.045) \\ 
                Mixed(1+A$\mid$c:cl) Sam. & -1.669 & (-1.775, -1.563) & 0.036 & (0.025, 0.047) \\ 
                \multicolumn{5}{l}{\textbf{Adjusted}}\\
				Na\"ive & -1.665 & (-1.743, -1.586) &  0.038 & (0.030, 0.047)  \\ 
				Fixed  & -1.650 & (-1.737, -1.564)  & 0.037 & (0.027, 0.047) \\ 
				Mixed(1$\mid$c:cl) & -1.659 & (-1.741, -1.576) & 0.038 & (0.029, 0.046) \\ 
                Mixed(1$\mid$c:cl) Sam. & -1.661 & (-1.758, -1.563) & 0.037 & (0.026, 0.048) \\ 
				Mixed(1+A$\mid$c:cl) & -1.659 & (-1.747, -1.572) & 0.037 & (0.027, 0.047) \\ 
                Mixed(1+A$\mid$c:cl) Sam. & -1.660 & (-1.757, -1.563)  & 0.037 & (0.026, 0.048) \\ 
				\midrule
				\multicolumn{5}{c}{\textit{The analysis is based on data from all arms}.}\\
                \multicolumn{5}{l}{\textbf{Unadjusted}}\\
                Na\"ive & -1.673 & (-1.755, -1.591) & 0.037 & (0.029, 0.045) \\ 
				Fixed & -1.668 & (-1.757, -1.579) & 0.037 & (0.028, 0.045)\\
				Mixed(1$\mid$c:cl) & -1.671 & (-1.751, -1.591) & 0.037 & (0.029, 0.045)\\
                Mixed(1$\mid$c:cl) Sam. & -1.677 & (-1.77, -1.585) & 0.037 & (0.028, 0.046) \\ 
				Mixed(1+A$\mid$c:cl) & -1.673 & (-1.762, -1.584) & 0.037 & (0.028, 0.045)\\ 
                Mixed(1+A$\mid$c:cl) Sam. & -1.675 & (-1.768, -1.583) & 0.037 & (0.028, 0.046) \\ 
                \multicolumn{5}{l}{\textbf{Adjusted}}\\
				Na\"ive & -1.661 & (-1.739, -1.583) &  0.038 & (0.030, 0.046)   \\ 
				Fixed & -1.658 & (-1.744, -1.573) & 0.037 & (0.028, 0.046) \\ 
				Mixed(1$\mid$c:cl) & -1.661 & (-1.74, -1.583) &  0.038 & (0.029, 0.046)\\ 
                Mixed(1$\mid$c:cl) Sam. & -1.664 & (-1.752, -1.577) & 0.037 & (0.028, 0.047) \\ 
				Mixed(1+A$\mid$c:cl) & -1.662 & (-1.747, -1.576) & 0.037 & (0.028, 0.046) \\ 
                Mixed(1+A$\mid$c:cl) Sam. & -1.664 & (-1.752, -1.577) & 0.037 & (0.028, 0.047)\\ 
				\bottomrule						
			\end{tabular}%
		}
	\end{center}
	\tiny
	\vspace{0.01cm} 
	\label{table:appe_WSH_N_CM_ch2}
\end{table}

\end{document}